\documentclass[twocolumn]{aastex631}
\usepackage{mathtools, cuted}
\usepackage{amsmath}
\usepackage{lipsum, color}

\usepackage{multirow}
\usepackage{comment}

\shorttitle{A universal equation to predict $\Omega_{\rm m}$}
\shortauthors{Shao et al.}

\graphicspath{{./}{figures/}}

\begin{document}

\title{A universal equation to predict $\Omega_{\rm m}$ from halo and galaxy catalogues}

\correspondingauthor{Helen Shao}
\email{hshao@princeton.edu}

\author[0000-0002-0152-6747]{Helen Shao}
%\email{hshao@princeton.edu}
\affiliation{Department of Astrophysical Sciences, Princeton University, 4 Ivy Lane, Princeton, NJ 08544 USA}
\affiliation{Center for Computational Astrophysics, Flatiron Institute, 162 5th Avenue, New York, NY 10010, USA}

\author[0000-0002-4728-6881]{Natalí S. M. de Santi}
\affiliation{Center for Computational Astrophysics, Flatiron Institute, 162 5th Avenue, New York, NY, 10010, USA}
\affiliation{Instituto de Física, Universidade de São Paulo, R. do Matão 1371, 05508-900, São Paulo, Brasil}

\author[0000-0002-4816-0455]{Francisco Villaescusa-Navarro}
\affiliation{Center for Computational Astrophysics, Flatiron Institute, 162 5th Avenue, New York, NY, 10010, USA}
\affiliation{Department of Astrophysical Sciences, Princeton University, 4 Ivy Lane, Princeton, NJ 08544 USA}

\author[0000-0001-7689-0933]{Romain Teyssier}
\affiliation{Department of Astrophysical Sciences, Princeton University, 4 Ivy Lane, Princeton, NJ 08544 USA}

\author[0000-0001-7899-7195]{Yueying Ni}
\affiliation{Harvard-Smithsonian Center for Astrophysics, 60 Garden Street, Cambridge, MA 02138, US}
\affiliation{McWilliams Center for Cosmology, Department of Physics, Carnegie Mellon University, Pittsburgh, PA 15213, US}

\author[0000-0001-5769-4945]{Daniel Angl\'es-Alc\'azar}
\affiliation{Department of Physics, University of Connecticut, 196 Auditorium Road, U-3046, Storrs, CT, 06269, USA}
\affiliation{Center for Computational Astrophysics, Flatiron Institute, 162 5th Avenue, New York, NY 10010, USA}

\author[0000-0002-3185-1540]{Shy Genel}
\affiliation{Center for Computational Astrophysics, Flatiron Institute, 162 5th Avenue, New York, NY, 10010, USA}
\affiliation{Columbia Astrophysics Laboratory, Columbia University, New York, NY, 10027, USA}

\author[0000-0001-8867-5026]{Ulrich P. Steinwandel}
\affiliation{Center for Computational Astrophysics, Flatiron Institute, 162 5th Avenue, New York, NY 10010, USA}

\author[0000-0002-1329-9246]{Elena Hern\'andez-Mart\'inez}
\affiliation{Universit\"ats-Sternwarte, Fakult\"at f\"ur Physik, Ludwig-Maximilians-Universit\"at M\"unchen, Scheinerstr. 1, 81679 M\"unchen, Germany}

\author{Klaus Dolag}
\affiliation{Universit\"ats-Sternwarte, Fakult\"at f\"ur Physik, Ludwig-Maximilians-Universit\"at M\"unchen, Scheinerstr. 1, 81679 M\"unchen, Germany}
\affiliation{Max-Planck-Institut f\"ur Astrophysik, Karl-Schwarzschild-Stra{\ss}e 1, 85741 Garching, Germany}

\author[0000-0001-7964-5933]{Christopher C. Lovell}
\affiliation{Institute of Cosmology and Gravitation, University of Portsmouth, Burnaby Road, 
Portsmouth, PO1 3FX, UK}
\affiliation{Centre for Astrophysics Research, School of Physics, Engineering \& Computer 
Science, University of Hertfordshire, Hatfield AL10 9AB, UK}

\author[0000-0002-9853-5673]{Lehman H. Garrison}
\affiliation{Center for Computational Astrophysics, Flatiron Institute, 162 5th Avenue, New York, NY 10010, USA}

\author[0000-0002-8365-0337]{Eli Visbal}
\affiliation{Department of Physics and Astronomy and Ritter Astrophysical
Research Center, \\University of Toledo, 2801 W Bancroft Street, Toledo, OH 43606, USA}

\author[0000-0002-9789-6653]{Mihir Kulkarni}
\affiliation{Department of Physics and Astronomy and Ritter Astrophysical Research Center,
University of Toledo, 2801 W Bancroft Street, Toledo, OH 43606, USA}

\author{Lars Hernquist}
\affiliation{Harvard-Smithsonian Center for Astrophysics, 60 Garden Street, Cambridge, MA 02138, USA}

\author[0000-0002-6292-3228]{Tiago Castro}
\affiliation{INAF-Osservatorio Astronomico di Trieste, Via G. B. Tiepolo 11, I-34143 Trieste, Italy}
\affiliation{INFN, Sezione di Trieste, Via Valerio 2, I-34127 Trieste TS, Italy}
\affiliation{IFPU, Institute for Fundamental Physics of the Universe, via Beirut 2, 34151 Trieste, Italy}

\author{Mark Vogelsberger}
\affiliation{Kavli Institute for Astrophysics and Space Research, Department of Physics, MIT, Cambridge, MA 02139, USA}
\affiliation{The NSF AI Institute for Artificial Intelligence and Fundamental Interactions, Massachusetts Institute of Technology, Cambridge MA 02139, USA}

\begin{abstract}

We discover analytic equations that can infer the value of $\Omega_{\rm m}$ from the positions and velocity moduli of halo and galaxy catalogues. The equations are derived by combining a tailored graph neural network (GNN) architecture with symbolic regression. We first train the GNN on dark matter halos from Gadget N-body simulations to perform field-level likelihood-free inference, and show that our model can infer $\Omega_{\rm m}$ with $\sim6\%$ accuracy from halo catalogues of thousands of N-body simulations run with six different codes: Abacus, CUBEP$^3$M, Gadget, Enzo, PKDGrav3, and Ramses. By applying symbolic regression to the different parts comprising the GNN, we derive equations that can predict $\Omega_{\rm m}$ from halo catalogues of simulations run with all of the above codes with accuracies similar to those of the GNN. We show that by tuning a single free parameter, our equations can also infer the value of $\Omega_{\rm m}$ from galaxy catalogues of thousands of state-of-the-art hydrodynamic simulations of the CAMELS project, each with a different astrophysics model, run with five distinct codes that employ different subgrid physics: IllustrisTNG, SIMBA, Astrid, Magneticum, SWIFT-EAGLE. Furthermore, the equations also perform well when tested on galaxy catalogues from simulations covering a vast region in parameter space that samples variations in 5 cosmological and 23 astrophysical parameters. We speculate that the equations may reflect the existence of a fundamental physics relation between the phase-space distribution of generic tracers and $\Omega_{\rm m}$, one that is not affected by galaxy formation physics down to scales as small as $10~h^{-1}{\rm kpc}$.

\end{abstract}


\section{Introduction} 
\label{sec:intro}

$\Lambda$CDM is the current standard model in cosmology that describes the evolution and expansion of the Universe, where CDM denotes cold dark matter and $\Lambda$ represents the cosmological constant. This model explains how primordial density perturbations in the early Universe were amplified by gravity and eventually lead to the formation of the large-scale structures that we observe today. To accomplish this, the model relies on several cosmological parameters that characterize the composition and other fundamental properties of our Universe. One of them is $\Omega_{\rm m}$, which quantifies the fractional energy density of total matter, and obtaining an accurate constraint for it is crucial for improving our understanding of the foundational physics that governs the Universe. 

Historically, the statistics used to analyze the density and velocity fields of matter and galaxies have been useful probes for $\Omega_{\rm m}$ \citep{peebles1980, davis1985, Angulo_2022}. This includes the analysis of redshift-space distortions of galaxy redshift surveys caused by virial and peculiar velocities that deviate from cosmic expansion \citep{kaiser}. Such distortions strongly affect the statistical properties of galaxy clustering because they break the symmetry in the line-of-sight direction. These anisotropies directly probe the growth factor, that depends on $\Omega_{\rm m}$ as described in \citet{sargent, Tonegawa_2020}. Another useful statistic is the pairwise velocity metric defined for  galaxies and galaxy clusters as the peculiar velocity difference of pairs along their radial separation vector. Its strong dependence on cosmology has allowed it to effectively provide constraints on various cosmological parameters including $\Omega_{\rm m}$ \citep{cen, yinzhe}. These methods demonstrate that valuable cosmological information is embedded on the small scales ($\lesssim 5 ~h^{-1}$Mpc). 

On large scales ($\gtrsim 10 ~h^{-1}$Mpc), methods that analyze cosmic flows \citep{cosmic_flows} such as the skewness in the divergence of galaxy velocity fields \citep{Bernardeau_1995} have led to constraints on $\Omega_{\rm m}$ independent of the biasing relation between the distribution of galaxies and the underlying matter density field. A similar method is using the Zel'dovich approximation to recover the initial density fluctuation field from observed galaxy peculiar velocity and density fields. With this, one can then compute the one-point probability distribution function (IPDF) which is sensitive to $\Omega_{\rm m}$. Thus, one can tune the value of $\Omega_{\rm m}$ assumed for the observed density fields to fit the IPDF of the observed velocity field \citep{1992ApJ...391..443N, 1993ApJ...405..437N}. 

In recent years, there have been significant advances in building detailed numerical simulations that accurately describe the distribution and dynamics of galaxies and dark matter. These include both $N$-body and state-of-the-art hydrodynamic simulations, and they have become powerful tools for constraining cosmological parameters such as $\Omega_{\rm m}$. However, the optimal method that can extract the maximum amount of information from this variety of data is still unknown for non-Gaussian density fields. Fortunately, the advent of revolutionary machine learning techniques provides an alternative way to extract information from large amounts of data. By training neural networks to learn cosmology directly from generic fields, one can achieve tight constraints on the values of cosmological parameters without relying on summary statistics \citep{Ravanbakhsh, Schmelzle_17, Gupta_18, Ntampaka_19, Ribli_19,  Fluri_19, Villaescusa-Navarro2020, Paco_2021b, pablo}.

In particular, Graph Neural Networks (GNNs), which are constructed to handle graph representations of irregular data structures, are especially useful for this purpose because of their unique ability to exploit relational knowledge between nodes in the graphs down to arbitrarily small scales \citep{hamilton, battaglia, bronstein}. Specifically, in our previous paper \citep{halos_gnn} we showed that GNNs are able to infer $\Omega_{\rm m}$ with a $6\%$ accuracy from halo catalogues of $N$-body simulations containing information about the spatial distribution and velocity modulus of the dark matter halos. More importantly, this network was shown to be robust across various $N$-body simulations that are run with different numerical codes, as well as various hydrodynamic simulations that each employ distinct sub-grid physics models and astrophysical processes. This suggests that the GNN is employing a fundamental relation between the halo properties and $\Omega_{\rm m}$ that is not affected by numerical errors from the $N$-body simulations or baryonic effects. Moreover, in our companion paper \citep{deSanti_2023}, we show that GNNs are able to perform robust inference of $\Omega_{\rm m}$ from the $3D$ positions and $1D$ velocities galaxies of five different hydrodynamic simulation codes while marginalizing over cosmologies, astrophysical effects, sub-grid physics models, and sub-halo definitions. These results demonstrate the abundance of robust information contained in the phase space distribution of halos and galaxies.

However, the learned relation is hard to understand because the GNN encodes information in high dimensional latent space representations that are not associated with obvious physical interpretations. On the other hand, one can use techniques in symbolic regression to reveal the physics underlying neural networks via mathematical formulae. Symbolic regression algorithms can be trained to approximate any learned network by fitting analytic expressions to the input and output of neural network components. Such approximations may also generalize better to data that exists outside the range of the data distribution used for training because they possess stronger extrapolation properties than neural networks, whose complex functional forms have the tendency to overfit and learn uninformative priors used during training \citep{Villaescusa-Navarro2020}. This method has been recently used to rediscover physical laws in planetary motion, uncover new relations in matter over-density fields, and more \citep{gn_sr, sr_inductive, Wadekar_2020,villaescusanavarro2020camels, pablolemos, sr_szflux, subhalo_relations, delgado_2022, ESR}.

Hence, in this paper, we attempt to understand the physical relations employed by the GNNs presented in \cite{halos_gnn} and \cite{deSanti_2023} by providing an explicit mathematical formula that approximates the learned networks. To achieve this, we follow a two-step method. First, we train a GNN on halo positions and velocity moduli to show that a model with reduced latent space dimensionality can recover the accuracy and robustness of the model discussed in \citet{halos_gnn}. Its compressed architecture will aid the use of symbolic regression and decrease the complexity of the approximating expressions. In the second step, we train a symbolic regressor to find mathematical equations that approximate each component of the GNN model. We show that the discovered analytic expressions are able to preserve the accuracy and robustness of the relation found by the GNN by testing them on halos from thousands of $N$-body and hydrodynamic simulations of varying cosmological and astrophysical parameters. More surprisingly, we also demonstrate that the equations are able to predict the value of $\Omega_{\rm m}$ from galaxy catalogues of five different hydrodynamic simulations. This suggests that the equations may be independent of the complex connection between the spatial and velocity distributions of halos and galaxies. Finally, we attempt to interpret the physical meaning of the equations. Since the expressions reveal that the network is exploiting rotationally-symmetric information encoded in the relative velocity modulus of the halo pairs on small scales $\sim 1.35 ~h^{-1}$Mpc, we draw connections to traditional techniques that rely on phase space distributions for galaxies and halos to constrain $\Omega_{\rm m}$.  

% This explicit dependence on the relative positions and velocity modulus of halo pairs also preserves the rotational and translational symmetries of the data. 

This paper is structured as follows. We first describe the data used for this project in Section \ref{sec:data}. In Section \ref{sec:methods}, we describe the architecture of our GNN models, the symbolic regression algorithm, and the methods used to train, validate, and test both models. In Section \ref{sec:results}, we present the results of our models and equations. We then provide a discussion of plausible physical interpretations of the equations in Section \ref{sec:discussion}. Finally, we summarize the main findings in Section \ref{sec:conclusion}. 

\section{Data\label{sec:data}}

We train our models using halo catalogues from high-resolution cosmological simulations that contain two halo properties. First, the halo positions, $\textbf{r}$, which are defined for the halo center using Cartesian coordinates in comoving-space. Second, the halo velocity modulus, $V$, which is defined as the modulus of the 3D peculiar velocity vector computed with respect to the velocity of the simulation box. In this work we focus on halo and galaxy catalogues at $z=0$. We describe the methods to generate the halo and galaxy catalogues we use to train, validate, and test the model in Section \ref{subsec:catalogues}.

% Following the method used in \citet{halos_gnn}, we construct multiple halo catalogues per simulation with a different number of halos; this step is important for training a robust GNN model. We refer the reader to that paper for more details about this construction.

% Rockstar identifies halos by first separating the simulation box into three-dimensional friends-of-friends (FoF) groups in which particles that are located within a certain linking threshold of each other are selected to be part of a common structure. The algorithm then normalizes the positions and velocities of all particles in each group, allowing for a phase-space linking length to be adaptively chosen. Particles that are linked constitute a subgroup, and this process is iteratively repeated for each constructed subgroup to generate more substructures hierarchically. At the end of this step, each subgroup is assigned a seed halo and particles are distinguished into their closest seed in phase space. Finally, any unbound particles are removed and the algorithm computes all desired properties for the identified halos.

\subsection{Simulations}

We follow the scheme used in \citet{halos_gnn} to test the accuracy and robustness of our models. This strategy is composed of two parts. First, we use cosmological $N$-body and hydrodynamic simulations that contain different $\Omega_{\rm m}$ values organized in Latin hypercubes and varying initial random seed conditions to quantify the percentage constraints and level of precision achieved by the models. Specifically, $\Omega_{\rm m}$ varies in the range
\begin{equation}
\label{eq:omegaM_range}
    0.1 \leq \Omega_{\rm m} \leq 0.5 
\end{equation} 
for both the $N$-body and hydrodynamic simulations. Note that these simulations also vary $\sigma_{\rm 8}$ in the range $0.6 \leq \sigma_{\rm 8}\leq 1.0$. Furthermore, for the hydrodynamic simulations, we vary several astrophysical parameters; most of them just alter four astrophysical parameters controlling the efficiency of supernova and active galactic nucleus (AGN) feedback but we also made use of a new set that varies 23 astrophysical parameters controlling most of the free parameters in the considered hydrodynamic code. The hydrodynamic simulations have been run with five different codes that not only solve the hydrodynamic equations using different methods, but they made use of different subgrid models. These simulations are part of the CAMELS project and we refer the reader to \citet{villaescusanavarro2020camels, CAMELS_public} for further details.

Second, we use simulations that are generated with the same cosmologies and initial seeds for a control set-up in which we can determine the robustness of the models when evaluated on halos generated with different codes. For this, we run 6 $N$-body simulations that have the same initial random seed and value of $\Omega_{\rm m}=0.3175$ (all other cosmological parameters are shared among codes), but each is run with a different code. Additionally, we run 4 hydrodynamic simulations that have the same value of $\Omega_{\rm m}=0.3$, initial random seed, and employ their fiducial sub-grid physics model using 4 distinct codes. 

For these two above steps, we employ thousands of $N$-body and hydrodynamic simulations that have volumes of (25 $h^{-1}$Mpc)$^3$ and have been run with 11 different codes. We briefly describe these codes below, but for more detailed information, we refer the reader to \citet{halos_gnn} and the listed paper(s) for each code. Note that at the end of the descriptions for each code, we include the number of simulations generated to contain the same cosmology and initial random seed as the other codes, and the number of simulations that contain varying cosmologies, initial seeds, and/or astrophysical parameters arranged in a Latin-hypercube (or Sobol sequence), respectively.  

\subsubsection{$N$-body codes}\label{subsubsec:nbody_codes}
The different $N$-body codes follow the evolution of dark matter particles (that represent the cold dark matter plus baryonic fluid) under the effect of self-gravity in a given expanding cosmological background using different numerical techniques and approximations. The six codes we use to run the $N$-body simulations are described briefly below.

\begin{enumerate}
     \item \textbf{Abacus}. This code computes the long-range gravitational potential by decomposing the near-field and far-field forces in which the near-field forces are reduced to a $r^{-2}$ summation (or an appropriately softened form) and the far-field forces to a discrete convolution over multipoles \citep{2021MNRAS.508..575G}. We run 51 simulations with Abacus: 1 simulation with a shared cosmology and initial random seed among codes and 50 simulations in a Latin-hypercube with varying values of $\Omega_{\rm m}$ and $\sigma_{\rm 8}$.
    
    \item \textbf{CUBEP$^3$M}. This code employs a particle-particle particle-mesh (P$^3$M) scheme, described in \citet{2013MNRAS.436..540H}, where long-range gravitational forces are computed via a two-level particle mesh calculation. We ran 51 CUBEP$^3$M simulations: 1 simulation with shared cosmology and initial random seed among codes and 50 simulations in a Latin-hypercube. For the simulation sharing the cosmology and initial random seed, we used the exact same initial particles as in the other codes, whereas the CUBEP$^3$M initial conditions, generated using the Zeldovich approximation, were used for the 50 simulations in the Latin-hypercube.
    
    \item \textbf{Enzo}. This is an Adaptative Mesh Refinement (AMR) code, as described in \citet{Enzo}, that solves the Poisson equation via a fast Fourier technique \citep{Hockney_1998} on the root grid and a multigrid solver on the individual sub-mesh. We only have one Enzo simulation which shares the same cosmology and initial random seed with the other codes. 
    
    \item \textbf{Gadget}. This code utilizes a TreePM algorithm to compute short-range forces and Fourier techniques to calculate long-distance forces, as described in \citet{Gadget}. We use the halo catalogues from these simulations to train the models. We run 1,001 of the Gadget simulations: 1 simulation with shared cosmology and initial random seed among codes and 1,000 simulations that have different values of $\Omega_{\rm m}$, $\sigma_{\rm 8}$, and initial random seed. We use the halo catalogues from these simulations to train the models.
    
    \item \textbf{PKDGrav3}. This code computes forces using Fast Multipole Method (FMM, \citealt{Greengard1987}) as described in \citet{PKDGrav}. We run 1,001 $N$-body simulations with this code: 1 simulation with shared cosmology and initial random seed among codes and 1,000 simulations with different values of $\Omega_{\rm m}$, $\sigma_{\rm 8}$, and initial random seed that are organized in a Latin-hypercube. 
    
    \item \textbf{Ramses}. This code uses the Adaptive Particle Mesh technique described in \citet{Ramses}. It solves Poisson's equation level by level using Dirichlet boundary conditions and a Multigrid relaxation solver. We have run 1,001 Ramses simulations: 1 simulation with shared cosmology and initial random seed among codes, and 1,000 simulations with different values of $\Omega_{\rm m}$, $\sigma_{\rm 8}$, and initial random seed that are organized in a Latin-hypercube.
\end{enumerate}

\subsubsection{Hydrodynamic codes}\label{subsubsec:hydro_codes}
The hydrodynamic simulations have been run using codes that solve the hydrodynamic equations with different numerical methods and employ distinct models to describe astrophysical processes such as star formation and feedback from supernova and AGN. The hydrodynamic simulations have been run with the codes MP-Gadget, Arepo, OpenGadget, Gizmo, and SWIFT-EAGLE. %Each of these codes employs a different subgrid physics model. 
In these simulations, which are part of the CAMELS project \citep{villaescusanavarro2020camels}, we vary the values of $\Omega_{\rm m}$, $\sigma_{\rm 8}$, the initial random seed, and several astrophysical parameters that we describe below. Instead of referring to these simulations by the name of the code used to run them, we will call them by name of the flagship simulations associated with them and their sub-grid model; i.e. ASTRID, IllustrisTNG, Magneticum, SIMBA, and SWIFT-EAGLE respectively. We note that the SB28 simulations have been run with the Arepo code and employ the IllustrisTNG sub-grid model, but since they vary 28 parameters we use a special name for them. Below, we briefly describe the simulations from the different codes:

\begin{enumerate}
 \setcounter{enumi}{6}
    \item \textbf{ASTRID}. These simulations employ the MP-Gadget  code to solve the gravity (with TreePM), hydrodynamics (with the pressure-entropy formulation of SPH), and astrophysical processes \citep{Astrid1, Astrid2}. We have run 1,001 simulations with this code: 1 simulation with shared cosmology and initial random seed among codes, and 1,000 simulations with different values of $\Omega_{\rm m}$, $\sigma_{\rm 8}$, four astrophysical parameters that control the efficiency of supernova and AGN feedback, and initial random seed that are organized in a Latin-hypercube.
    
    \item \textbf{IllustrisTNG}. These simulations have been run with the Arepo code \cite{Arepo, Arepo_public}, making use of a TreePM plus moving-mesh finite volume (MMFV) method \citep{WeinbergerR_16a,PillepichA_16a}. We have run 1,029 simulations with this code: 1 simulation with shared cosmology and initial random seed among codes, and 1,000 simulations with different values of $\Omega_{\rm m}$, $\sigma_{\rm 8}$, four astrophysical parameters that control the efficiency of supernova and AGN feedback, and initial random seed that are organized in a Latin-hypercube. We also run 27 simulations using this code that only differs in the value of their initial random seed to study the effect of cosmic variance, which we refer to as the CV set. Finally, we have 1 simulation of this code containing a periodic comoving volume of $(205~h^{-1}{\rm Mpc})^3$. This simulation is part of the IllustrisTNG-300 set \citep{Marinacci_2018,nelson_18,NaimanJ_17a,IllustrisTNG_public,SpringelV_17a,Pillepich_2018} and we use it to quantify how our analytic expressions behave in the presence of super-sample covariance effects.
    
    \item \textbf{Magneticum}. This simulation is run with the code OpenGadget3 and implements the SPH-scheme following \citet{Beck2016}. For more details, see \citet[][]{Dolag2004, Jubelgas2004, Hirschmann2014, Groth2023}. We have run 51 Magneticum simulations: 1 simulation with shared cosmology and initial random seed among codes, and 50 simulations with different values of $\Omega_{\rm m}$, $\sigma_{\rm 8}$, four astrophysical parameters that control the efficiency of supernova and AGN feedback, and initial random seed that are organized in a Latin-hypercube.

    \item \textbf{SIMBA}. These simulations have been run with the GIZMO code \citep{Gizmo} with a TreePM plus Meshes Finite Mass method (MFM), see \citet{Dave2019_Simba}. We have run 1,001 SIMBA simulations: 1 simulation with shared cosmology and initial random seed among codes, and 1,000 simulations with different values of $\Omega_{\rm m}$, $\sigma_{\rm 8}$, four astrophysical parameters that control the efficiency of supernova and AGN feedback, and initial random seed that are organized in a Latin-hypercube.

    \item \textbf{SB28}. These simulations have been run with Arepo and employ the IllustrisTNG model. They contain 1,024 simulations and we place them in a different category as they vary the value of 5 cosmological parameters ($\Omega_{\rm m}$, $\Omega_{\rm b}$, $h$, $n_s$, $\sigma_{\rm 8}$), and 23 astrophysical parameters controlling most of the code free parameters. The values of the 28 parameters are organized in a Sobol sequence \cite{SOBOL1967}.

    \item \textbf{SWIFT-EAGLE}. These simulations have been run with the SWIFT-EAGLE code \citep{schaller_swift_2016,2018ascl.soft05020S} and employ a sub-grid physics model that aims at mimicking the original Gadget-EAGLE model \citep{schaye_eagle_2015, crain_eagle_2015}, with some parameter and implementation differences \citep{borrow_impact_2022}. The full model will be described in \cite{Borrow-prep}. 
    The suite contains $64$ simulations varying eight subgrid parameters that control stellar and AGN feedback on a Latin-hypercube (parameter ranges are given in square brackets):
\begin{itemize}
    \item $f_{\mathrm{E,min}}$, the minimal stellar feedback fraction, [0.18, 0.6]
    \item $f_{\mathrm{E,max}}$, the maximal stellar feedback fraction, [5, 10]
    \item $N_{\mathrm{H,0}}$, pivot point in density that the feedback energy fraction plane rotates around, [$10^{-0.6}$, $10^{-0.15}$]
    \item $\sigma_{\mathrm{n}}$ and $\sigma{\mathrm{Z}}$, energy fraction sigmoid width, controlling the density and metallicity dependence, [0.1, 0.65]
    \item $\varepsilon_{\mathrm{f}}$, coupling coefficient of radiative efficiency of AGN feedback, [$10^{-2}$, $10^{-1}$]
    \item $\Delta T_{\mathrm{AGN}}$, AGN heating temperature, [$10^{8.3}$, $10^{9.0}$]
    \item $\alpha$, black hole accretion suppression / enhancement factor, [0.2, 1.1].
\end{itemize}

     % \item \textbf{IllustrisTNG-300}. This simulation has been run with the cosmology $\Omega_{\rm m} = 0.3089$ using the AREPO code which uses a moving mesh \citep{Arepo_public}. The simulation contains a periodic comoving volume of $(205~h^{-1}{\rm Mpc})^3$ at $z=0$, with a resolution of 2,500$^3$ dark matter particles and fluid elements. We refer the reader \citet{PillepichA_16a, WeinbergerR_16a} for details regarding the subgrid galaxy formation model employed by this simulation. We use this simulation to quantify how our analytic expressions behave in the presence of super-sample covariance effects.
    
\end{enumerate}

%%%%%%%%%%%%%%%%%%%%%%%%%%%%%%%%%%%%%%%%%%%%%%%
%%%%%%%%%%%%%%%%%%%%%%%%%%%%%%%%%%%%%%%%%%%%%%%
\section{Methods} \label{sec:methods}

\begin{figure*}
 \centering
 \includegraphics[width=0.75\textwidth]{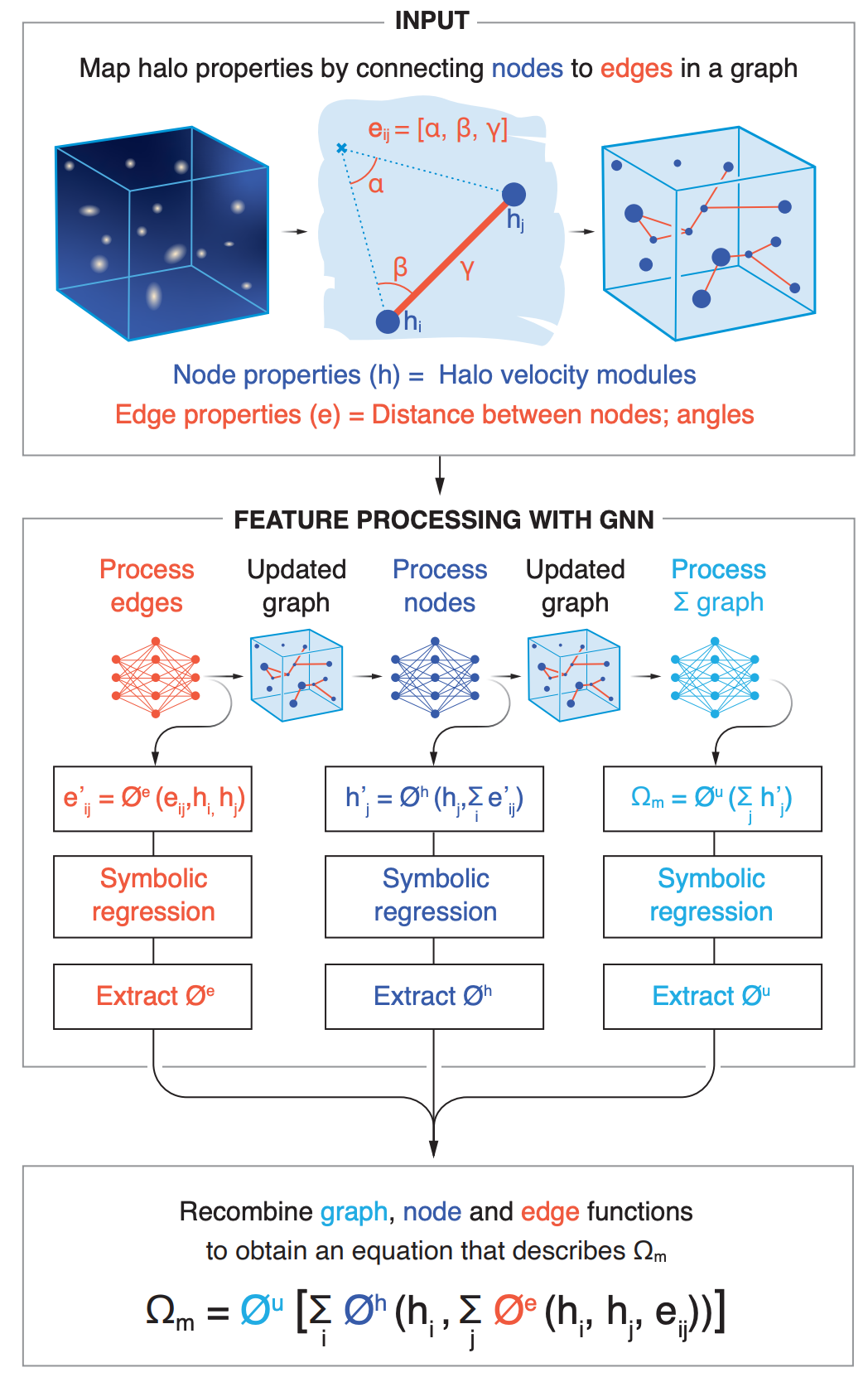}
 \caption{This is a schematic of our methodology which is explained in Section \ref{sec:methods}. We begin by constructing a graph from a halo catalogue using halo positions and velocity moduli. We then feed the graphs to a GNN and train it to perform parameter inference for $\Omega_{\rm m}$. After training the model, we use symbolic regression to extract the equations from each component of the GNN architecture. Finally, we assemble the equations into one expression and use it to predict $\Omega_{\rm m}$ from halos and galaxies of various $N$-body and hydrodynamic simulations.}
 \label{fig:scheme}
\end{figure*} 

In \citet{halos_gnn} we found that GNNs are not only able to infer $\Omega_{\rm m}$ with a $5.6 \%$ precision but are also robust across different $N$-body and hydrodynamic codes, suggesting that the learned relation might be physically fundamental. In this work, we build upon this previous study to understand the found relation and search for an analytic formula that can approximate the mapping from the halo positions and velocities, $\textbf{r}$ and $V$, to the cosmological parameter, $\Omega_{\rm m}$. To accomplish this, we make use of both GNNs and symbolic regression algorithms. We refer the reader to \citet{gn_sr, symbolic_physics} for similar methodologies devised to extract symbolic relations from trained neural networks. 

We begin by training a GNN with the goal of obtaining a low-dimensional latent space network to learn a relation between the input halo properties and $\Omega_{\rm m}$ that can approximate the previously found model. We can do this by fixing certain hyperparameters of the GNN so that it has a reduced architecture depth and width. This step is key to aiding the search for analytic expressions when we use symbolic regression to approximate the GNN, as we later explain. We then evaluate this GNN model on halo catalogues from the $N$-body and hydrodynamic codes described in the previous section to ensure that the sparse architecture is able to achieve comparable precision and accuracy to the model obtained in \cite{halos_gnn}. Finally, we use symbolic regression to fit mathematical formulae to each component of the architecture in the trained GNN model to obtain approximate analytic equations. To improve their interpretability, we also make modifications motivated by physical principles, such as preserving the symmetries present in the data and model and simplifying the found expressions. We refer the reader to Fig.~\ref{fig:scheme} which depicts this methodology schematically.

In the following sections, we describe in detail the ingredients we use to perform this procedure: 1) the method for constructing the halo (training, validating, and testing) and galaxy catalogues (testing), 2) the graph data used to train the GNN, 3) the GNN architecture and training procedure, 4) the data and procedure used to train the symbolic regressor, and finally, 5) the metrics used to evaluate the accuracy and precision of the models.

%\Natali{I know that it is not so usual but I like to mention the validation procedure too. So, I would include it in the above paragraph. But it is just a question of ``style''.}

\subsection{Halo and Galaxy Catalogues}\label{subsec:catalogues}
Here, we describe the procedures for constructing the halo and galaxy catalogues that we use to train, validate, and test the GNN and symbolic expressions.

% For training and testing the models, the halo catalogues are generated by running \textsc{Rockstar} \citep{rockstar} on snapshots from the numerical simulations described below. However, we also run \textsc{Subfind} to generate halo catalogues that we use to test the models to gauge their robustness to different halo definitions, as described in Section \ref{subsec:gnn_results}. We also test our models on galaxy catalogues generated from the six hydrodynamic simulation suites as described below. To generate our galaxy catalogues, we run \textsc{Rockstar} 

\begin{itemize}
    \item \textbf{Halo Catalogues for Training and Validating.} For training and validation, we use halo catalogues from the Gadget simulations. For each simulation, we generate 10 halo catalogues by taking all halos with masses larger than $M_{\rm X}$, where $M_{\rm X}$ is a randomly chosen number between $100\,m_{\rm p}$ and $500\,m_{\rm p}$. Here, $m_p$ is the mass of a single dark matter particle. As explained in \citet{halos_gnn}, using different dark matter particle thresholds is key to achieving a model that is robust to different simulations. These halo catalogues are generated by running \textsc{Rockstar} \citep{rockstar} on snapshots from the numerical simulations described above.

    \item \textbf{Halo Catalogues for Testing.} We use all $N$-body simulations described in the previous section and two hydrodynamic simulations: IllustrisTNG and SIMBA. For each simulation, we generate 5 halo catalogues for the five different dark matter particle thresholds: $\{100, 200, 300, 400, 500\}$. Note that for hydrodynamic simulations, the mass of a dark matter halo contains contributions from various mass sources. Hence, instead of considering only the amount of dark matter mass to make our mass cuts, we define $m_p$ as the effective particle mass: $m_p = \frac{1}{N_c}\Omega_{\rm m}V\rho_c$, where $V$ is the volume of the simulation, $\rho_c$ is the Universe's critical density today, and $N_c=256^3$ is the effective number of particles. These halo catalogues are generated by running \textsc{Rockstar} \citep{rockstar} on snapshots from the numerical simulations described above. However, for one test where we gauge the robustness of the train models to different halo definitions, we run \textsc{Subfind} \citep{Dolag2009} to generate halo catalogues from the Gadget $N$-body and Illustris-TNG simulations. %\Paco{how you do this for hydro? Is a threshold in total mass or number of DM particles?} 

    \item \textbf{Galaxy Catalogues for Testing.} We use galaxy catalogues from all the hydrodynamic simulations described in the previous section. We define a galaxy as a subhalo (can either be a central or satellite) that contains a stellar mass of at least $N \times m_*$ where $N \in {3,4,5,6}$ and $m_* = 1.3 \times 10^{7}~ h^{-1}M_\odot$. For each simulation, we construct 4 catalogues, each using a different $N$. We limit the range of the stellar mass thresholds to be no larger than $6\times m_*$ because we find that using larger cuts result in catalogues with galaxy number densities that are smaller than the number densities (from the halo catalogues) used to train the network and equations. We find that using catalogues with number densities that are outside the training range can lead to inaccurate predictions. These galaxy catalogues are generated by running \textsc{Rockstar} \citep{rockstar} on snapshots from the six hydrodynamic simulations described above, with the exception of the catalogues from the SWIFT-EAGLE simulations which were generated using the halo finder \textsc{VELOCIraptor} \citep{velociraptop1, velociraptor2}. %\Paco{Perhaps check that the equations still work for large $N$ when modifying $a$ and $b$}

\end{itemize}

\subsection{GNNs\label{subsec:gnns}}

The methods described in this section closely follow those presented in \cite{halos_gnn} to infer $\Omega_{\rm m}$. We emphasize the key changes that we implement in this work are: 1) using only the summation operator as the aggregation function and 2) reducing the depth and width of the GNN architecture with constrained hyperparameter optimization. These steps decrease the complexity of the model and allow for easier interpretation of the learned relations.

% \Natali{What about including a footnote specifying from which to which interval you are referring while doing the reduction of the depth and the width of the GNN architecture?} 

\subsubsection{Model input: Halo Graphs\label{subsubsec:graphs}}

The input of the GNN is a graph defined as $\mathcal{G} = (\mathcal{V}, \mathcal{E})$, where $\mathcal{V}$ is the set of nodes and $\mathcal{E}$ is the set of edges. The nodes represent the halos (or galaxies) and an edge is created between two nodes if their distance is smaller than the \textit{linking radius}, $r_{\rm link}$. This property is considered a hyperparameter that we optimize during training, as we explain later. Thus, two nodes $i,j \in \mathcal{V}$ are referred to as neighbors if they are connected via an edge, $(i,j) \in  \mathcal{E}$. As in \citet{halos_gnn}, we do not consider self-loops and account for periodic boundary conditions when computing distances and angles between nodes.

The nodes and the edges can have different properties associated with them, that we denote as  $\textbf{v}_i^{(n)}$ and $\textbf{e}_{ij}^{(n)}$, respectively. Since the architecture of the GNN models may consist of multiple layers that take a graph as the input and outputs an updated graph. For this reason, we denote the node and edge features at the $n^{th}$ layer with the superscript $n$.

The initial node feature, represented by $\textbf{v}_i^{(0)}$, that we use is the halo velocity modulus, $V$. Since the velocities are defined with respect to the simulation box, the node features preserve Galilean invariance. The edge features between nodes $i$ and $j$ at the $n^{th}$ layer are represented by $\textbf{e}_{ij}^{(n)}$ and they contain information about the spatial distribution of halos. To ensure that the model preserves the rotational and translational invariance of the data, we use the following vector for the edge features: 
\begin{equation}\label{eq: initial_edge}
    \textbf{e}^{(0)}=[\alpha_{ij}, \beta_{ij}, \gamma_{ij}]
\end{equation}
where
\begin{eqnarray}    
\alpha_{ij} &&= \frac{\textbf{r}_i-\textbf{c}}{|\textbf{r}_i-\textbf{c}|} \cdot \frac{\textbf{r}_j-\textbf{c}}{|\textbf{r}_j-\textbf{c}|}\label{Eq:edge_features1}\\
\beta_{ij} &&= \frac{\textbf{r}_i-\textbf{c}}{|\textbf{r}_i-\textbf{c}|} \cdot \frac{\textbf{d}_{ij}}{|\textbf{d}_{ij}|}\label{Eq:edge_features2}\\
\gamma_{ij}&&=\frac{|\textbf{d}_{ij}|}{r_{\rm link}}\label{Eq:edge_features3}, 
\end{eqnarray}
with $\textbf{d}_{ij}=\textbf{r}_i-\textbf{r}_j$ being the relative distance between the nodes $i$ and $j$ and $\textbf{c}$ is the centroid of the halo/galaxy distribution. $\alpha_{ij}$ defines the angle between the positions of node $i$ and its neighbor node $j$, while $\beta_{ij}$ describes the angle between positions of node $i$ and the separation between nodes $i$ and $j$. Note that we have normalized the distance, $\textbf{d}_{ij}$, by dividing it with the linking radius, $r_{\rm link}$, to have dimensionless edge features. We refer the reader to \cite{pablo} for more details on this construction.

% \Natali{Helen, because the above pieces of text are a single paragraph I removed the lines that you have skipped, to get the proper indentation.}

\subsubsection{Architecture\label{subsubsubsec:architetcure}}

The architecture of our GNN model closely follows  \textsc{CosmoGraphNet}\footnote{ \href{https://github.com/PabloVD/CosmoGraphNet}{https://github.com/PabloVD/CosmoGraphNet}} \citep{pablo_villanueva_domingo_2022_6485804}, presented in \cite{pablo} and used in \citet{halos_gnn}. However, our model only includes one message-passing layer and a final aggregation layer. We arrived at this architecture by experimenting with different numbers of hidden layers to optimize the simplicity of the model while maintaining the precision and accuracy of its predictions. We explain this in more details in Section \ref{subsubsec:training_procedure}.

In the message-passing layer, information from the input node and edge features are encoded with multilayer perceptrons (MLP) and recursively exchanged and aggregated between each node's neighbors and edges. Afterwards, the node and edge features are updated. This creates hidden feature vectors that are ultimately used to predict the target parameter. For this reason, we denote the edge and node features that are input to the message-passing layer (the initial halo properties) with the superscript $(0)$ and output (hidden) features by the message-passing layer with the superscript $(1)$. 

For our \textit{compressed} GNN, we restricted to two hidden features for each node and edge because the number of hidden features scales in proportion to the number of analytic expressions needed to approximate the network, as we explain later.

For the message-passing layer, the input of the edge model are the initial features of the node $i$, the neighboring node $j$, and their shared edge. In this case, the initial node features are $V$ as defined in Section \ref{sec:data} and the initial edge features are ${e}^{(0)}$ as defined in Equation \ref{eq: initial_edge}. This information is passed through an MLP, denoted by $\phi ^e$ and the output are the updated hidden edge features:
\begin{equation}
    \textbf{e}_{ij}^{(1)}=\phi^e\left(\left[\textbf{v}_i^{(0)},\textbf{v}_j^{(0)},\textbf{e}^{(0)}\right]\right). 
    \label{eq:edgelayer}
\end{equation}
This hidden edge feature, along with the initial node feature of node $i$, is then passed to the node model, where another MLP, denoted by $\phi ^v$, outputs the hidden node features:
\begin{equation}
    \textbf{v}_i^{(1)} = \phi^v \left(\left[\textbf{v}_i^{(0)}, \sum_{j \in \mathcal{N}_i} \textbf{e}^{(1)}\right]\right)~.
    \label{eq:nodelayer}
\end{equation}
Here, we use a permutationally invariant aggregation function - the summation - to aggregate the node features of the neighbor nodes $j \in\mathcal{N}_i$ that are connected to node $i$. In \citet{halos_gnn}, the aggregation function used was a concatenation of the maximum, summation, and mean operators. In this work, we reduce this function to just the summation to decrease the complexity of the learned relations. This choice is motivated by the fact that the summation can serve as a proxy for the other two operators. Using only one aggregation operator as opposed to three decreases the number of hidden channels by a factor of three and thus reduces the number of equations we for our model. 

The final layer in the architecture aggregates the hidden node features output by the message passing layer to make the prediction $\textbf{y}$:
\begin{equation}
    \textbf{y} = \phi ^u \left( \left[\sum_{i \in \mathcal{G}} \textbf{v}_i^{(1)}\right] \right),
    \label{eq:final_mlp}
\end{equation}
where $\sum_{i \in \mathcal{G}}$ operates over all nodes in the graph and $\phi ^u$ is another MLP that extracts the target information.

% \Natali{I have removed the additional line here (above) too.}

\subsubsection{Training procedure}\label{subsubsec:training_procedure}
We train and test the models using graphs constructed from halo catalogues of the Gadget simulations. For each simulation, we construct 10 catalogues using the procedure described in Section \ref{subsec:catalogues} to marginalize over the halo number density. Once trained, the model is tested using catalogues from all simulations. For Gadget, we split the simulations into training (80\%), validation (10\%), and testing (10\%) data sets before creating halo catalogues for each simulation. For the other codes, we use the entirety of the dataset for testing. 

We standardize the values of input node features as
\begin{equation}
    \tilde{x}=\frac{x - {\mu}}{{\delta}},
\end{equation}
where $\mu$ and $\delta$ denote the mean and standard deviation of the feature $x$. However, we explain in later sections that the value of $\delta$ must be tuned for when evaluating the symbolic equations. We also normalize the values of the target cosmological parameter, $\Omega_{\rm m}$:

\begin{equation}
    \bar{\Omega}_{\rm m}=\frac{\Omega_{\rm m} - {\rm min}(\Omega_{\rm m})}{{\rm max}(\Omega_{\rm m}) - {\rm min}(\Omega_{\rm m})},
\end{equation}
where the minimum and maximum values of the ranges of $\Omega_{\rm m}$ are listed in Eq.~\ref{eq:omegaM_range}.

As we did in \citet{halos_gnn}, we train the GNN to perform likelihood-free inference so the output of the model is $\mathbf{y} = [\mu_i, \sigma_i]$, where $\mu_i$ is the posterior mean  and $\sigma_i$ is the posterior standard deviation of $\Omega_{\rm m}$. To achieve this, we employ the following loss function:

\begin{equation}
\begin{split}
    \mathcal{L} & = \log{\bigg(\sum_{j\in {\rm batch}}(\theta_{i,j} - \mu_{i,j})\bigg)^2} + \\
    & \log{\bigg(\sum_{j\in {\rm batch}}\big((\theta_{i,j} - \mu_{i,j})^2 - \sigma_{i,j}^2\big)\bigg)^2}
\end{split}
\end{equation}
where the sums are performed over the halo catalogues in the batch. Further details on this can be found in  \citet{jeffrey_wandelt} and \cite{CMD}.

Our model is implemented in PyTorch \citep{Pytorch} and PyTorch Geometric \citep{Fey_Fast_Graph_Representation_2019}. We use the AdamW optimizer \citep{AdamW} with beta values equal to 0.9 and 0.999. We train the network using a batch size of 8 for 500 epochs. The hyperparameters for our model are: 1) the learning rate, 2) the weight decay, and 3) the linking radius. We use the \textsc{optuna} code \citep{Optuna} to perform Bayesian optimization and find the best value of these hyper-parameters for each model. As mentioned earlier, we aim to reduce the depth and width of our GNN architecture to obtain a \textit{compressed} network so we restrict to only one layer and two hidden neurons. For each model, we run 100 trials, where each trial consists of training the model using selected values of the hyper-parameters. We perform the optimization of the hyper-parameters required to achieve the lowest validation loss possible and use early stopping to save only the model with a minimum validation error. 

%\Natali{``What about including a footnote specifying from which to which interval you are referring too while doing the reduction of the depth and the width of the GNN architecture?''. This comment can be made here, as a footnote, instead of at the beginning of the section, where I originally included the comment.}

%Note that we did not optimize the number of hidden layers in the GNN model. This is because the accuracy of symbolic regressors is limited to low-dimensional input data, as explained later. Also, the number of equations scales proportionally to the number of hidden features so reducing the model size allows for better interpretability. 

%In \cite{halos_gnn}, the found model contained 2 message-passing layers and 88 hidden features. Motivated by the fact that this model is not very deep, we first fix the number of message-passing layers of the GNN to 1 and use a bound of 10 on the number of hidden features during the hyperparameter optimization. Once trained, we test the model on halo catalogues from all $N$-body and hydrodynamic simulations. We find that this model does not have significantly worse accuracy and robustness so we gradually decrease the maximum number of hidden features. Ultimately, we obtain a GNN architecture that consists of only 1 message-passing layer and 2 hidden neurons that can achieve comparable accuracy and robustness to the model found in \citep{halos_gnn}. 

\subsection{Symbolic Regression \label{subsec:SR_training}}

While neural networks can provide precise and accurate approximations of complex relations in the data, interpreting them is often challenging because they employ a large number of parameters to make predictions. Therefore, it is desirable to extract mathematical expressions that characterize, or approximate, the relation learned by the neural network because it is easier to understand the physics of the found relationships in such forms. Moreover, analytic equations have been found to generalize better, than neural networks, to data with characteristics not presented in the training set, which can give us more robust predictions and possibly illuminate fundamental properties of the model \citep{subhalo_relations}.

%\Natali{Can you explain, or give an example (citations, as one of your previous works) while analytic equations can generalize better?}

For this purpose, we first train a symbolic regression algorithm designed to approximate functions with analytic formulae. We then modify the expressions using reasoning based on physical principles - such as that the model should preserve rotational and translational symmetries of the data - to improve the interpretability of the equations and reduce their complexity. In this section, we describe the symbolic regression algorithm we use and the procedure for fitting functions to components of the learned GNN.

We use the package \textsc{pysr} \citep{pysr} to train a symbolic regression algorithm with the ability to fit mathematical formulas to the learn GNN relations. This package implements genetic programming which searches for the optimal analytic expression creating combinations between the sets of given operators and input variables. The found expressions of each so-called generation are evaluated, and the most accurate ones \textit{survive} to the next generation. Throughout this iterative process, \textit{mutations} and \textit{crossovers} take place to explore the entire equation space and find an accurate expression.

%\Natali{``To train a symbolic regression algorithm to fit mathematical formulas onto the learned GNN relations using the package \textsc{pysr} \citep{pysr}.'' is sounding a bit strange to me. What about: ``To train a symbolic regression algorithm to fit mathematical formulas onto the learned GNN relations we have used the package \textsc{pysr} \citep{pysr}.''?}

However, a key limitation of symbolic regression is that its tractability and accuracy are restricted to low-dimensional spaces of input data. To circumvent this, we limit the size of the latent space produced by the GNN, as described in Section \ref{subsubsubsec:architetcure}. Using the learned parameters and relations from the low-dimensional GNN architecture, we search for equations that characterize the model by approximating the individual MLPs used in the node model, edge model, and final layer described in Equations \ref{eq:edgelayer}, \ref{eq:nodelayer}, and \ref{eq:final_mlp}, respectively. We emphasize that since there is only one message-passing layer, we only need to approximate one node model MLP and one edge model MLP. Moreover, for each of the node and edge models, we search for two equations because there are two hidden features. The data and procedure used to obtain these equations are described below.

\begin{itemize}
    \item \textbf{Approximating Edge Model:} To approximate the edge model, we train a symbolic regressor to map from the input variables, $\textbf{x}^e$, to the target variables, $\textbf{y}^e$, defined as: 
\begin{eqnarray}
\textbf{x}^e&&=\left(v_i^{(0)}, v_j^{(0)}, \alpha_{ij}, \beta_{ij}, \gamma_{ij} \right)\\
\textbf{y}^e &&=\left(e_{1}^{(1)}, e_{2}^{(1)}\right).
\end{eqnarray}
The input variables are the initial features of the nodes and their neighbors, as well as the initial edge features as described in Section \ref{subsubsec:graphs}. The corresponding target variables are the edge features of the MLP in the edge model defined in Equation \ref{eq:edgelayer}. Since the GNN employs only two hidden features for each message-passing layer, we denote the first component of the edge feature as $e_1^{(1)}$ and the second component as $e_2^{(1)}$. To obtain this data, we randomly select 10 $(\textbf{x}^e, \textbf{y}^e)$ pairs from each graph in the training set. This selection is done to ensure that we have a representative sample of the training set without using every node pair of all graphs which would result in too large of a dataset.

    \item \textbf{Approximating Node Model:} Similarly, to approximate the node model, the input variables, $\textbf{x}^n$, and the target variables, $\textbf{y}^n$, of the symbolic regressor, are:
\begin{eqnarray}
\textbf{x}^n &&= \left(v_i^{(0)}, \sum_{j \in \mathcal{N}_i}{e}_{1}^{(1)}, \sum_{j \in \mathcal{N}_i}{e}_{2}^{(1)}\right)\\
\textbf{y}^n &&= \left(v_{1}^{(1)}, v_{1}^{(1)}+v_{2}^{(1)}\right).
\end{eqnarray}
As seen above, the inputs are the initial node feature and the neighborhood-wise sums of the hidden edge features because the output of the edge model is aggregated using the summation operator before being passed onto the node model. The corresponding target variables are the hidden node features of the MLP in the node model defined in Equation \ref{eq:nodelayer}. We denote the first and second hidden node features as $v_{1}^{(1)}$ and $v_{2}^{(1)}$, respectively. However, instead of directly finding an equation for the second node feature, $v_{2}^{(1)}$, we instead search for a formula for the sum $v_{1}^{(1)}+v_{2}^{(1)}$. This is because we find that the change of variables allows us to obtain more accurate approximations than with the original target variable. Ultimately, to obtain the expression of $v_{2}^{(1)}$, we subtract from it $v_{1}^{(1)}$. To obtain this data, we randomly sample 10 $(\textbf{x}^n, \textbf{y}^n)$ pairs from each graph in the training set as we did with the edge model data.

    \item \textbf{Approximating Final MLP:} Lastly, to approximate the MLP in the final aggregation layer, the input and target variables are: 
\begin{eqnarray}
\textbf{x}^u &&= \left( \sum_{i \in \mathcal{G}} v_{1}^{(1)}, \sum_{i \in \mathcal{G}} v_{2}^{(1)} \right)\\
\textbf{y}^u &&= \mu_i.
\end{eqnarray}
Here, the inputs are the graph-wise sums of the hidden node features because the output of the node model is aggregated using the summation operator before being passed onto the final MLP. The corresponding target is the mean posterior. We do not attempt to find an expression for the posterior standard deviation as it is solely a component of the parameter inference methodology and does not contribute additional physical understanding. We obtain this data from each graph in the training set. Note that this time there is no need to select a sub-sample of nodes from each graph because $x^u$ and $y^u$ are global properties of the graph so we can use every graph in the training set.
\end{itemize}

%\Natali{I have included ``.'' after the equations because you start new sentences after the equations.}

In each of the above approximation steps, the symbolic regression algorithm searches for analytic expressions that can map from the given input variables to the desired target. For the training, the regressor is allowed to employ the following binary operators: \textsc{"add", "sub", "mult", "div", "pow"}\footnote{The listed operators perform addition, subtraction, multiplication, and division. \textsc{"pow"} takes the power of $X$ to the input variable, where $X$ is any number.} and the following unary operators: \textsc{"1/x"} (the inverse of a variable), \textsc{"abs", "log", "log10", "sqrt"}. We employ a standard Mean Squared Error (MSE) loss function to optimize the fitting defined as,

%\Natali{Do you think it is ok to write ADD, MULT and etc without explain? I think so, but maybe should be interesting say, in parentheses, what each operation is really doing. Also I would like to include ``Mean Squared Error (MSE)'' instead of only saying MSE in the first appearence in the text.}

\begin{equation}
    {\rm MSE}=\frac{1}{N}\sum_{i=1}^N \left(y_{\rm true}-y_{\rm pred}\right)^2,
\end{equation}
where $\textbf{y}_{\rm pred}$ denotes the predicted value of the target variable and $\textbf{y}_{\rm true}$ is corresponding true value. The model was trained for 100,000 trials with a batch size of 64. 

During training, the algorithm outputs a list of equations found by the regressor. For each equation, \textsc{PYSR} provides three values to quantify the fit of the equation: its \textit{complexity}, MSE, and \textit{score}. The \textit{complexity} of the equation takes into account the number of operators, constants, and variables used. The MSE and the complexity are combined into an overall metric that gives the equation's \textit{score}, akin to Occam's Razor \citep{pysr}. Specifically, the algorithm sorts the found equations from the least to the most complex, and for each equation, it computes the fractional decrease in MSE relative to the next (more complex) equation. The \textit{score} is maximized if this fractional decrease is large. We evaluate several candidate equations on a test set for each hidden feature before selecting one that optimizes the tradeoff between complexity and accuracy with these metrics in mind. 

\subsection{Performance Metrics\label{acc_metrics}}

For the graph $i$, with the true value of the considered parameter $y_{\rm truth,i}$, our models output the posterior mean, $y_{\rm infer,i}$, and standard deviation $\sigma_i$. To evaluate the accuracy and precision of our models, we follow \cite{pablo} and \cite{halos_gnn}, and employ four different metrics:

\begin{enumerate}
    \item \textbf{Mean relative error},  $\epsilon$, defined as
    \begin{equation}
    \epsilon = \frac{1}{N} \sum_i^N \frac{|y_{{\rm truth}, i} - y_{{\rm infer}, i}|}{y_{{\rm truth}, i}},
\end{equation}
where $N$ is the number of halo catalogues in the test set.

    \item \textbf{Coefficient of determination}, $R^2$, defined as
    \begin{equation}
    R^2 = 1 - \frac{\sum_i^N (y_{{\rm truth}, i} - y_{{\rm infer}, i})^2}{\sum_i^N (y_{{\rm truth}, i} - \overline{y}_{{\rm truth}})^2},
    \end{equation}
    
    \item \textbf{Root mean squared error}, RMSE, defined as:
    \begin{equation}
    {\rm RMSE} = \sqrt{\frac{1}{N}\sum_{i=1}^N \left(y_{{\rm truth}, i} - y_{\rm infer}\right)^2}
    \end{equation}

    \item \textbf{Chi squared}, $\chi^2$, defined as:
    \begin{equation}
\chi^2=\frac{1}{N}\sum_{i=1}^N \frac{(y_{\rm truth,i} - y_{\rm infer,i})^2}{\sigma_{i}^2}~.
\label{Eq:chi2}
\end{equation} 

Note that a value of $\chi^2$ that is close to one suggests that the standard deviations are accurately predicted. On the other hand, a larger or lower value indicates that the uncertainties are under - or overestimated, respectively. 
\end{enumerate}

Note that the sums in all expressions above run over the graphs in the test set.

%%%%%%%%%%%%%%%%%%%%%%%%%%%%%%%%%%%%%%%%%%%%%%%%%
%%%%%%%%%%%%%%%%%%%%%%%%%%%%%%%%%%%%%%%%%%%%%%%%%
\section{Results} \label{sec:results}

In this section, we present the results we obtain from training the GNN model. We then show the analytic approximations that were found using  symbolic regression.

\subsection{GNN Results \label{subsec:gnn_results}}

\begin{figure*}
    \centering
    \includegraphics[width=0.47\textwidth]{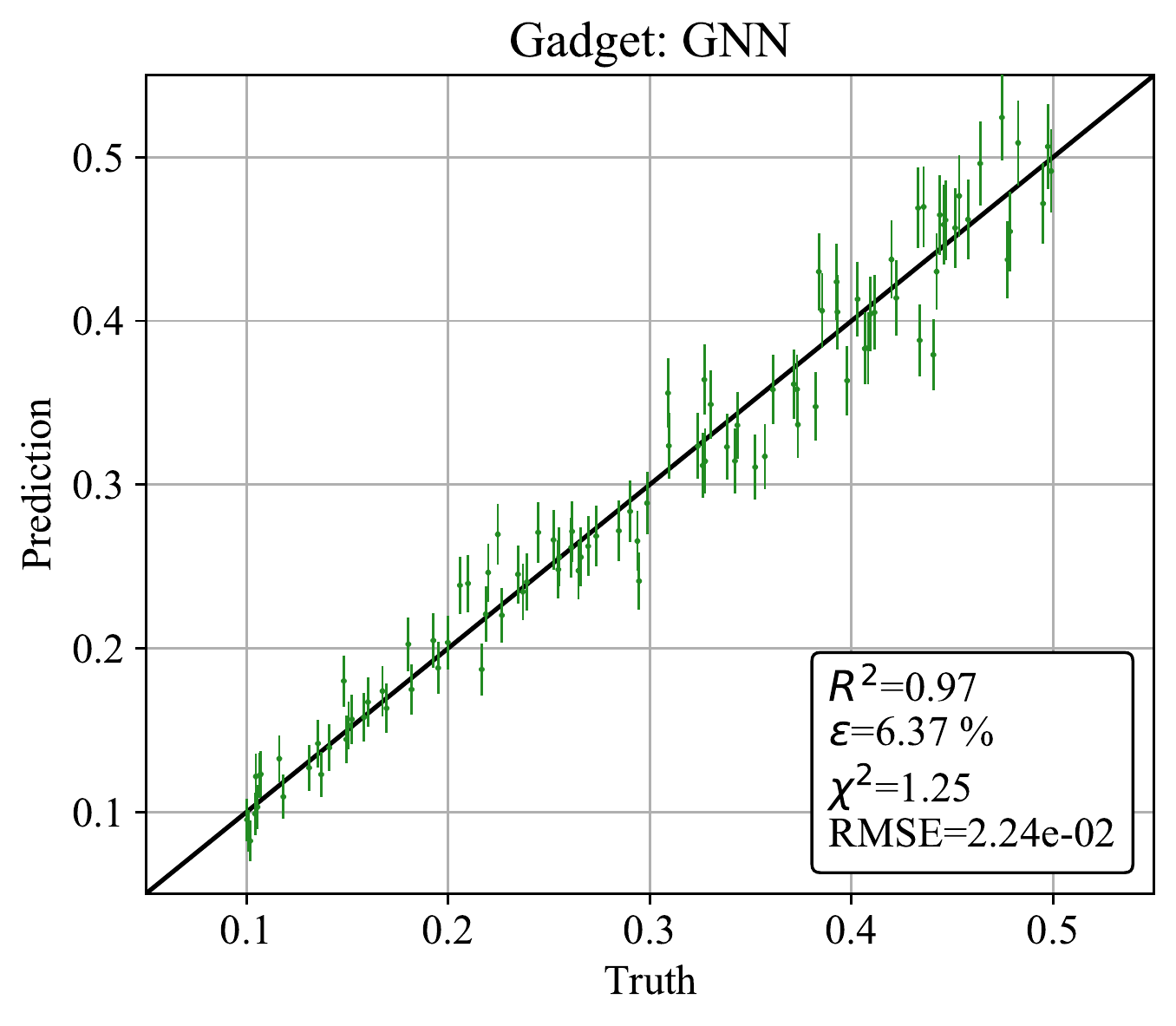}
    \includegraphics[width=0.47\textwidth]{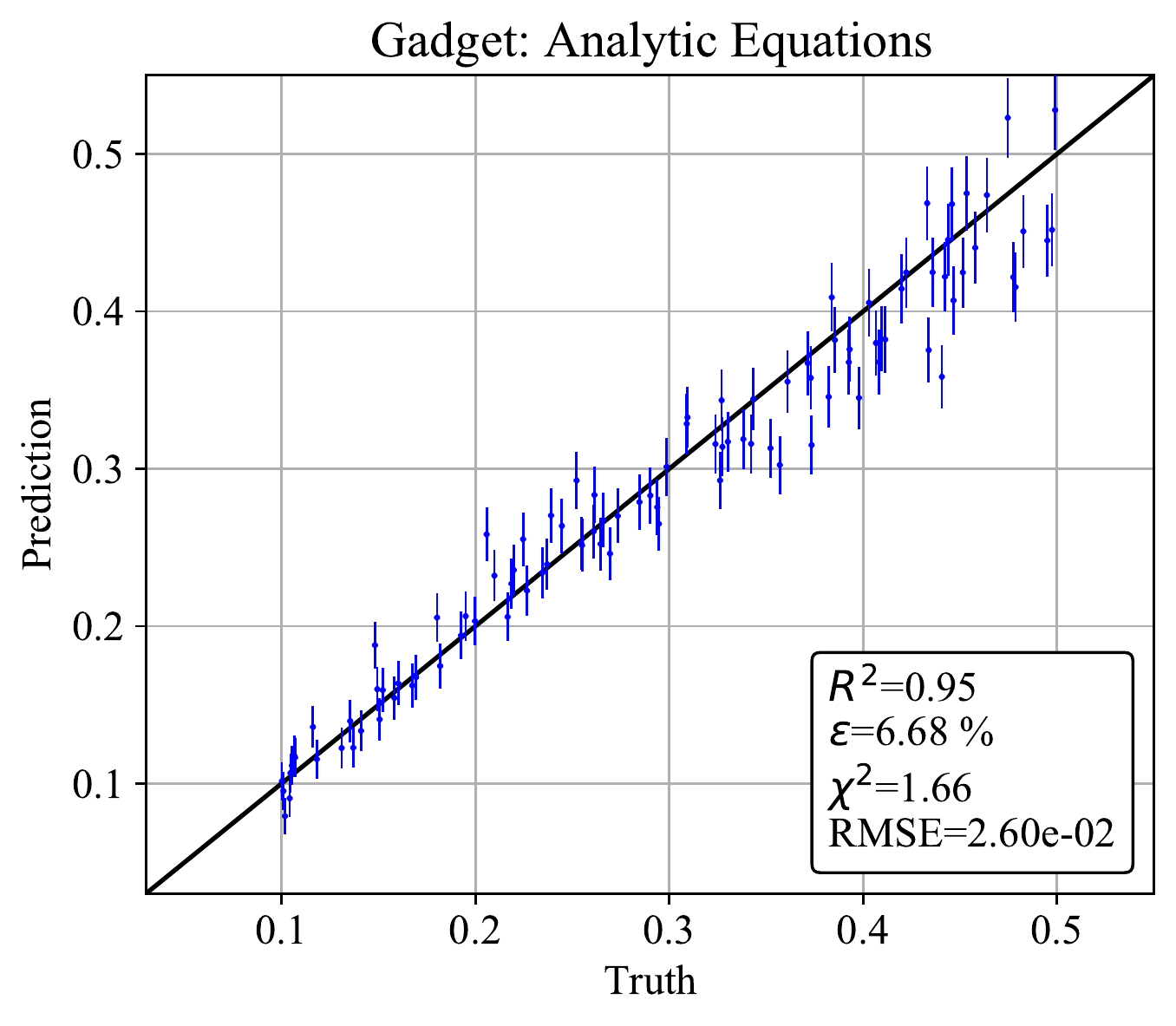}
    \caption{This figure compares the accuracies of the GNN model with the analytic formulae obtained from symbolic regression and the modified equations using physical principles. \textbf{Left:} We first train a GNN with a compressed latent space representation to perform likelihood-free inference for the cosmological parameter $\Omega_{\rm m}$ with halo catalogues containing the positions and velocity moduli of the halos. Evidently, the model is able to achieve very high accuracy with a mean relative error of only $\sim 6.4 \%$ when evaluated on the test set of Gadget simulations. Despite its reduced dimensionality, this accuracy is comparable to the model found in \citet{halos_gnn}. \textbf{Right:} We then use symbolic regression to extract analytic expressions for each MLP in the message-passing and final aggregation layers of the GNN. After modifying them to reduce their complexity and to preserve the symmetries of the model, we evaluate the expressions on the Gadget test set. As shown, the expressions are able to maintain the accuracy of the GNN, with an error of only $\sim 6.7 \%$, indicating that the equations are close approximations for the learned GNN relations.}
    \label{fig:gadget_individual}
\end{figure*}  

We first train a GNN with a single message-passing layer and fix the number of hidden features to two. Using Bayesian optimization of the hyperparameters, we find that the optimal linking radius is $\sim 1.35 ~h^{-1}{\rm Mpc}$ which describes the characteristic length scale of the model. When we evaluate the trained model on a test set of Gadget simulations, we find that it is able to attain very accurate predictions of $\Omega_{\rm m}$ with a mean relative error of ~$6 \%$ and a $\chi^2$ of 1.37. This indicates that both the posterior mean and standard deviations are accurately inferred. These results are depicted in the left panel of Fig. \ref{fig:gadget_individual}. Hence, we see that the accuracy of the model is not significantly compromised by the reduction in the dimensions of its latent space with respect to the model used in \cite{halos_gnn}, which was $\sim 5.6\%$. In the following two sections, we present the results for testing the equations on halos from the six different $N$-body simulations and four hydrodynamic simulations. We then present the predictions for the model tested on galaxies from six different hydrodynamic simulation suites.

\subsubsection{Halos}\label{gnn_halos}

\begin{figure*}
    \centering
    \includegraphics[width=1\textwidth]{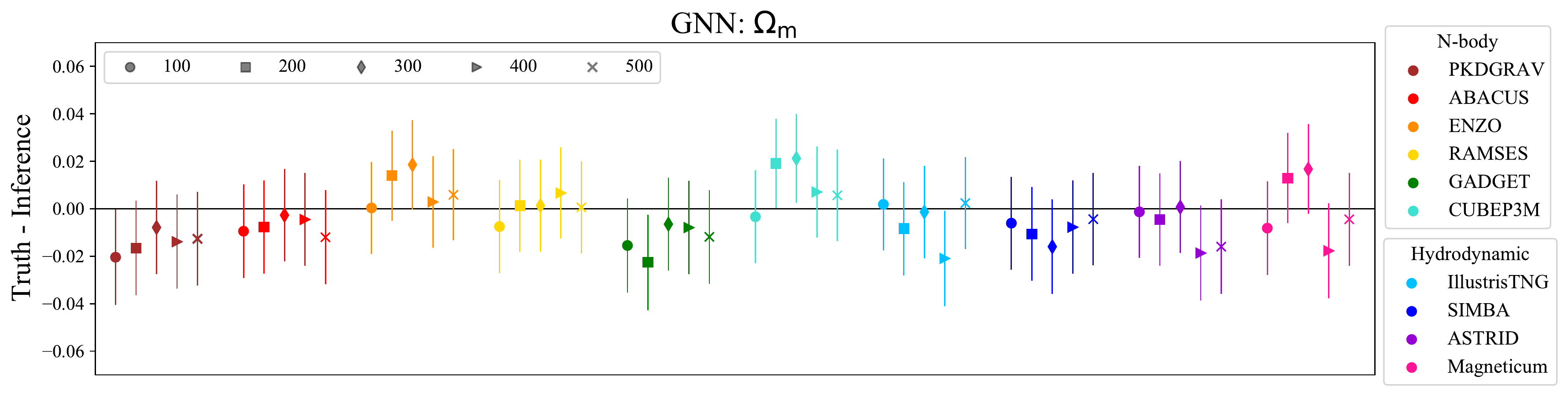}
    \includegraphics[width=1\textwidth]{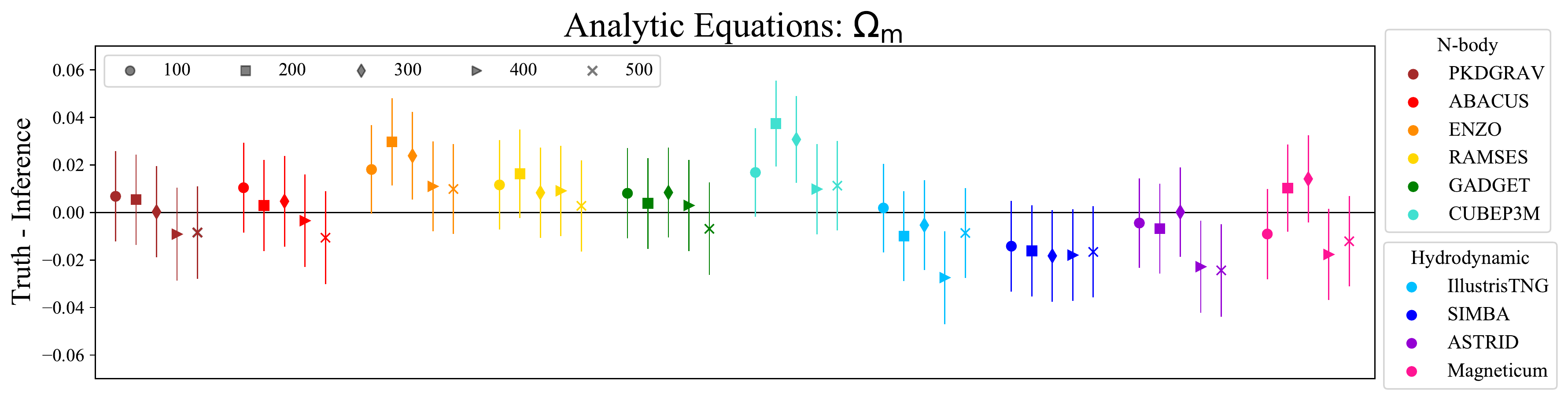}
    \caption{\textbf{Top:} We train a GNN model with a compressed latent space to perform likelihood-free inference for the cosmological parameter $\Omega_{\rm m}$. The input to the model are halo catalogues from Gadget that only carry information about halo positions and peculiar velocity moduli. Once trained, we test the model on halo catalogues from different $N$-body and hydrodynamic simulations as indicated in the legend.  We note that simulations of the same type, either $N$-body or hydrodynamic, are run with the same initial conditions, cosmology (and fiducial astrophysics for the hydrodynamic simulations). For each simulation, we generate 5 catalogues. Each halo catalogue contains all halos with masses above $Nm_p$, where $m_p$ is the particle mass and $N$ can be 100, 200, 300, 400, or 500 (see legend). The y-axis represents the difference between the truth and the inference. As can be seen, this model exhibits surprising extrapolation properties and is robust to all simulation codes despite only containing one message-passing layer and two latent features. \textbf{Bottom:} Same as above but for the analytic equations obtained using symbolic regression and modified to preserve rotational and translational symmetries in the data, as described in Section \ref{subsec:SR_training}. As can be seen, the formulae maintain the robustness of the GNN model and achieve a very similar accuracy compared to the GNN.}
    \label{fig:same_seed}
\end{figure*}

We first find that the model is robust to different $N$-body codes despite being trained on halo catalogues from only the Gadget simulations, agreeing with the results discussed in \citet{halos_gnn}. The simulations we use for this test are Abacus, Ramses, PKDGrav3, Enzo, and CUBEP$^3$M which share the same cosmology and initial conditions but employ different numerical methods, as described in Section \ref{sec:data}. As shown in the top panel of Fig. \ref{fig:same_seed}, the model obtains similar constraints on $\Omega_{\rm m}$ for these simulations. We present more detailed results of this test in Appendix \ref{sec:additional plots}, where the figures depict the accuracies of the model when tested on 50 catalogues of different cosmologies from each simulation code. 

Moreover, the model is robust to different hydrodynamic codes. When tested on halo catalogues from the IllustrisTNG, SIMBA, Astrid, and Magneticum simulations, the GNN is able to achieve similar precision and accuracy compared to the predictions for the $N$-body codes, as seen in the top panel of Fig. \ref{fig:same_seed}. This demonstrates that the model is robust even to hydrodynamics, varying astrophysical parameters, and different subgrid physics models. This agrees with the results from \citet{halos_gnn} and shows that even with a reduced latent space dimensionality, the model could possibly still be learning a fundamental relation between the halo properties and $\Omega_{\rm m}$. However, note that by reducing the size of the latent space the precision of the predictions decreases slightly, which is expected.

Another test that we performed to gauge the extent of the robustness of the GNN is evaluating our model on halos generated using a halo finder that is different (\textsc{Subfind}) from the one used during training (\textsc{Rockstar}). We find that the model is able to extrapolate to these halos and we present the details of this test in Appendix \ref{sec:subfind}.

\subsubsection{Galaxies}\label{subsubsec:galaxies_gnn}
We also asked if the network would extrapolate to galaxy distributions after being trained on only the positions and velocities of $N$-body halos. Hence, we test the GNN on galaxy catalogues from the following hydrodynamic simulations: Astrid, IlustrisTNG, Magneticum, SB28, SIMBA, and SWIFT-EAGLE. As per the halo catalogues employed in the previous sections, the galaxy catalogues used to perform the following tests contain the galaxy positions and velocity moduli.

We find that the GNN is unable to accurately predict $\Omega_{\rm m}$ for all galaxy catalogues of each simulation. We include the results in Fig.~\ref{fig:galaxies_gnn} of Appendix \ref{sec:gnn_galaxies_appendix}. This is not surprising given that the GNN was trained on $N$-body simulations and hence was not given any information regarding the intricate astrophysical and baryonic processes in galaxy distributions. Moreover, the halo-galaxy connection is known to be a complex and challenging relation \citep{moster18, behroozi19}.

\subsection{Analytic Approximations}\label{subsec:analytic_eqns}

Here we present the equations extracted from the trained GNN model using the symbolic regression method explained in Section \ref{subsec:SR_training}. The formulae for each of the hidden edge and node features, as well as for the predicted posterior mean from the final MLP, are listed in Table \ref{table:modified_eqns}. The listed RMSE values are computed by individually replacing the corresponding component in the GNN architecture with each expression while keeping all other components of the GNN unchanged and evaluating them on halo catalogues of the Gadget test set. The computed RMSE values are used to gauge the error that each approximate equation introduces. 

It is important to note that the variables $v_i$ and $v_j$ in the equations represent the initial edge features or velocity moduli. As explained in Section \ref{subsubsec:training_procedure}, these variables were normalized by the mean and standard deviation of the velocity modulus for the halos from the training set to ensure that all terms in the equations are dimensionless. Hence, the velocity modulus terms in the equations are $v_i = \frac{v_i - \mu}{\delta}$ and $v_j = \frac{v_j - \mu}{\delta}$ where $\mu = 189$ km$\rm s^{\rm -1}$ is a fixed value that was the computed mean velocity modulus for all halos in the training set and $\delta$ is treated as a free parameter. For testing on halo catalogues, we set $\delta = 129$ km$\rm s^{\rm -1}$ which is equal to the value used during training and was the standard deviation computed for all halos in the training set. On the other hand, for testing on galaxy catalogues, we tune $\delta$ to fit to each hydrodynamic simulation set as listed in Section \ref{sec:data} because we find that using the value $\delta = 129$ km$\rm s^{\rm -1}$ leads to inaccurate predictions. This is not surprising given that this value was computed for $N$-body halos which would not be expected to extrapolate to galaxies. Hence, it is possible that tuning it for different simulations can account for the halo-galaxy bias. We discuss this in more detail in Section \ref{subsubsec:galaxies_eqns}.

We also note that the presented edge equations were modified to include terms that depend only on the relative velocity moduli of the halos and their neighbors. This was done to simplify the equations and improve their interpretability. Moreover, including only the relative velocity modulus as opposed to arbitrary linear combinations of $v_i$ and $v_j$ (see equations in Table \ref{tab:sr_eqns}) enforces the symmetry between the information from the velocity of a halo and its neighbor. Furthermore, as described in Section \ref{sec:data}, the halo velocity moduli that appear in all the equations are defined with respect to the simulation box, implying that the equations also preserve Galilean invariance. We note that this modification improves the accuracy of the equations compared to the original expressions found by the symbolic regression algorithm. We include more details on this result, as well as the original equations found by the symbolic regression algorithm, in Appendix \ref{sec:sr_eqns}. In the following discussions, we only refer to the modified equations.

% We also introduce two free parameters which are the coefficient $a$ and the bias $b$ found in the equation for the final MLP layer. As we will explain later, we find that these two parameters can be tuned to improve the accuracy of the predicted $\Omega_{\rm m}$ values when evaluating the analytic equations on galaxies from hydrodynamic simulations. However, when testing the equations on halos from both $N$-body and hydrodynamic simulations, we fix their values to $a = 4.0 \times 10^{-4}$ and $b = -0.13$ which were found by the symbolic regressor. 

\begin{deluxetable*}{lccc}\label{table:modified_eqns}
\tablewidth{2pt}
\tablecaption{This table lists the analytic formulae obtained using symbolic regression for each component of the learned GNN model: the edge model, node model, and the MLP in the final aggregation layer. The last column lists the RMSE values of the analytic expressions when they are individually substituted into the GNN architecture. This evaluation is done by replacing the corresponding MLP in the edge model, node model, or final aggregation layer with the symbolic approximation while keeping all other components of the GNN unchanged. When these approximations replace all components of the GNN architecture, the RMSE of the predictions is 0.026, as shown in Fig. \ref{fig:gadget_individual}. We note that the edge equations have been modified based on physical motivations to preserve the symmetries of the data. Specifically, we modified the edge equations to depend only on relative velocity moduli $v_i - v_j$, rather than individual halo velocity modulus terms. This is done to enforce the parity between the information from the velocity of a halo and its neighbor. Compared to the predictions shown in Fig. \ref{fig:sr_eqns}, we see that using these modified equations improves the overall accuracy of the predictions. \\ The way to use these equations is as follows. First, given a halo/galaxy catalogue, a mathematical graph is constructed by considering the halos/galaxies as nodes and linking nodes by edges if their distance is smaller than $r_{\rm link}=1.35~h^{-1}{\rm Mpc}$ (see Sec. \ref{subsubsec:graphs} for details). Second, the feature of node $i$ is defined as $v_i=(|\vec{v}_i| - \mu)/\delta$, where $|\vec{v}_i|$ is the velocity modulus of halo/galaxy $i$, $\mu = 189~{\rm km~s^{-1}}$, and $\delta$  is a free parameter with units of km $\rm s^{-1}$ that needs to be adjusted for galaxy catalogues (see Section \ref{subsubsec:galaxies_eqns} and table \ref{tab:param_values} for more details). Third, the edge features $\beta_{ij}$ and $\gamma_{ij}$ between nodes $i$ and $j$ are computed using Eqs. \ref{Eq:edge_features2} and \ref{Eq:edge_features3}, respectively. Fourth, the updated edge features of the graph are computed using the below first two equations. Fifth, the updated node features are computed using the below third and fourth equations. Finally, from the updated graph we can estimate $\Omega_{\rm m}$ by using the below fifth equation.}
\tablehead{
GNN Component & Formula & RMSE}
\startdata
Edge Model: $e^{(1)}_1$  & $1.32|v_i - v_j + 0.21| + 0.12(v_i - v_j) - 0.12(\gamma_{ij} + \beta_{ij} - 1.73)$ & 0.03 \\
Edge Model: $e^{(1)}_2$ & $|1.62(v_i - v_j) + 0.45| + 1.98(v_i - v_j) + 0.55$ & 0.04\\
\hline
Node Model: $v^{(1)}_1$  & $1.21^{v_i}(0.77^{3.29 \sum_{j\in\mathcal{N}_j} e^{(1)}_1 + \sum_{j\in\mathcal{N}_j} e^{(1)}_2}) + 0.12$ & 0.02 \\
Node Model: $v^{(1)}_1 + v^{(1)}_2$ & $0.78 - \sqrt{\log(0.16 ^ {\sum_{j\in\mathcal{N}_j} e_2+\sum_{j\in\mathcal{N}_j} e_1 - 0.41v_i - 1.05})} + 1.45$ & 0.03 \\
\hline
Final MLP: $\mu_{\Omega_{\rm m}}$ & $ 4\times10^{-4} \cdot (-5.5 \sum_{i\in\mathcal{G}} v^{(1)}_2 + 2.21\sum_{i\in\mathcal{G}} v^{(1)}_1 + |0.96 \sum_{i\in\mathcal{G}} v^{(1)}_2 + 0.82\sum_{i\in\mathcal{G}} v^{(1)}_1|) - 0.103\ $ & 0.03\\
\hline
\enddata
\end{deluxetable*}

The accuracy of the equations when evaluated on the halo catalogues of the Gadget simulations is shown in the right panel of Fig. \ref{fig:gadget_individual}. As can be seen, these analytic approximations achieve similar mean relative error  ($6.7 \%$) and RMSE ($2.6 \times 10^{-2}$) as the GNN, suggesting that they are accurate representations of the trained network. We emphasize that our analytic formula predicts the posterior mean while the error bars (posterior standard deviation) are obtained from the GNN discussed in Section \ref{subsec:gnn_results}.

In the following two sections, we present the results for testing the equations on halos from the six different $N$-body simulations and four distinct hydrodynamic simulation codes, as well as galaxies from six different hydrodynamic simulation sets.

% Finally, we modified the found expressions using inductive biases motivated by the fact that the model preserves symmetries found in the data to arrive at simplified equations. These equations are evaluated on the same test set, and it can be seen in the bottom plot that they actually perform with higher accuracy than the originally found symbolic regression formulas.

\subsubsection{Halos}\label{subsubsec:halos_eqns}

We first test the robustness of the analytic equations by evaluating them on halos of the different $N$-body simulations, as we did with the GNN. We find the analytic formulae to be accurate across all simulations, with predictions of comparable mean relative errors as depicted in the lower panel of Fig. \ref{fig:same_seed}. We note that in some cases, the analytic expressions are able to extrapolate better than the GNN due to their known improved generalization abilities (e.g see \citet{subhalo_relations}). For instance, certain numerical artifacts that appear in the predictions made by the GNN for boundary cases such as halo catalogues generated with 100 or 500 minimum particle thresholds, are not present in the predictions made by the analytic expressions. We elaborate on this in Appendix \ref{sec:additional plots}. Again, this suggests that the found formulae might represent fundamental relations between the halo properties and the cosmological parameter, $\Omega_{\rm m}$, as they are not affected by the additional astrophysical processes such as gas cooling and AGN feedback. Similar to the GNN, we perform a second robustness test using halo catalogues generated with SUBFIND and find that the equations reach comparable accuracies. See Appendix \ref{sec:subfind} for more details and plots. For all these tests, the depicted errorbars are represent the inferred posterior standard deviation values obtained by the GNN model trained on the halo catalogues since we do not find an expression for this value, as discussed in Section~\ref{subsec:SR_training}.

\subsubsection{Galaxies}\label{subsubsec:galaxies_eqns}

We also test the equations on galaxy catalogues from the six hydrodynamic simulation suites: Astrid, IlustrisTNG, Magneticum, SB28, SIMBA, and SWIFT-EAGLE. We emphasize that this is not a trivial task as the GNN and the corresponding equations were trained using dark matter halos from $N$-body simulations that do not contain any information about the intergalactic dynamics or baryonic processes present in hydrodynamic simulations. There is also a complex galaxy-halo connection which can, for instance, be reflected in the relative abundances of halos and galaxies where larger halos can contain multiple galaxies while smaller halos may not contain any. These biases can possibly leave a significant imprint in the relations between the relative position and velocity terms of the equations found for halos. For these tests, we follow the definitions of galaxies and stellar mass thresholds discussed in Section \ref{subsec:catalogues} in constructing the galaxy catalogues where we include both central and satellite galaxies. 

\begin{figure*}
    \centering
    \includegraphics[width=1.0\textwidth]{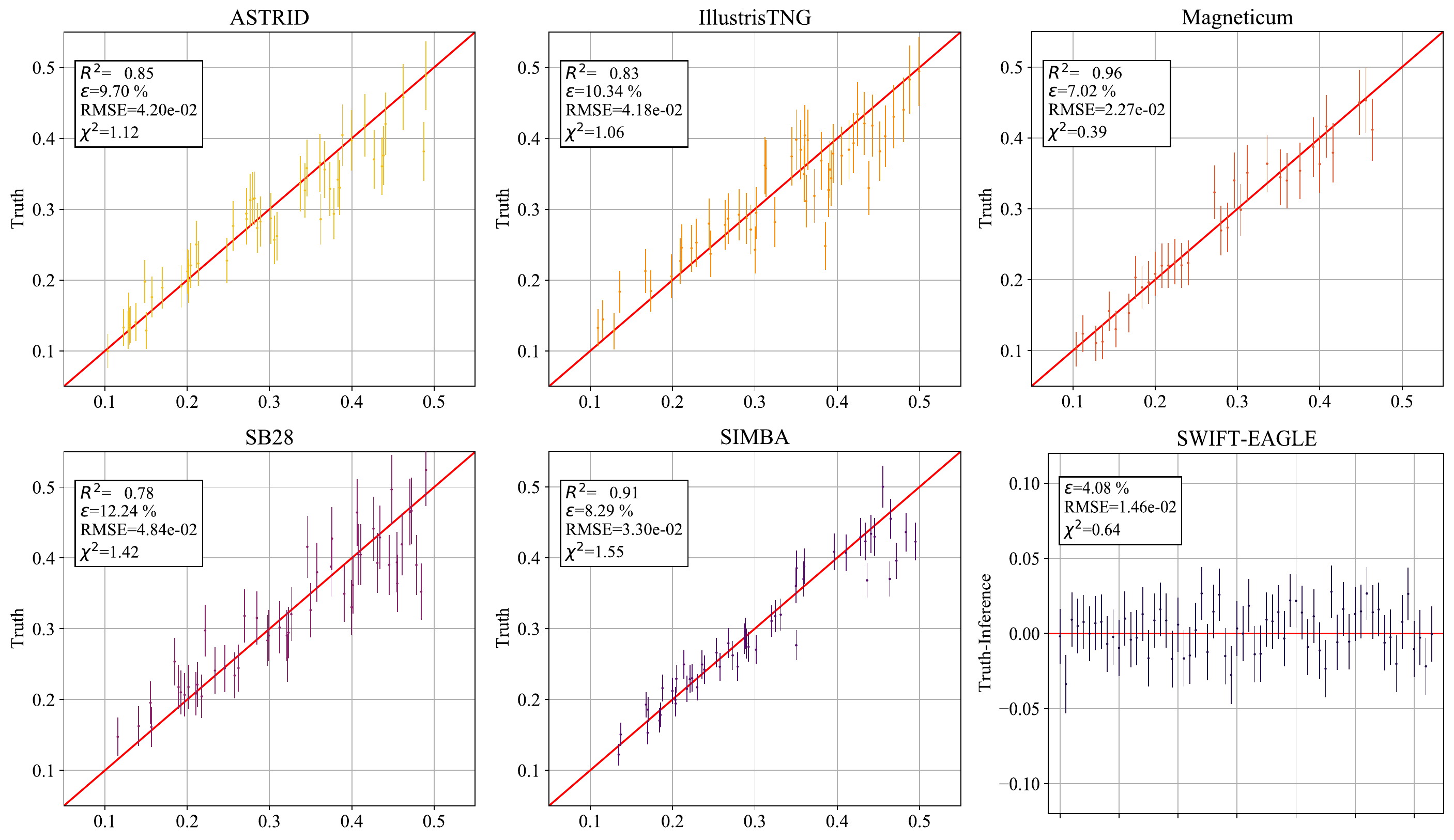}
    \caption{We test the analytic equations that were trained for halo catalogues of $N$-body simulations on thousands of galaxies from 6 different hydrodynamic simulation sets: Astrid, IllustrisTNG, Magneticum, SB28, Simba, and SWIFT-EAGLE, to predict the value of $\Omega_{\rm m}$ and plot the predicted against truth for each simulation. To conserve space, we only present results for the tests performed on catalogues constructed with a stellar mass threshold of  $4 \times m_*$ where $m_*$ is a fixed mass for an individual stellar particle as described in Section \ref{subsec:catalogues}, but we reach similar accuracies for catalogues constructed with other mass cuts. We also include only 50 randomly selected catalogues for each simulation set for the clarity of the figures but the reported metrics were computed for all simulations in the suites. Note that for the bottom right panel, which depicts the predictions for the SWIFT-EAGLE simulation set, we use simulations that are generated with the same value of $\Omega_{\rm m} = 0.3$. Thus, we plot the difference between the truth and the prediction on the y-axis for these catalogues. As depicted in Fig.~\ref{fig:eqns_galaxies_color}, a large fraction of the catalogues, particularly for the ASTRID, IllustrisTNG, and SB28 simulations, contain galaxy number densities that are outside the range of the number densities exhibited by the halo catalogues used during training of the network and equations. Hence, in this plot we remove these outliers and find that the mean relative errors of the predictions significantly decrease (see Fig.~\ref{fig:eqns_galaxies_color} for comparison). These results exhibit a relatively high accuracy with mean errors that average around $\epsilon \sim 9.4\%$, comparable to the accuracies obtained by our companion paper \citep{deSanti_2023} with model trained on galaxy properties. This further demonstrates the robustness of the equations as well as their ability to use halo properties to extrapolate to galaxy distributions. This is a surprising result given the various astrophysical processes exhibited by the hydrodynamic simulations and the complex mapping between galaxies and halos.}
    \label{fig:eqns_galaxies_errors}
\end{figure*}

\begin{figure*}\label{fig:eqns_galaxies_color}
    \centering
    \includegraphics[width=1.0\textwidth]{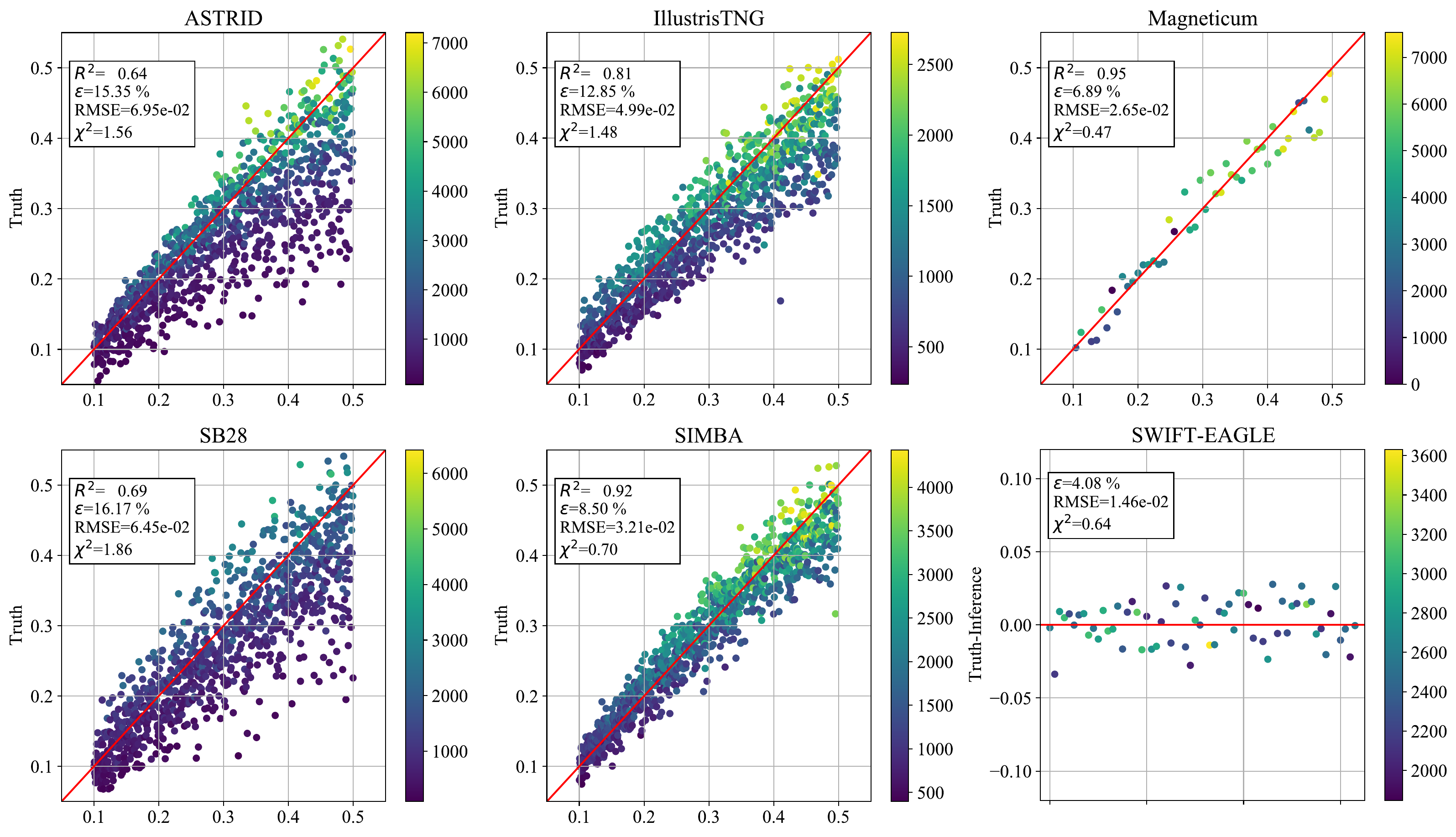}
    \caption{This figure follows the format of Fig.~\ref{fig:eqns_galaxies_errors}. Here, each scatterpoint (representing one galaxy catalogue) is colored according to the number of galaxies the catalogue contains. The colorbar depicts the range of galaxy number density present in the catalogues for the corresponding stellar mass threshold of each column. As it can be seen, a significant portion of the galaxy catalogues from simulations such as Astrid, Illustris-TNG, and SB28 contain much smaller or larger galaxy number densities than the number densities seen during training, which were within the range of $(1000, 6000)$. These catalogues account for the the relatively larger errors in these predictions is expected because the halo number density acts as an uninformative prior during the training of the GNN and equations. When we omit these outlier catalogues, we obtain smaller scatter in the results, as shown in Fig.~\ref{fig:eqns_galaxies_errors}.}
\end{figure*}

We present the results for evaluating the equations on galaxy catalogues from the different hydrodynamic simulations in Fig.~\ref{fig:eqns_galaxies_errors}. Each panel is labeled with the corresponding simulation suite. For simplicity, we present the predictions for only the galaxy catalogues generated with the stellar mass threshold of $4 \times m_*$ for a fixed $m_*$ denoting the mass of an individual stellar particle as described in Section \ref{subsec:catalogues}. However, we find that the equations are able to perform with similar accuracies for catalogues constructed with different mass cuts, which we discuss further in Appendix~\ref{sec:varying_cuts}. Moreover, since the simulations from the SWIFT-EAGLE suite are run with the same value of $\Omega_{\rm m}$, we plot the difference between the true ($\Omega_{\rm m} = 0.3$) and the predicted values on the y-axis for these catalogues. We note that the presented errorbars for all simulations are the inferred posterior standard deviation values obtained by the model trained and tested on galaxy catalogues discussed in \citet{deSanti_2023}, since the equations predict only the first moment of the posterior for $\Omega_{\rm m}$ (see Section~\ref{subsec:SR_training}).

There are are several important features to note for evaluating the equations on galaxy catalogues from the different hydrodynamic simulations. First, for each simulation we tune the parameter $\delta$ to improve the accuracy of the predictions. As discussed in Section \ref{subsec:analytic_eqns}, this parameter appears in the equation as a normalization of the velocity modulus terms $v_i$ and $v_j$, and its value varies for different hydrodynamic simulations when testing on galaxies. We tune this normalization because we noticed that using the original value $\delta = 129$ km$\rm s^{\rm -1}$, the standard deviation of the velocity moduli for all halos in the training set, resulted in predictions that deviated from the truth in terms of a slope and bias, which varies for each simulation. Thus, in Table \ref{tab:param_values} we list the values of $\delta$ that we optimize for each simulation using non-linear least squares with \textsc{scipy-optimize}\footnote{\url{https://docs.scipy.org/doc/scipy/reference/optimize.html}} for the catalogues constructed using the $4 \times m_*$ stellar mass threshold. We also compare these found values with the $\delta$ used to evaluate on halo catalogues in the table and in later discussions.

Second, after tuning this parameter, we find that the equations are able to predict $\Omega_{\rm m}$ with mean relative errors of $15.35 \%$ for ASTRID, $12.85 \%$ for Illustris-TNG, $6.89 \%$ for Magneticum, $16.17 \%$ for SB28, $8.50 \%$ for SIMBA, and $4.08 \%$ for SWIFT-EAGLE, across the four stellar mass thresholds. Evidently,  the predictions for the galaxy catalogues from ASTRID, SB28, and Illustris-TNG exhibit significantly larger error than for the halo catalogues. This can be explained by two reasons. One, there are additional astrophysical processes and dynamics present in the thousands of hydrodynamic simulations that can interfere with the equations' extrapolation ability. Given that the equations can only encode information regarding the gravitational interactions between halos from $N$-body simulations, the effects of these various astrophysical parameters may impede on the accuracy of the predictions. Moreover, there are likely to be significantly more outliers for simulations such as SB28, where we vary 28 cosmological and astrophysical parameters at a time. This is also true for the ASTRID simulations which encompass a wider range of galaxy properties and are able to encapsulate the variations found in the other simulation suites. A more detailed discussion of the wide range of characteristics in the ASTRID simulations can be found in our companion papers, \citet{deSanti_2023} and in \citet{Yueying-prep} (in preparation). 

Two, there is a large fraction of the galaxy catalogues that contain galaxy number densities outside the scope of the halo number densities seen by the GNN and equations during training. For instance, the number of halos in catalogues from the Gadget simulations used for training ranges from $\sim$1,000 to 6,000. However, there are galaxy catalogues that contain fewer than 500 galaxies at this stellar mass threshold. These outliers are particularly dominant in the IllustrisTNG, Astrid, and SB28 simulations, which leads to under-predicted values of $\Omega_{\rm m}$. This effect can be seen in Fig.~\ref{fig:eqns_galaxies_color} which contains the same plots as Fig.~\ref{fig:eqns_galaxies_errors} but with each scatter-point colored according to the galaxy number density that the catalogue contains. The colorbars accompanying each plot indicate the range of the galaxy number densities present in the catalogues. As it can be seen, in catalogues with significantly lower (higher) galaxy number densities compared to those seen in training, the value of $\Omega_{\rm m}$ is often under- (over-) predicted that contributes to the large scatter. On the other hand, if one removes these outliers, the mean relative errors significantly decrease. Hence, Fig.~\ref{fig:eqns_galaxies_errors} depicts the results for only the catalogues with galaxy number densities that fall within the range of $(1000, 6000)$. Restricting to these catalogues decreases the mean relative errors to: $9.76 \%$ for Astrid, $10.34 \%$ for IllustrisTNG, $7.02 \%$ for Magneticum, $12.24 \%$ for SB28, $8.29 \%$ for SIMBA, and $4.08 \%$ for SWIFT-EAGLE. Thus, we conclude that the equations are able to extrapolate to galaxies with with accuracies that are comparable to those attained for the halo catalogues from hydrodynamic simulations. These results are also comparable to those  obtained by our companion paper \citep{deSanti_2023} where we trained a model directly on galaxy properties. We note that the effect of the number density being an uninformative prior during the learning process can be diminished by broadening the range of halo number densities used to train the network and equations, but we leave this for future work. 

% While these predictions exhibit slightly larger scatter compared to those for $N$-body halos, it is important to acknowledge the challenging mapping between the distribution of halos to the distribution of galaxies. 

\begin{table}[ht]
\centering
\caption{In this table, we list the optimized values of the free parameter, $\delta$, which is the normalization of the velocity modulus terms used in the analytic expressions (see \ref{subsec:analytic_eqns}). We list the values for the six different hydrodynamic sets, ASTRID, Illustris-TNG, Magneticum, SB28, SIMBA, and SWIFT-EAGLE. These values were obtained using linear least squares optimization with \textsc{scipy-optimize} as described in Section \ref{subsubsec:galaxies_eqns} to achieve robustness across various simulation codes. We also include the $\delta$ used for testing on $N$-body halos for comparison.}
\begin{tabular}[t]{lccc}
\hline
Simulation &$\delta$& Simulation & $\delta$\\
\hline
$N$-body codes & $129.2$  & SB28 & $100.0$ \\
ASTRID & $126.5$ & SIMBA & $122.5$\\
Illustris-TNG & $99.6$ & SWIFT-EAGLE & $114.5$\\
Magneticum & $147.2$ \\
\hline
\label{tab:param_values}
\end{tabular}
\end{table}%

After accounting for the aforementioned details, we conclude that the equations are able to accurately predict $\Omega_{\rm m}$ for galaxy catalogues. We emphasize that the ability of the equations to achieve a reasonable inference of $\Omega_{\rm m}$, being trained on halo catalogues from $N-$body codes, is a surprising result because it is expected that baryonic effects will affect the abundance and clustering of galaxies in a complex and unknown manner. This is particularly astounding for simulations such as those from the SB28 suite that covers a vast volume in parameter space with many regions not covered by the training set (e.g. cosmological parameters like $h$, $n_s$, and $\Omega_{\rm b}$). Furthermore, the equations work really well for SWIFT-EAGLE catalogues that were created running a different halo/subhalo finder than the one used for training. Furthermore, the equations are robust to the nontrivial galaxy-halo connection as they can map the information learned about the halo position and velocity fields to those for galaxies. The ability of the equations to remain robust to these variations provide strong indication that they may be relying on fundamental relations in the galaxy and halo phase-space distribution that encodes effective information on $\Omega_{\rm m}$. Another possibility is that the equations are extracting information on scales unaffected by astrophysical dynamics. In the next section, we explore possible interpretations of these equations in more detail.

\section{Discussion}\label{sec:discussion}

Here, we discuss some speculative interpretations of the found equations. We attempt to only explain the formulae for the edge models because their functional forms are simpler than those for the node models. The edge model also solely employs physical information about the halo positions and velocity moduli so they are responsible for directly leveraging the clustering and distribution of the halos. This aligns with the analysis from \citet{gn_sr}, where it was argued that the relations used in the edge models of GNNs are analogous to describing the force laws between pairs of particles in physical systems. We will elaborate on how the edge equations found in this work may also reflect physical relations pertaining to the halo and galaxy populations. The node model, on the other hand, exhibits a more complex form because it introduces non-linearities to the formulae and makes use of information pertaining to the aggregate features from all neighboring halos. However, this should not suggest that the equations for the node model contain information that is less important than those for the edge model.

\subsection{Relative Peculiar Velocity Modulus}\label{sec:interp_velocities}
In both edge model equations, $e^{(1)}_1$ and $e^{(1)}_2$, the information regarding the velocities of the halos appear in terms in the form of $(v_i-v_j)$, which indicates that the model is taking advantage of the relative velocity moduli of the halos and their neighbors. This dependence also preserves the parity between the information content of a halo and that of its neighbor. The ability for the edge model in the GNN to employ relational information between pairs of bodies of a system has been a recognized advantage \citep{gn_sr, sr_inductive} towards understanding the physical principles underlying the model predictions. We believe that in this case, using the relative velocities allows the models to gauge the local gravitational forces where the relative velocity moduli between two halos can serve as a proxy for the depth of the potential wells in the bound system. This is reasonable since larger relative speeds of interacting bodies can result from the presence of stronger attractive forces between them. From this, the model may be learning a representation of the masses of the halos. An analogous discussion in \citet{cen} reached similar conclusions pertaining to the pairwise peculiar velocities and speeds which were found to have strong dependence on $\Omega_{\rm m}$ at the same small scale as that used by the models in this work ($\lesssim 5~h^{-1}$Mpc). 

We also speculate that the presence of these terms reflect the strong dependence of $\Omega_{\rm m}$ on the information available in the cosmic velocity fields \citep{Bernardeau_1995, cosmic_flows}. For instance, \citet{Bernardeau_1995} discusses a derived relation between the moments of the scalar field of the peculiar velocity divergence and $\Omega_{\rm m}$ that is independent of the biasing between the distribution of galaxies and the underlying dark matter density field. It is possible that the found expressions in this work reflect a similar relationship because our models have been trained using the scalar halo velocity modulus and demonstrate an accuracy that is not significantly affected by the presence of astrophysical and baryonic effects. We speculate that the network and equations may be correcting for the non-linearities of the galaxy velocity fields on smaller scales by considering the galaxy distribution and number densities. Specifically, the equations may be obtaining stochastic velocities from the relative positions of galaxies using the baryonic physics present in the hydrodynamic simulations. This information, coupled with the input pairwise velocity moduli, may then be used to compute the contribution of the galaxy velocities from the bulk flows that trace the large-scale structure of the Universe. Since the bulk flows are a consequence of the mass continuity equation which relates the large-scale density and growth rate, the equations are able to extract cosmological information on $\Omega_{\rm m}$. A similar argument was made in the formulation of the cosmic virial theorem \citet{cosmic_virial, peebles1980} which constructs a relation between the mean square relative peculiar velocity computed for galaxy pairs and the galaxy correlation functions. Hence, we emphasize the importance of leveraging both the positions and velocities of the halos/galaxies in the analytic expressions. This aligns with previous findings that using only the positions or only the velocities fails to achieve accurate inference \citep{halomass}. Our companion paper, \citet{deSanti_2023}, also reaches similar conclusions about the amount of information contained in the galaxy phase-space. Moreover, we have found that introducing additional halo properties such as the halo mass and maximum circular velocity eliminates the generalization of the expressions to various simulation codes \citep{halos_gnn}, which further indicates the robustness of the information contained in peculiar velocities for inferring $\Omega_{\rm m}$. 

%\Paco{{Perhaps you can cite your previous paper saying than using either mass of internal properties cancels the robustness of the model even with N-body halos.}}

\subsection{Velocity normalization}

Here we also discuss the implications of tuning the normalization of the velocity modulus terms, $\delta$, for galaxies from each simulation set. Previous findings in \citet{Juszkiewicz_1999, Juszkiewicz_2000} indicate that the halo-galaxy distribution bias can induce biases in pairwise velocity statistics defined using the radial separation between galaxies. Thus, we speculate that the normalization of the velocity modulus terms $v_i$ and $v_j$ in our equations reflect a similar correction to account for the fact that the spatial clustering of galaxies may not trace that of the matter field. In that case, it would expected for the values of $\delta$ to differ for various galaxy populations. Since the optimal value of $\delta$ varies across different hydrodynamic codes, we hypothesize that this parameter relates the kinematics of the galaxy velocities to their abundances. For instance, as seen in Table~\ref{tab:param_values}, the value of $\delta$ is largest for the Magneticum simulations which have been found to contain significantly higher galaxy number densities compared to the other codes \citep{deSanti_2023}. Consequently, the disparity in optimal $\delta$ values can possibly reflect the variations in the abundances of satellites in simulations of difference codes since the peculiar motion of satellites are more sensitive to small scale dynamics and their presence would thus contribute to a larger spread in the dispersion of the peculiar velocity. On the other hand, the mean galaxy number densities are smallest for IllustrisTNG and SB28, which can explain why $\delta$ is smallest for these two simulations (see Table~\ref{tab:param_values}). We leave for future work to further investigate the role of $\delta$ in the context of galaxy abundances, populations, and cosmological inference.

% Explain why we do not approximate sigma8: Not robust and Note that Sigma8 does not depend on velocities as strongly as omegaM 

\subsection{Spatial distribution and clustering}\label{sec:halo_clustering}

Next, we discuss the implications of halo clustering and spatial distribution in the found edge equations. In the first edge equation, $e_{1}$, the presence of the terms $\beta$ and $\gamma$ reflect the spatial distribution of the halos in the catalogues. Specifically, the variable $\gamma \in (0,1]$ describes the distance between two halos where its range is restricted due to its normalization by the linking radius, $r_{\rm link} \sim 1.35 ~h^{-1}{\rm Mpc}$, as described in \ref{subsubsec:graphs}. Thus, a smaller $\gamma$ would indicate a denser distribution of halos. Meanwhile, the variable $\beta \in [-1, 1]$ describes the angular orientation of a halo with respect to its neighbor and can provide information about the shape of the distribution, e.g. the filamentary structure of the cosmic web. Both parameters are used by the model to learn about the presence of large scale structures such as superclusters and filaments. 

%\Natali{What about say something more general too, that people have been pursuing the correlation of $\Omega_m$ and peculiar velocities since Peebles \cite{peebles1980} and Strauss \cite{Strauss1995} works? I have included a paragraph related to these works and that these ideas were the trigger to the development of the peculiar velocities surveys, such as SLOAN one \cite{SLOAN_catalog-2022}.}

% While we do not attempt to explain the direct relationship between the spatial distribution of halos and $\Omega_{\rm m}$, we show that the  parameters governing this information possesses a distinct and strong correlation with the value of $\Omega_{\rm m}$. As seen in Fig. \ref{fig:varying_prefactor}, when we decrease the value of the pre-factor associated with $\beta$ and $\gamma$ in $e_{1}$ to values lower than its optimal one as found by the symbolic regressor ($-0.12085$), the predictions for $\Omega_{\rm m}$ begin to exbhit larger biases at the tail values. Specifically, the predictions for $\Omega_{\rm m}$ increase in value for cosmologies of lower $\Omega_{\rm m}$ while the predictions for $\Omega_{\rm m}$ decrease in value for cosmologies of higher $\Omega_{\rm m}$. This may indicate that 

% \begin{figure*}
%     \centering
%     \includegraphics[width=0.99\textwidth]{varying_beta_prefactor.pdf}
%     \caption{}
%     \label{fig:varying_prefactor}
% \end{figure*} 

\section{Conclusions} \label{sec:conclusion}
In this work we have found an analytic expression that approximates the relation employed by a GNN that was trained to infer $\Omega_{\rm m}$ from dark mater halo catalogues. This was motivated by the results of \citet{halos_gnn} which found that GNNs are able to perform accurate field-level inference of $\Omega_{\rm m}$ using halo catalogues from various $N$-body and hydrodynamic simulations. These results imply that the found relation could be a fundamental one as it is not affected by varying numerical errors, astrophysical processes, subgrid physics, or even halo definitions. This motivates us to gain a better understanding of the learned relation by approximating it with symbolic equations that are more physically interpretable than a neural network. 

%\Natali{What about say ``analytic expressionS'', because you found more than one, right?} \Helen{Yes there are multiple expressions (because of the multiple components of the GNN), but they assemble into a single equation that predicts $\Omega_{\rm m}$}

To derive the analytic approximations, we followed a two-step approach. We first simplified the model that was used in the previous work to obtain a GNN with reduced latent space dimensionality. The intention for this step was to maintain the accuracy and precision of the model discussed in \citet{halos_gnn} while building a less complex, and hence more easily interpretable, architecture. We train our \textit{compressed} model on catalogues that only contain the positions and peculiar velocity moduli of dark matter halos from $N$-body simulations. Next, we trained a symbolic regressor to fit equations to each component of the trained GNN (see Scheme \ref{fig:scheme}).

We summarize the main results of this work below:

\begin{itemize}
    \item We train a compressed GNN architecture composed of only 1 message passing layer and 2 hidden features on halo catalogues from the Gadget $N$-body simulations. We find that it is able to achieve precise constraints on $\Omega_{\rm m}$ with a mean relative error of $\epsilon\sim6.5 \%$, similar to the GNN model with larger latent space dimensionality as discussed in \citet{halos_gnn}, which achieved a mean relative error of $\epsilon\sim5.6 \%$. 

    \item The compressed GNN model, trained on Gadget simulations, is also robust across thousands of halo catalogues generated from five different $N$-body codes --Abacus, CUBEP$^3$M, Enzo, PKDGrav3, Ramses-- and four different hydrodynamic codes that employ different galaxy formation implementations: Astrid, IllustrisTNG, Magneticum, SIMBA. This model reproduces the results of \cite{halos_gnn} where the non-triviality of this robustness was discussed.

    \item We use symbolic regression to find equations that approximate the different MLPs that our GNN model is comprised of. These analytic equations can approximate the learned relation between $\Omega_{\rm m}$ and the input halo properties with a mean relative error of $\epsilon\sim6.7 \%$ when evaluated on halos from Gadget $N$-body simulations. We then evaluate the equations on thousands of $N$-body and hydrodynamic simulations run with the different codes listed above. Thus, we demonstrate that the equations are able to reproduce the preciseness and robustness of the GNN, concluding that they are successful approximations of the learned network.

    \item We further find that the equations are able to extrapolate better than the GNN in certain cases. Specifically, we test on galaxy catalogues from six different hydrodynamic simulation suites and find that while the equations are able to predict the value of $\Omega_{\rm m}$ accurately while the GNN is unable to. This is a surprising feat given that the equations were trained only on halo properties from $N$-body simulations and were not given any information regarding the complex baryonic effects and astrophysical feedback processes present in galaxy interactions. This also demonstrates that the equations may be exploiting a relation between positions, velocities, and $\Omega_{\rm m}$ that is independent of the halo-galaxy connection. %However, we emphasize that one parameter need to be tuned in the equations to obtain good results. %may encode information regarding the nontrivial galaxy-halo connection as they are able to use the position and velocity fields of halos to extrapolate to galaxy distributions. \Paco{I think this sentence is not correct. I think the network/equation may have learned a relation between positions and velocities that is independent of the halo-galaxy connection}
    
    \item To obtain good accuracies in the galaxy catalogues we need to tune one single free-parameter, $\delta$, which is the normalization of the velocity modulus terms used in the analytic expressions. The value of $\delta$ appears to be sensitive to the characteristics of the considered galaxy population. We leave for future work in studying its physical role as well as the best strategy to constrain it - such as fitting it using a subset of data, marginalizing over its values, or others.
    
    %Tuning this parameter to each hydrodynamic simulation is key for obtaining robust results when evaluating on galaxy catalogues. 
    %Since the value of $\delta$ varies across simulations, this parameter depends not only on galaxy abundance but also on the inherent characteristics of the galaxy populations. Thus, one would need to employ an independent constraint on $\Omega_{\rm m}$ to tune the fitting. 

    \item As in our companion paper \citep{deSanti_2023}, we find some robustness to super-sample covariance effects, although further work is needed to properly assest it taking into account the setup we used to train our models. Further details are presented in Appendix \ref{sec:supersample_cov}.
    
    \item We attempt to provide physical interpretation of the equations for the edge component of the GNN, which could reflect physical laws and forces between interacting objects represented by the nodes of the graph. Specifically, the equations demonstrate an explicit dependence on the pairwise velocity modulus and relative positions of halos/galaxies at separation distances $\lesssim 1.35 ~h^{-1}$Mpc. These dependencies illustrate how the rotational and translational symmetries present in the data are maintained and exploited by the model. Moreover, the dual reliance on the spatial and velocity fields of the halos indicate that there is robust information embedded in the phase-space distribution of halos, perhaps reflecting some underlying physical law like the continuity equation. We draw speculative connections to past works that have analyzed similar information in observational fields at the same scales, such as the pairwise velocity and speed statistics as analyzed in \citet{cen, Juszkiewicz_1999, Juszkiewicz_2000}, and the use of cosmic velocity fields as seen in \cite{Bernardeau_1995, cosmic_flows}.
    %\Paco{I'll rephrase this paragraph.}
    
    %We hypothesize that the equations may be relying on scales larger than the halo splashback radius that are unaffected by intergalactic dynamics because
    
\end{itemize}

\section{Acknowledgments}

We thank Lucy Reading-Ikkanda for creating Fig. \ref{fig:scheme}. We thank Ravi Sheth, Oren Slone, David Spergel, Ben Wandelt, Michael Strauss, Oliver Philcox, Gigi Guzzo, Marina Silvia Cagliari, and Miles Cranmer for the enlightening discussions. NSMS acknowledges financial support from FAPESP, grants
\href{https://bv.fapesp.br/en/bolsas/187647/cosmological-covariance-matrices-and-machine-learning-methods/}{2019/13108-0} 
and \href{https://bv.fapesp.br/en/bolsas/202438/machine-learning-methods-for-extracting-cosmological-information/}{2022/03589-4}. The CAMELS project is supported by the NSF grant AST 2108078. EV is supported by NSF grant AST-2009309 and NASA grant 80NSSC22K0629. 
EH acknowledge supported by the grant agreements ANR-21-CE31-0019 / 490702358 from the French Agence Nationale de la Recherche / DFG for the LOCALIZATION project.
KD acknowledges support by the COMPLEX project from the European Research Council (ERC) under the European Union’s Horizon 2020 research and innovation program grant agreement ERC-2019-AdG 882679 as well as by the Deutsche Forschungsgemeinschaft (DFG, German Research Foundation) under Germany’s Excellence Strategy - EXC-2094 - 390783311.
TC is supported by the INFN INDARK PD51 grant and the FARE MIUR grant ‘ClustersXEuclid’ R165SBKTMA.
The research in this paper made use of the SWIFT-EAGLE open-source simulation code \citep[\url{http://www.swiftsim.com},][]{2018ascl.soft05020S} version 1.2.0.

\bibliography{references,gnn_papers}{}
\bibliographystyle{aasjournal}

\appendix

\section{Additional $N$-body and Hydrodynamic Simulations \label{sec:additional plots}}
In Section \ref{sec:results}, we presented the perfromance of the GNN and analytic expressions on the different $N$-body and hydrodynamic simulations run with the same cosmologies and initial conditions. Here, we present additional results demonstrating the accuracy and robustness of both model predictions for both $\Omega_{\rm m}$ and $\sigma_{\rm 8}$ for different minimum halo particle thresholds. For these plots, we evaluate the models on 50 simulations containing different cosmologies and initial conditions for four different $N$-body codes: Abacus, CUBEP$^3$M, PKDGrav, and Ramses (in Fig. \ref{fig:Nbody_SR}). We also test the models on 1,000 simulations from two hydrodynamic codes: IllustrisTNG and SIMBA, but plot the results for 50 randomly selected simulations to conserve space (in Fig. \ref{fig:omegaM_Hydro}). As before, we perform these tests using halo catalogues created with different minimum halo particle thresholds as indicated in the plots. Each of these plots depict the predictions plotted against the truth minus the inference. 

\begin{figure*}
    \centering
    \includegraphics[width=.45\textwidth]{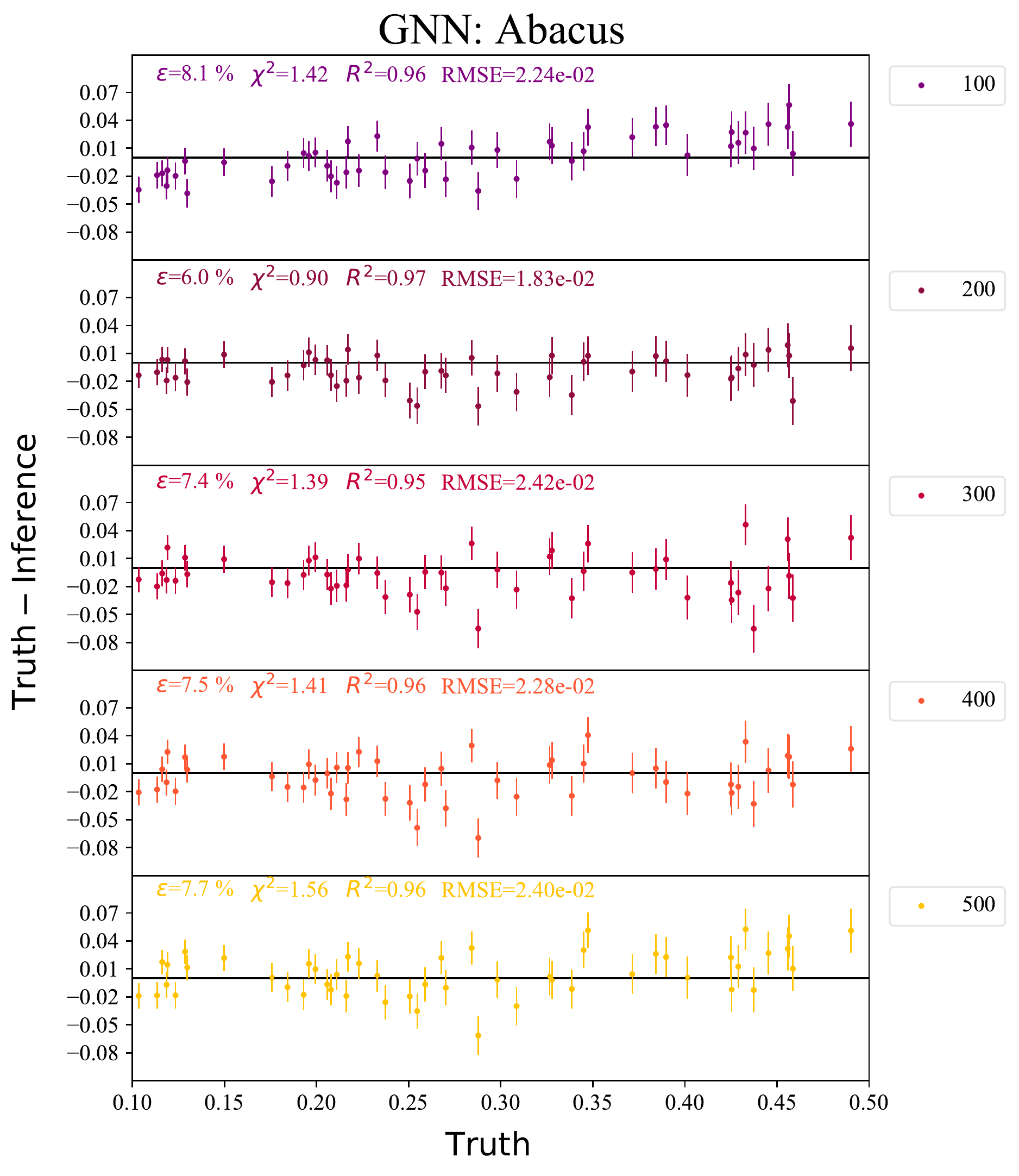}
    \includegraphics[width=.45\textwidth]{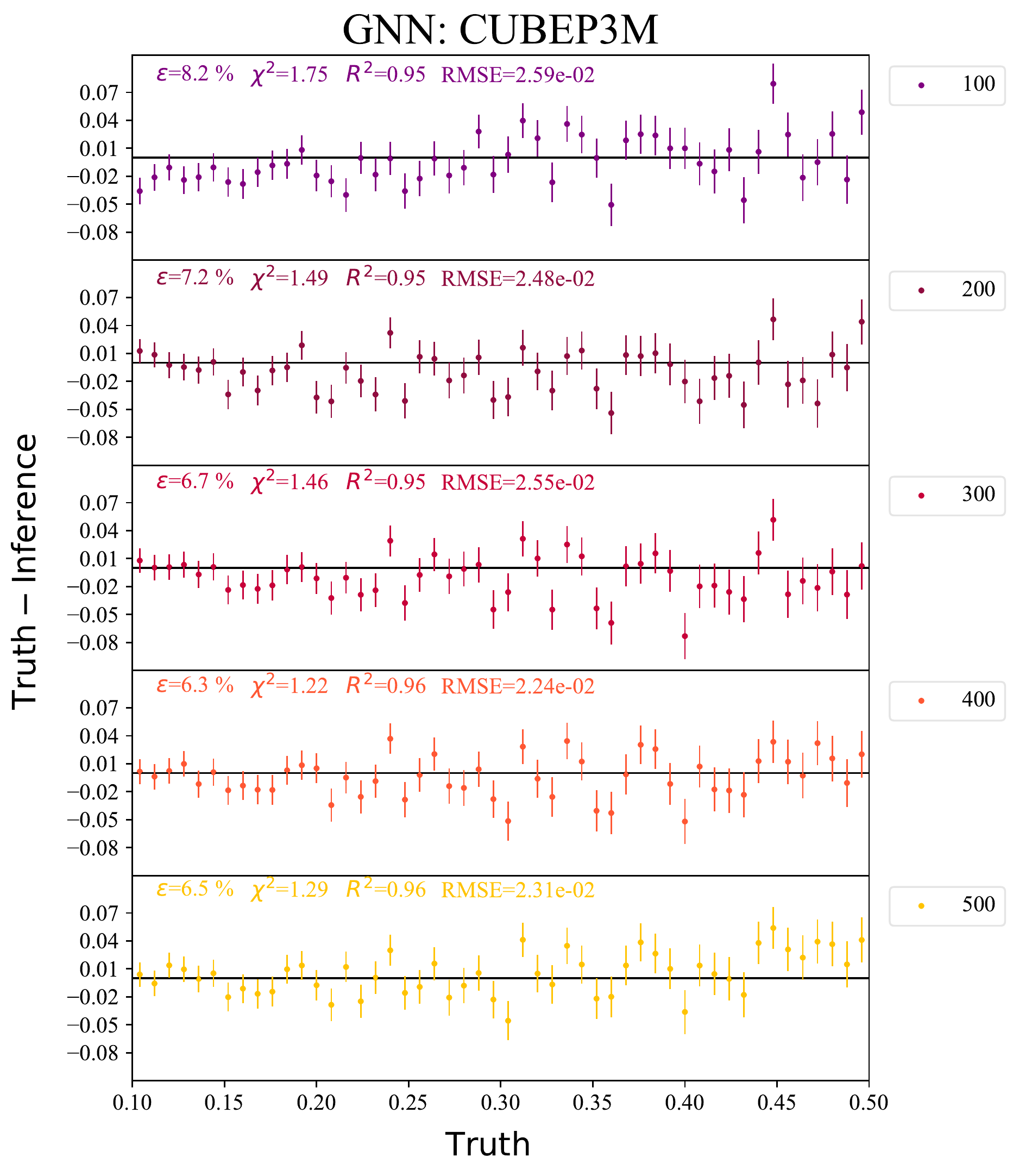}
    \includegraphics[width=.45\textwidth]{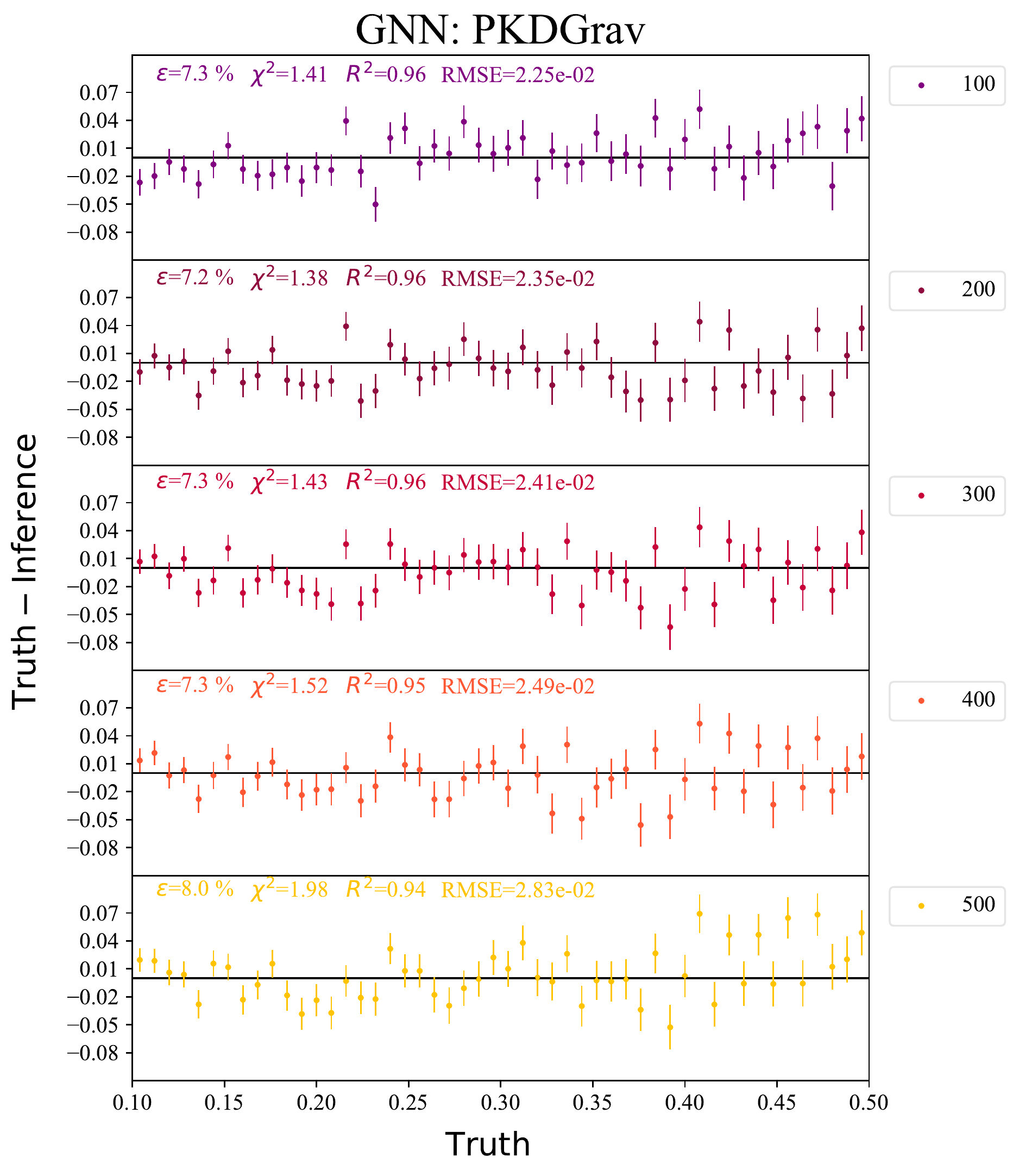}
    \includegraphics[width=.45\textwidth]{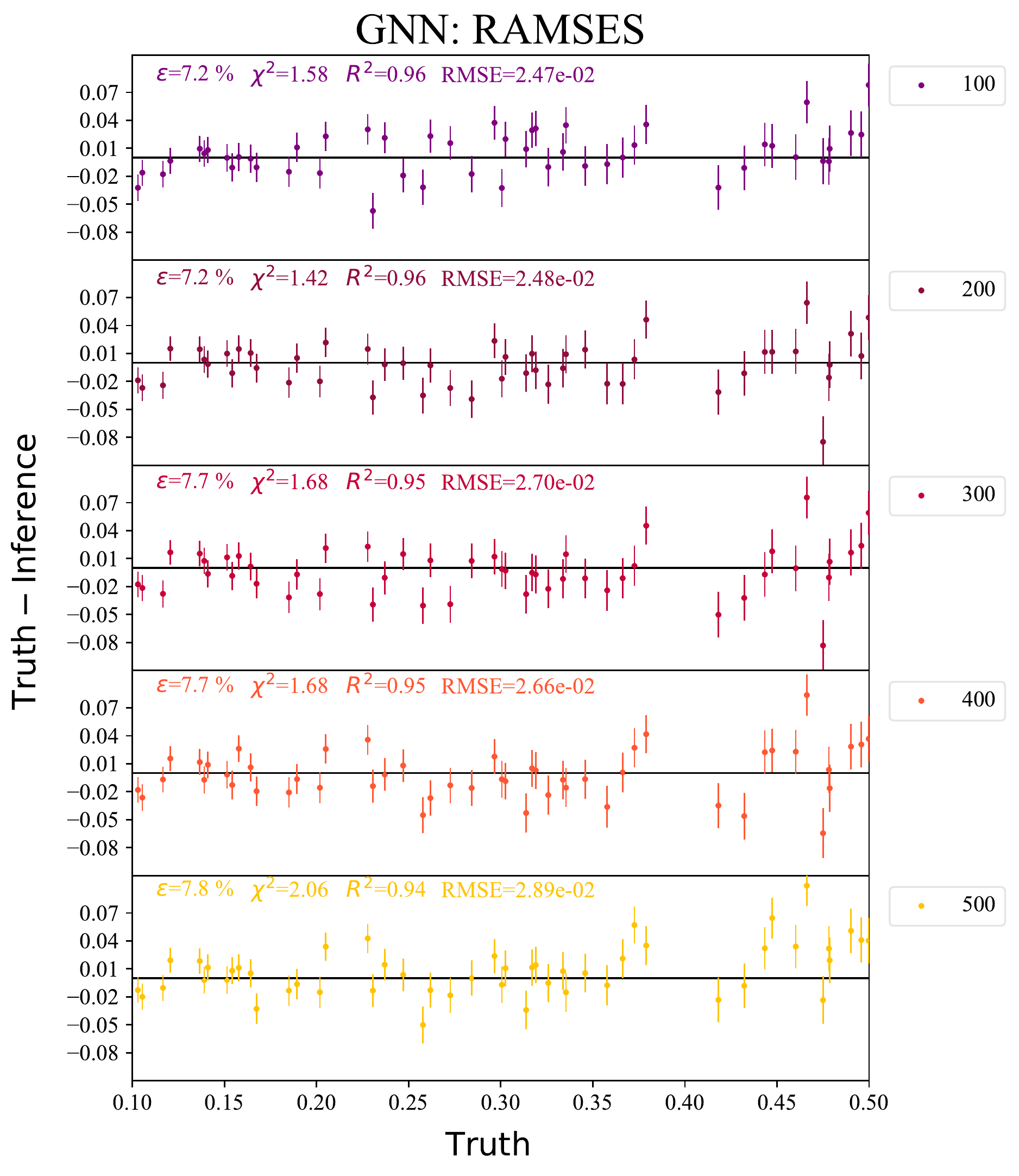}
    \caption{We train a GNN with a low-dimensional latent space to infer $\Omega_{\rm m}$ from catalogues of the Gadget $N$-body simulations using the halo relative positions and velocity moduli. We then evaluate this model on different $N$-body simulations: Abacus, CUBEP$^3$M, PKDGrav3, and Ramses using catalogues created with particle thresholds indicated next to the plots. As can be seen, the model is able to extrapolate well to different $N$-body codes and is able to predict with similar accuracy compared to that of the halo catalogues from Gadget.}
    \label{fig:Nbody_GNN}
\end{figure*}

\begin{figure*}
    \centering
    \includegraphics[width=.45\textwidth]{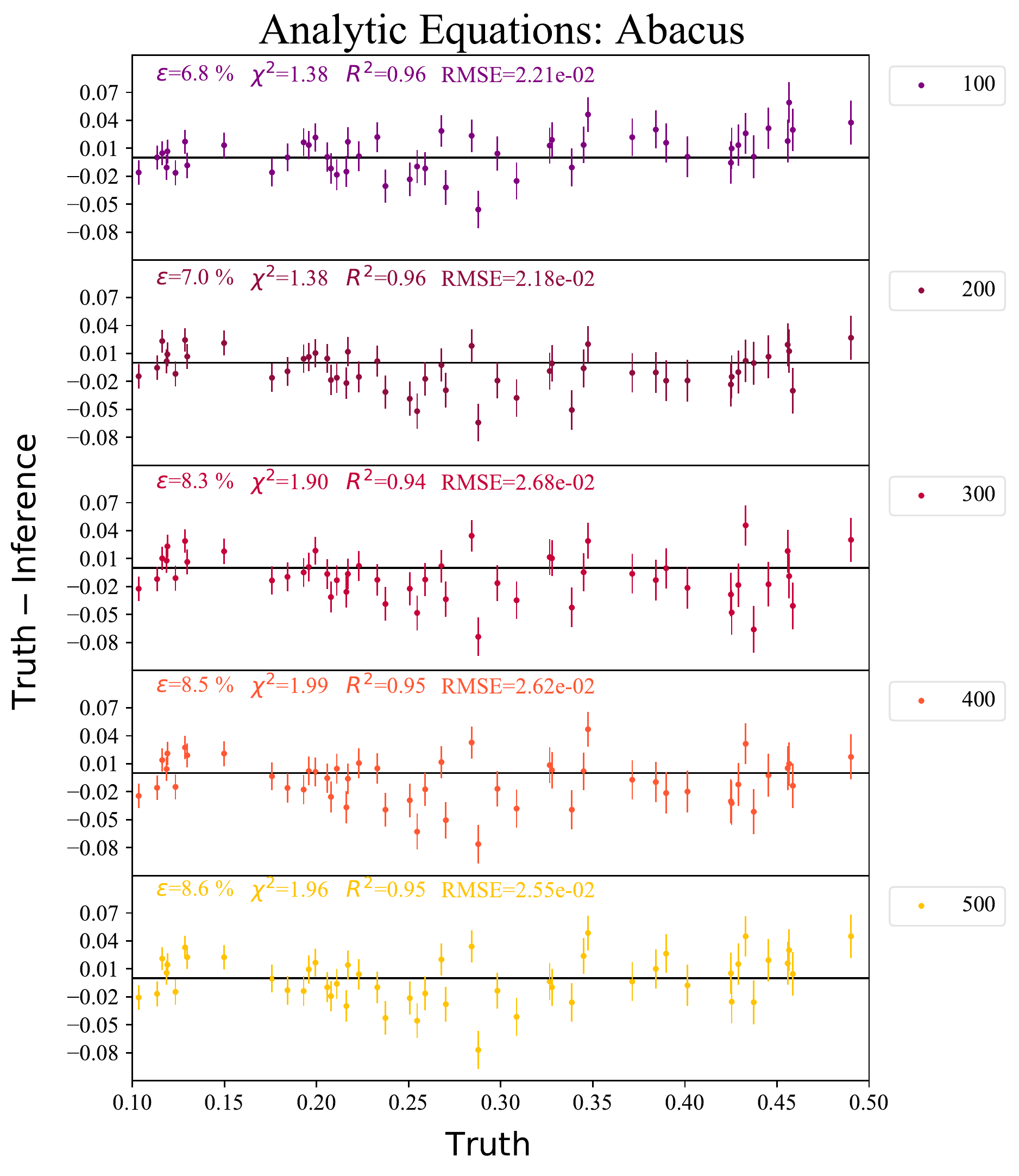}
    \includegraphics[width=.45\textwidth]{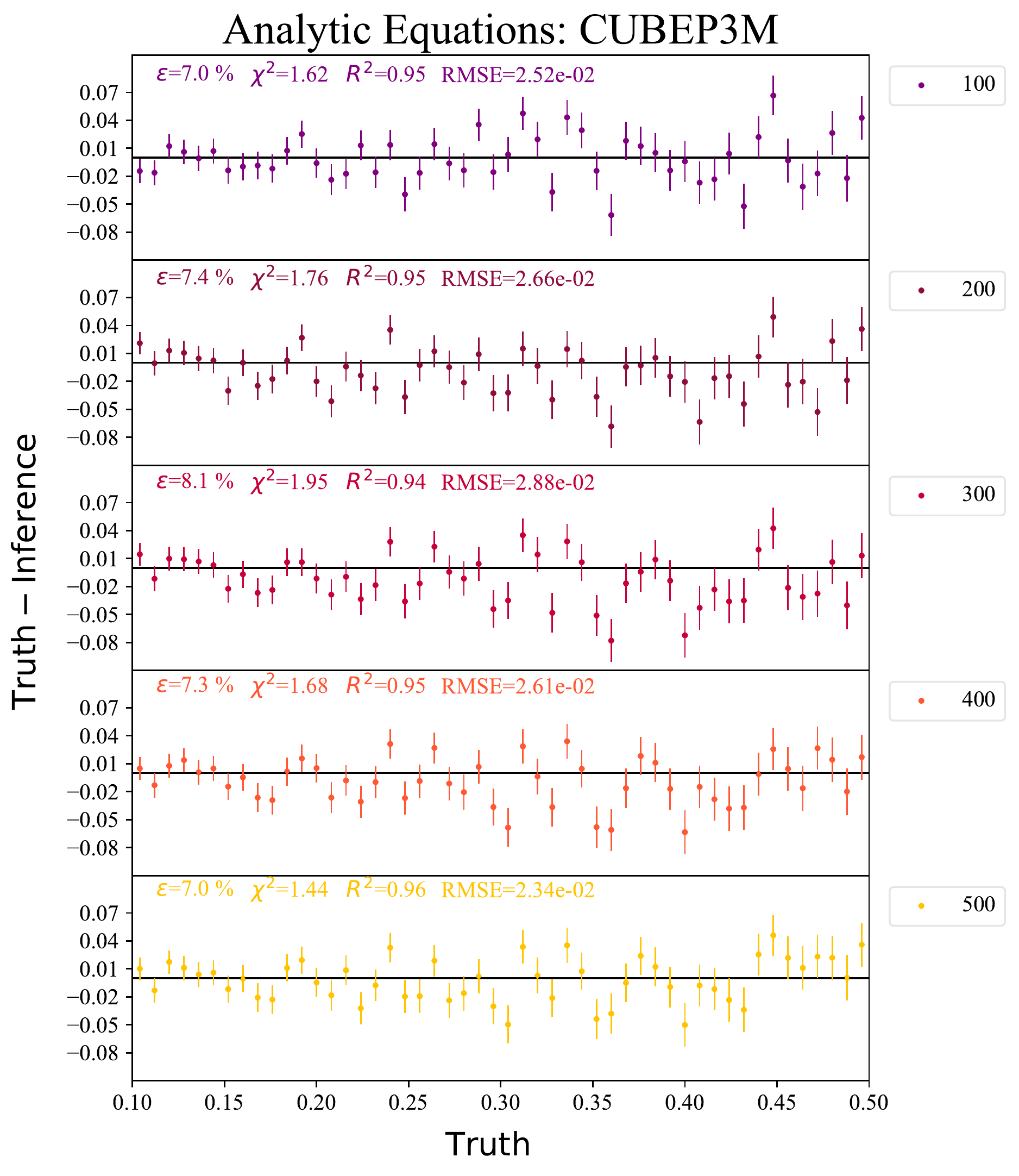}
    \includegraphics[width=.45\textwidth]{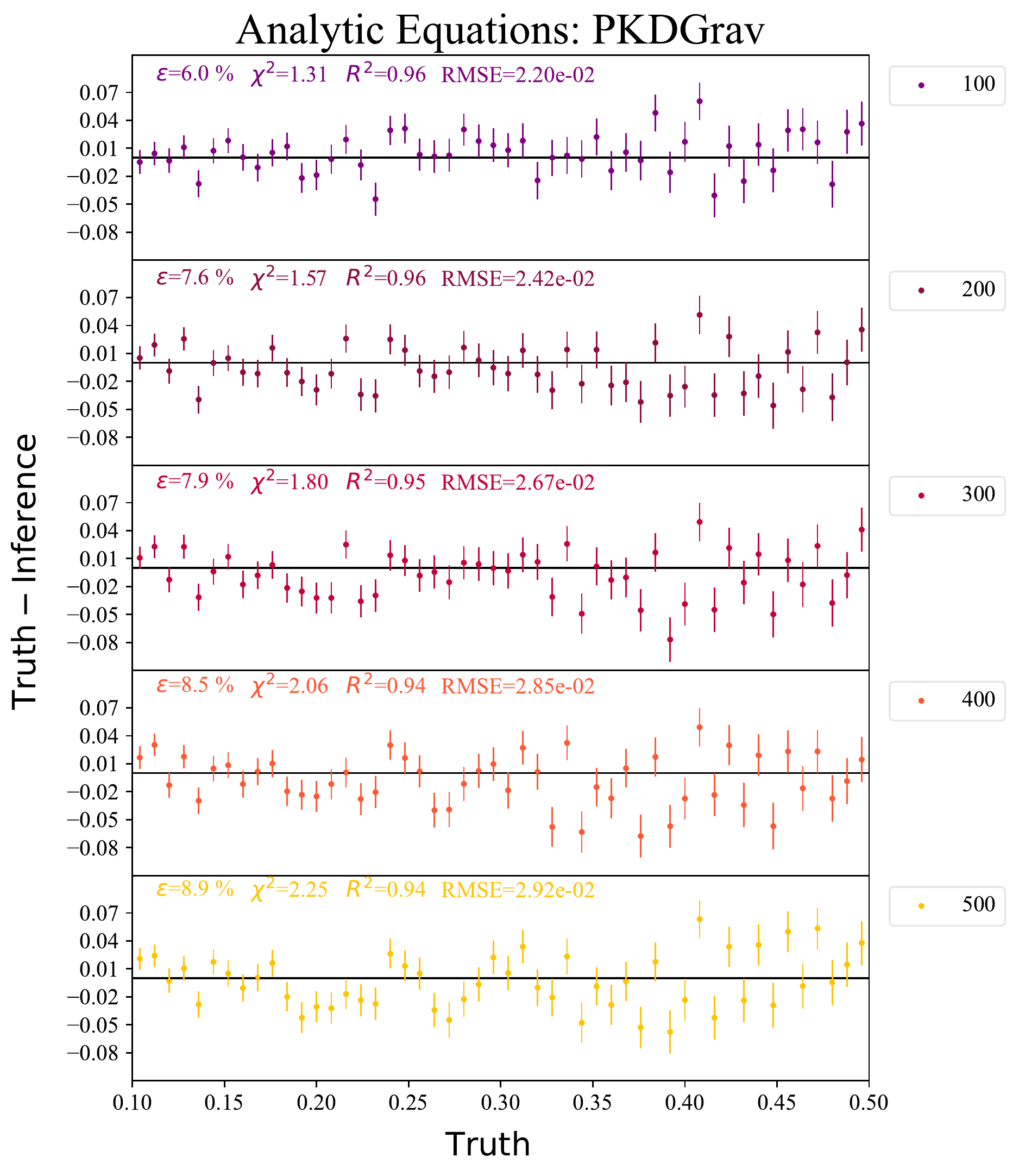}
    \includegraphics[width=.45\textwidth]{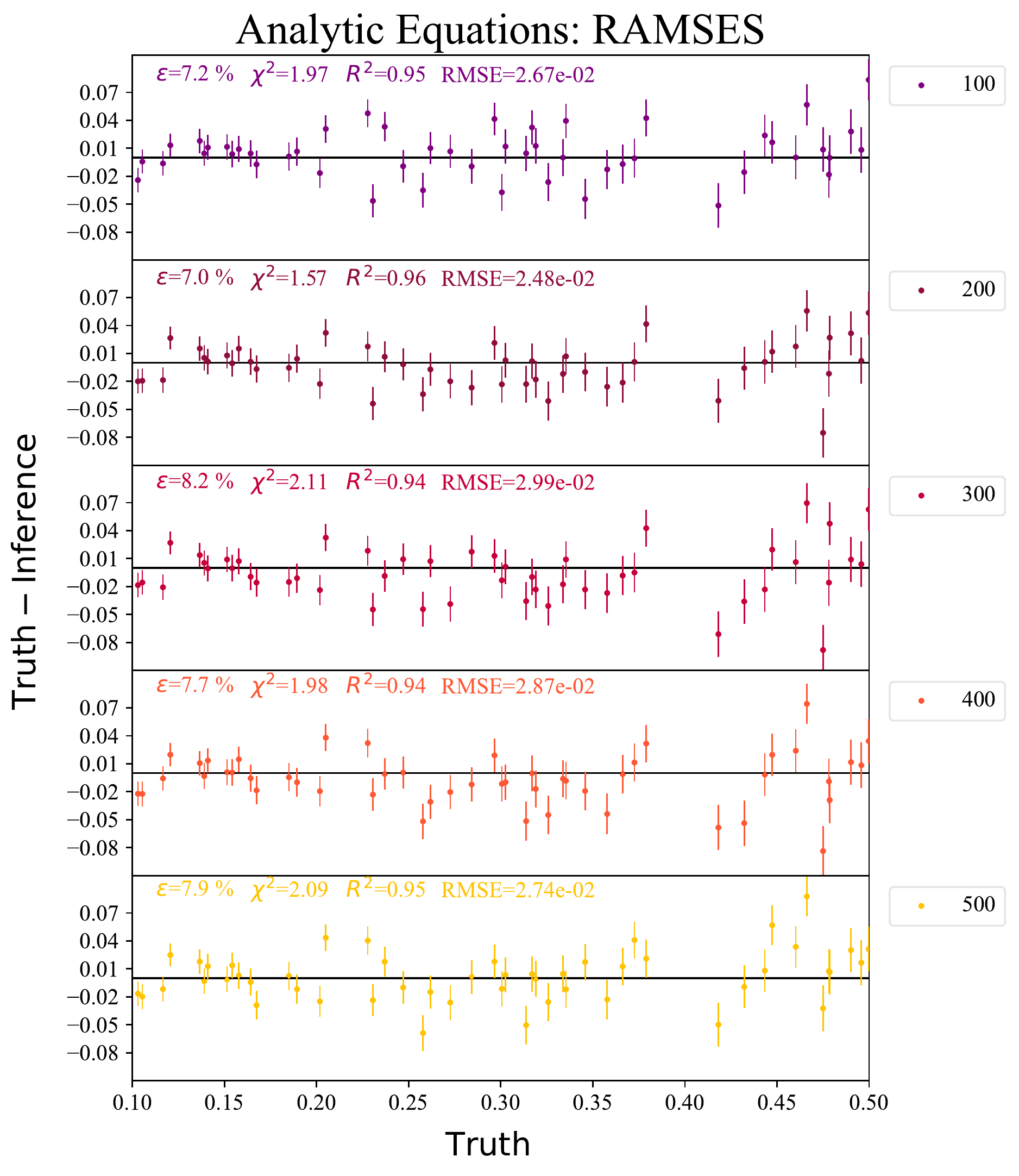}
    \caption{This follows the format as Fig. \ref{fig:Nbody_GNN} but for the analytic equations discussed in Section \ref{sec:sr_eqns}. The equations were found using symbolic regression and modified using physical principles to preserve the rotational and translational symmetries of the data. As can be seen, the equations maintain the accuracy and robustness exhibited by the GNN in Fig. \ref{fig:Nbody_GNN}, indicating that the formulae offer good approximations to the model.}
    \label{fig:Nbody_SR}
\end{figure*}

\begin{figure*}
    \centering
    \includegraphics[width=.45\textwidth]{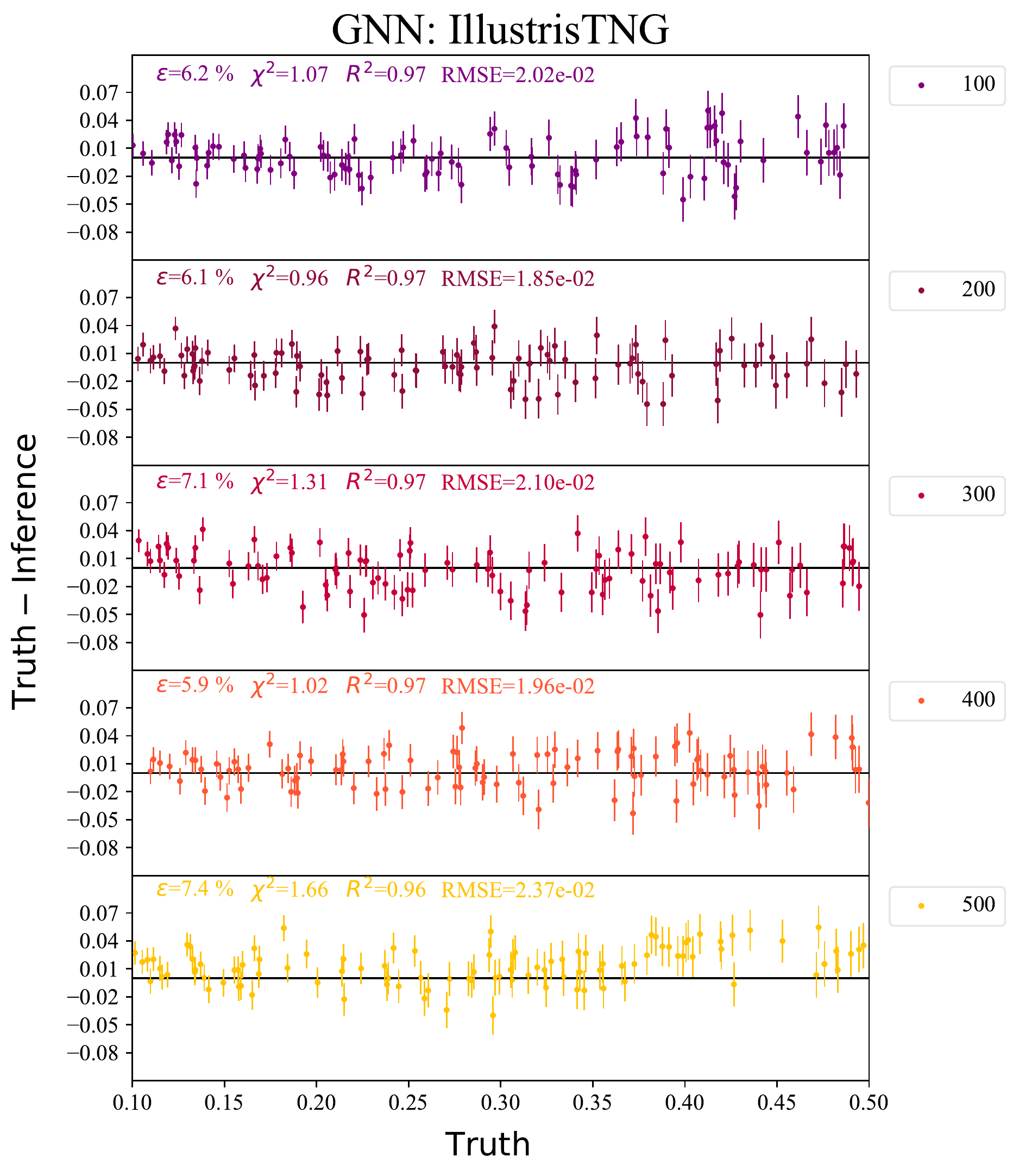}
    \includegraphics[width=.45\textwidth]{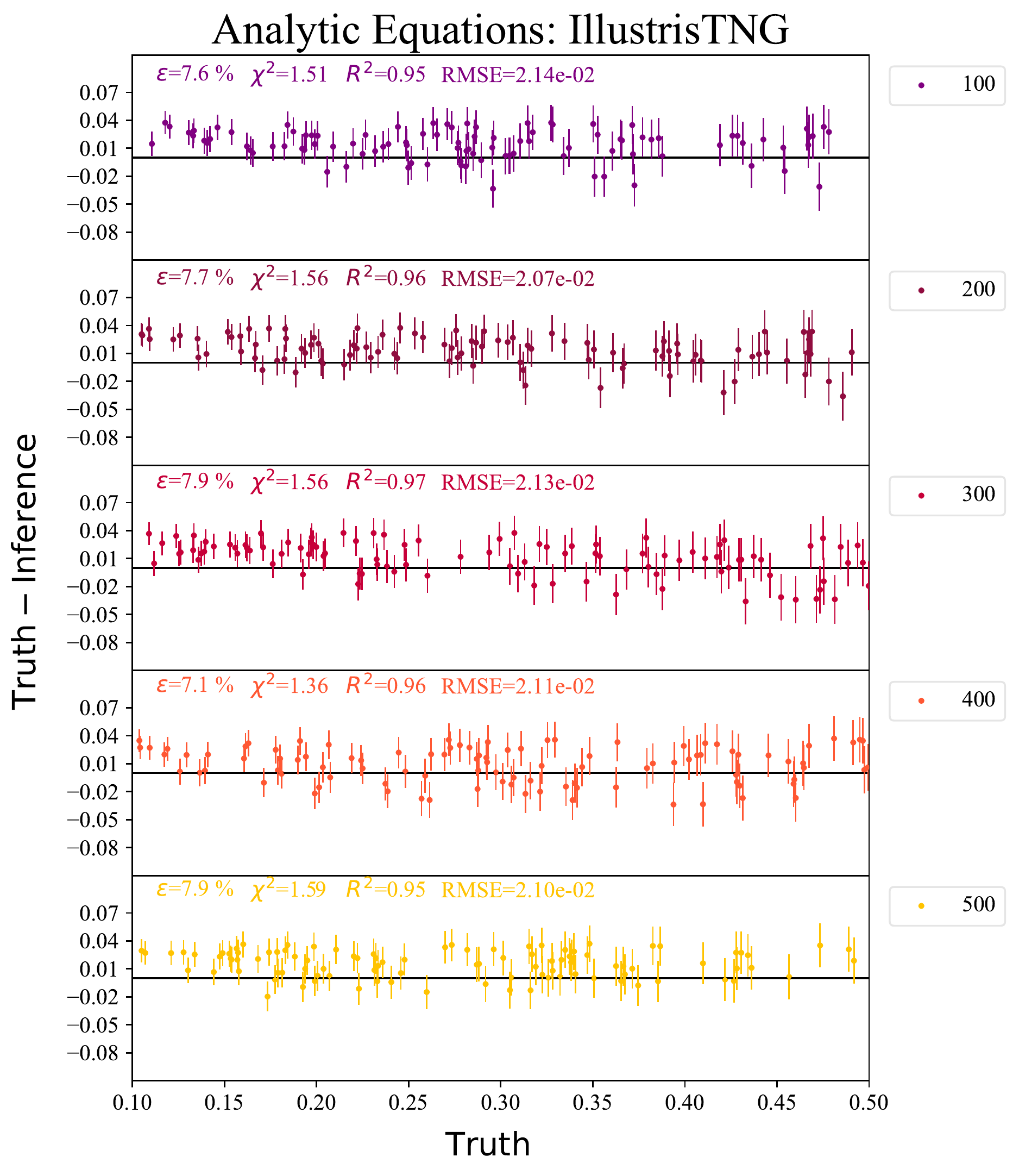}
    \includegraphics[width=.45\textwidth]{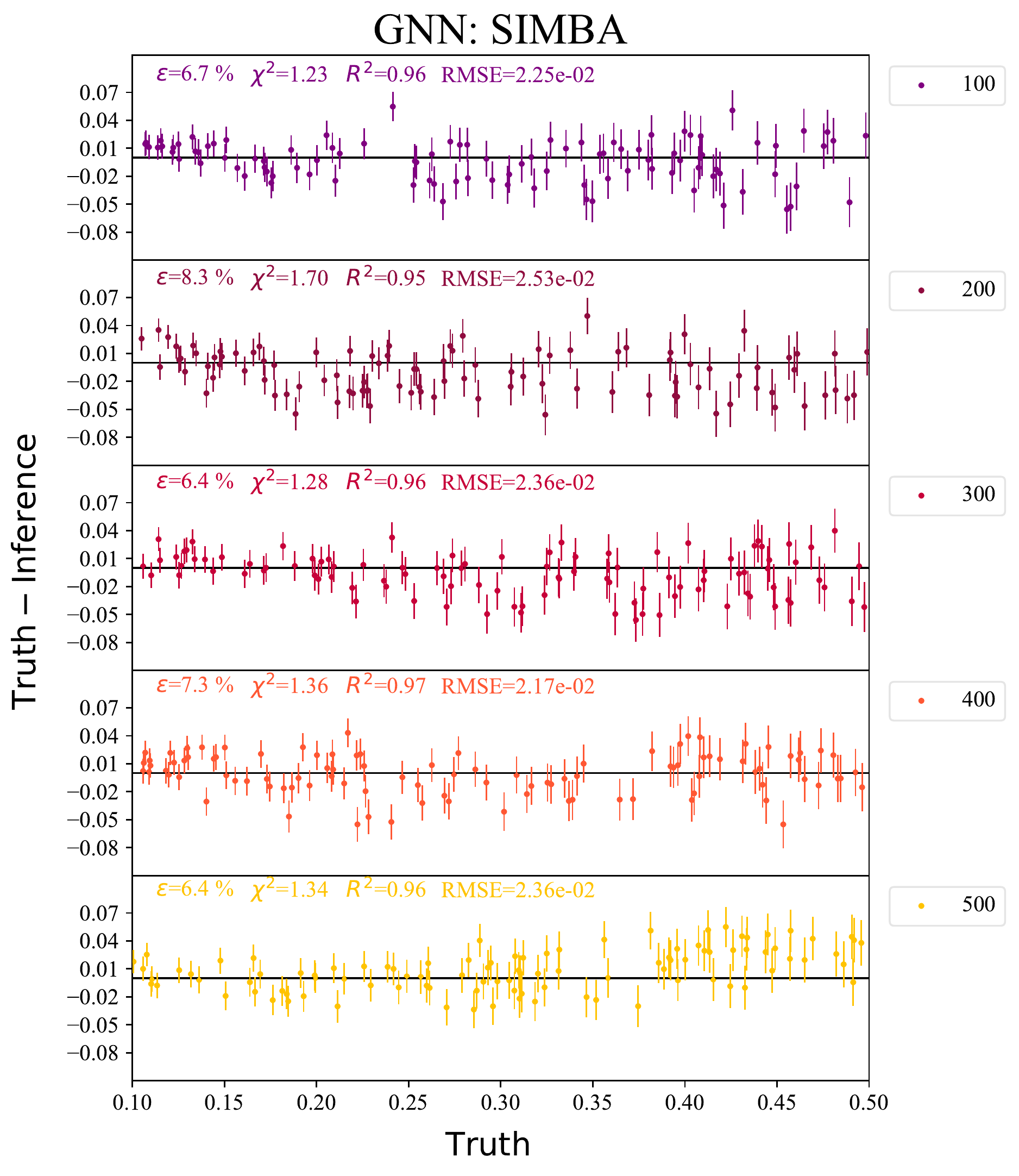}
    \includegraphics[width=.45\textwidth]{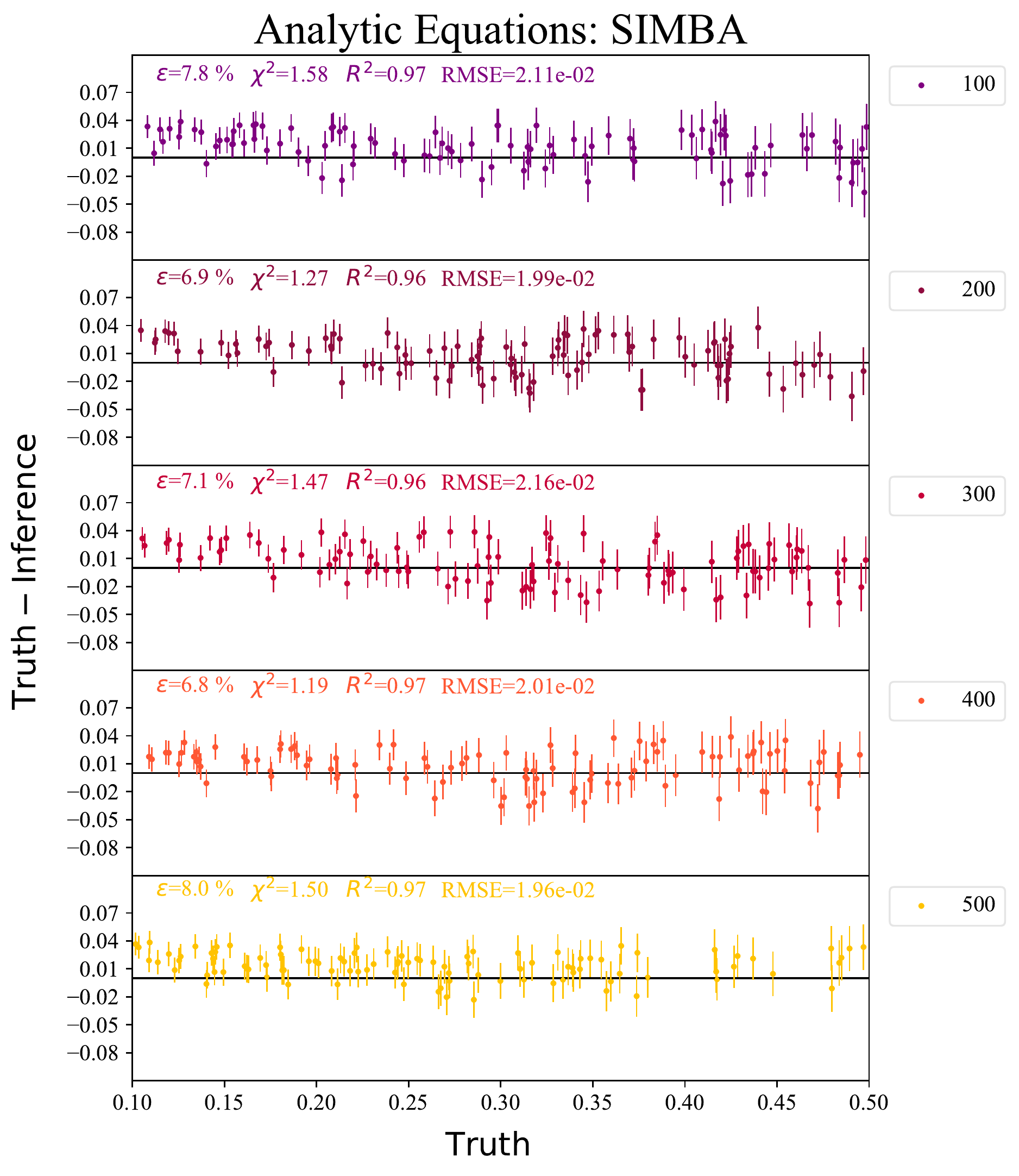}
    \caption{Similar to Figures \ref{fig:Nbody_GNN} and \ref{fig:Nbody_SR}, we test the GNN and the analytic equations on halo catalogues generated from the SIMBA and IllustrisTNG hydrodynamic simulations. For clarity, we plot the predictions for 50 randomly selected catalogues in each panel. It can be seen that both models remain robust to the additional astrophysical effects present in these simulations, indicating that they are employing a possibly fundamental relation between the relevant halo properties and $\Omega_{\rm m}$. Moreover, the analytic equations are able to capture this as its accuracies for all hydrodynamic simulations are similar to that of the GNN.}
    \label{fig:omegaM_Hydro}
\end{figure*}

As it can be seen, the GNN is able to infer $\Omega_{\rm m}$ accurately for all $N$-body simulations with similar mean relative errors of $\sim 7\%$ and the analytic expressions have comparable accuracies of $\sim 8\%$ (see Figs. \ref{fig:Nbody_GNN} and \ref{fig:Nbody_SR}).

For the hydrodynamic simulations IllustrisTNG and SIMBA, we obtain concurring results where both the GNN and analytic expressions are able to attain mean relative errors of $\sim 7\%$ (Fig. \ref{fig:omegaM_Hydro}). An interesting note is that the GNN predictions for halo catalogues constructed with 100 or 500 minimum particle thresholds exhibit tail biases due to the effects of the prior distribution, as seen in the right panels of \ref{fig:omegaM_Hydro}. These numerical artifacts are not present in the inferences made by the analytic expressions due to their known better generalization capabilities.

\section{Robustness to different halo finder: \textsc{SUBFIND}}\label{sec:subfind}

In Section \ref{sec:results}, we discussed the accuracy and robustness of the GNN and analytic expressions when evaluated on various simulation codes. Here, we present another test for the robustness of these models where we evaluate the GNN and the analytic approximations on halo catalogues generated using a different halo finder (\textsc{Subfind}) than the one used for training (\textsc{Rockstar}). \textsc{Subfind} identifies halos by determining local peaks in the three-dimensional density field and separating them using saddle points. The overdense regions and their surroundings are then examined for subhalos, which are gravitationally self-bound regimes. Those that are not bound are attached to their neighboring overdensities with whom they share saddle points. \textsc{Subfind} operates on all particle types in the simulations, dark matter and baryonic alike \cite{Dolag2009} .

To perform these tests, we consider the total mass of the halo contained in a sphere with a mean density that is 200 times the mean density of the Universe at redshift $z=0$. Same as the previous tests, we construct halo catalogues with varying minimum particle thresholds in the range of $[100, 500]$, as explained in Section \ref{sec:data}. 

\begin{figure*}
    \centering
    \includegraphics[width=.45\textwidth]{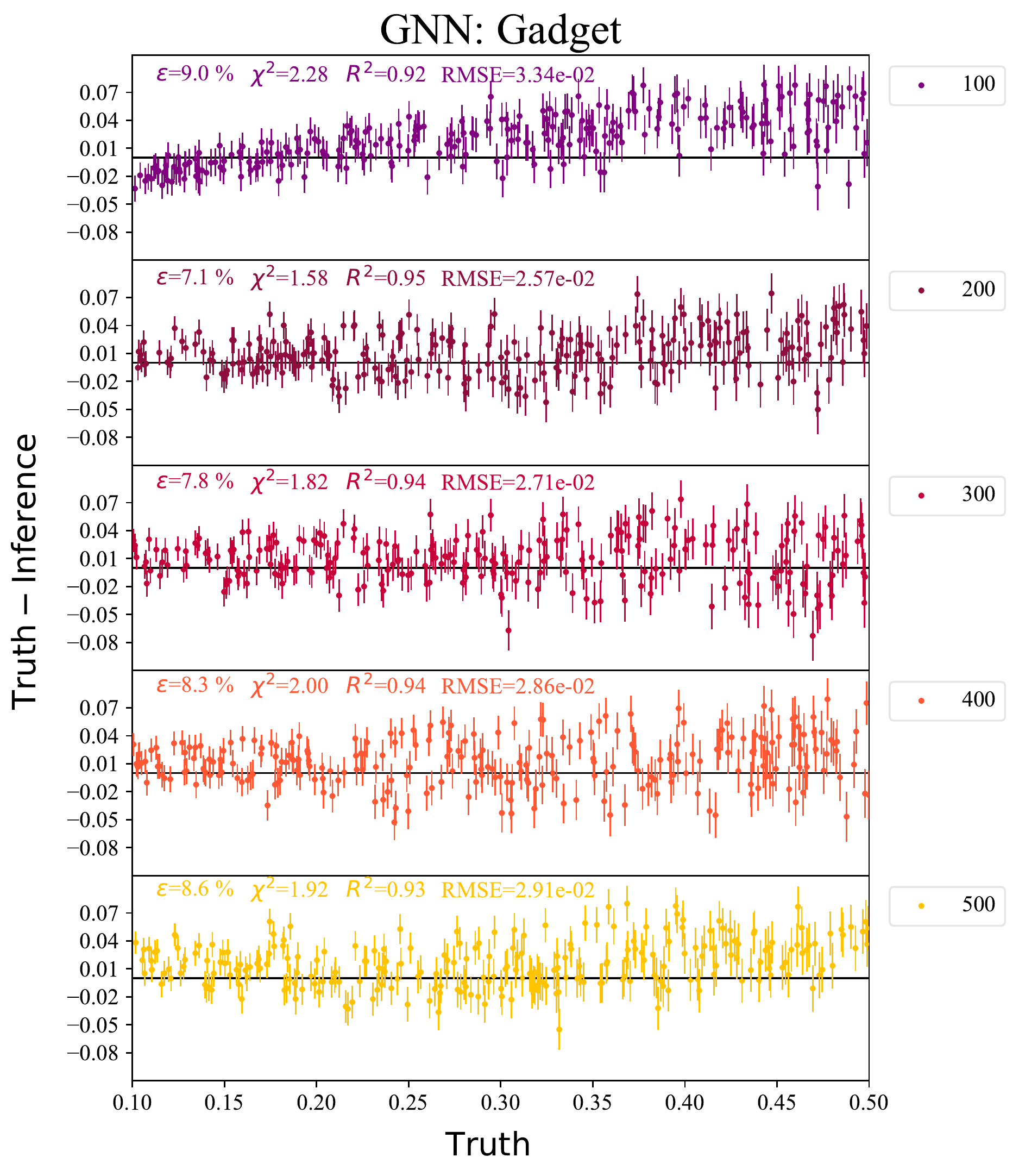}
    \includegraphics[width=.45\textwidth]{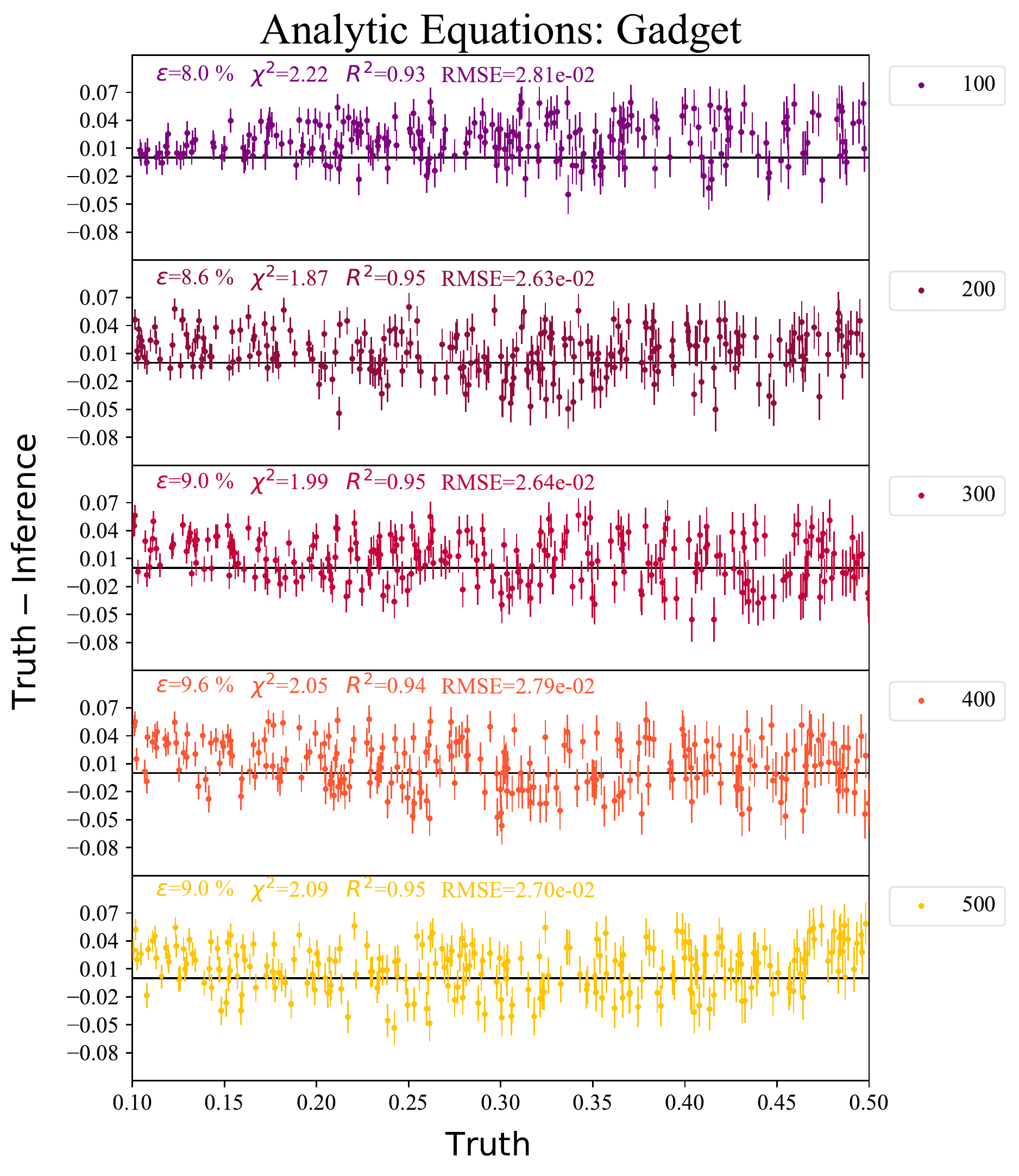}
    \includegraphics[width=.45\textwidth]{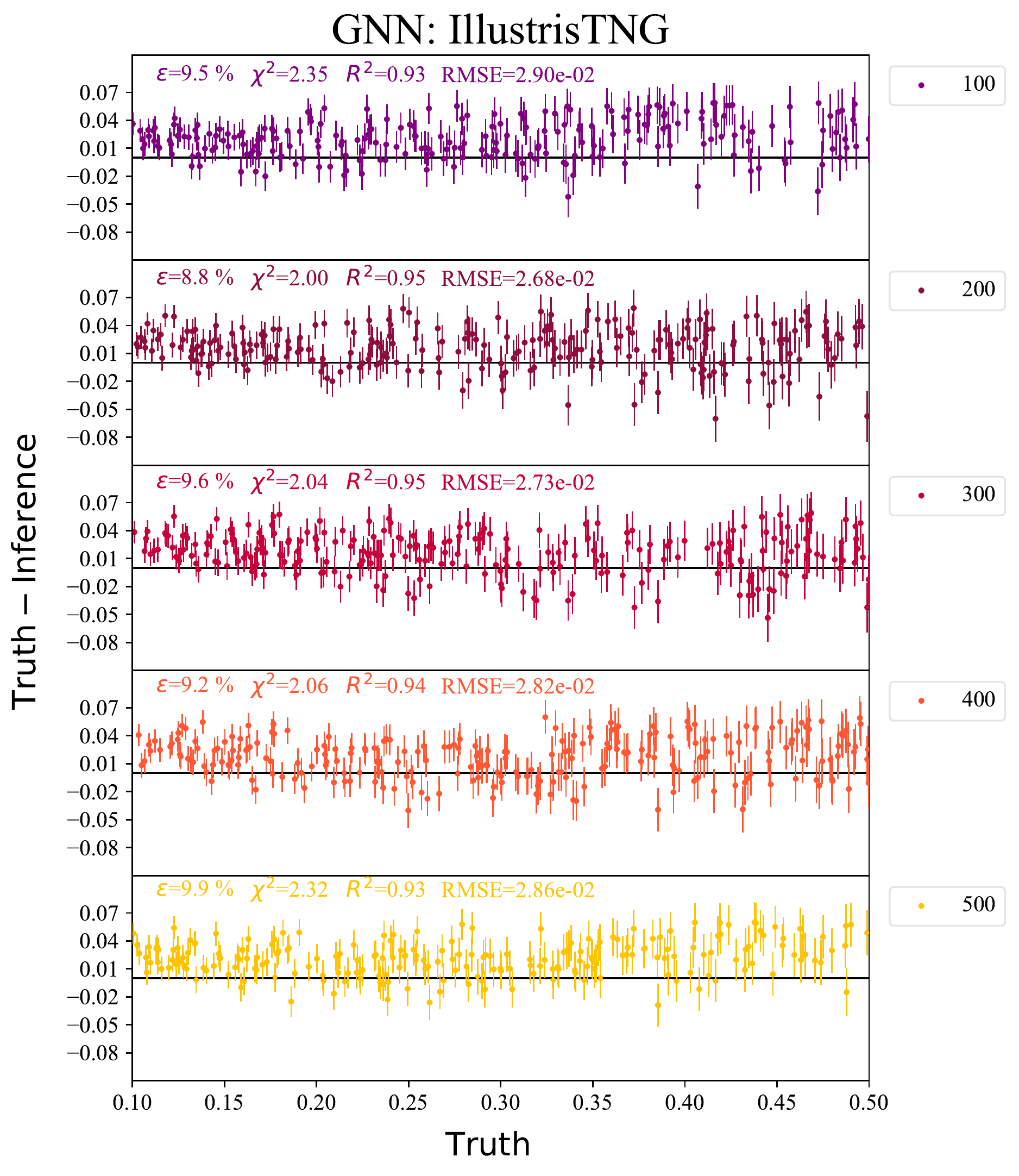}
    \includegraphics[width=.45\textwidth]{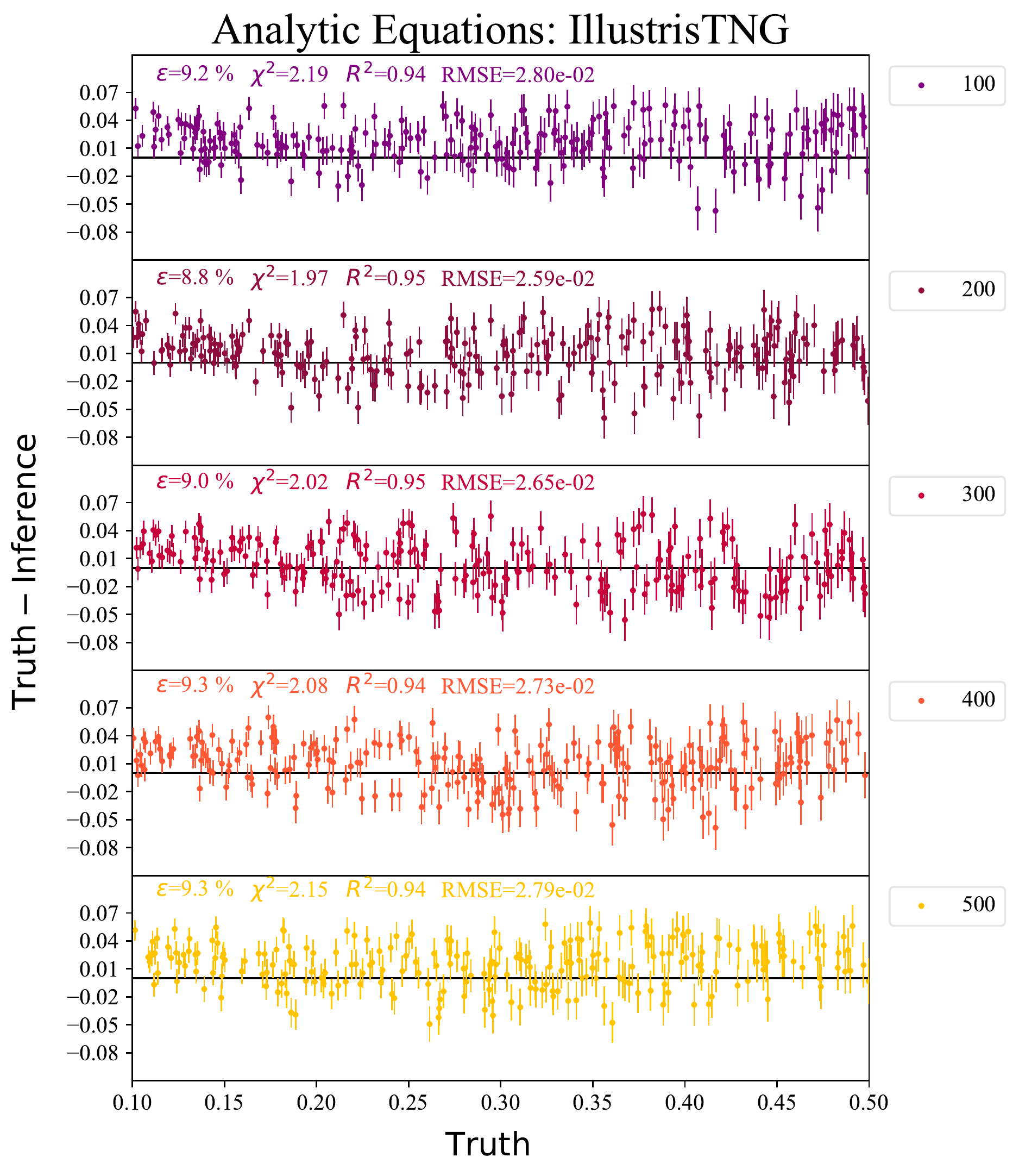}
    \caption{We trained a GNN using halo catalogues generated with the halo finder \textsc{Rockstar} to infer $\Omega_{\rm m}$, and approximated the learned model with analytic equations using symbolic regression. The top plots show the accuracy of the model and the analytic approximations when evaluated on halo catalogues from the $N$-body Gadget simulations using a different halo finder - \textsc{Subfind} - constructed with the varying minimum particle thresholds as described earlier. It is overall able to accurately extrapolate to the different halo finder with $\sim 8 \%$ mean relative error across the different catalogues. On the other hand, while the analytic expressions have a slightly larger error of $\sim 9 \%$, they do not exhibit the noticeable biases present in the predictions from the GNN, demonstrating the known improved extrapolation properties of analytic expressions over neural networks. The bottom plots depict the same test as above but for halo catalogues from the IllustrisTNG hydrodynamic simulations.}
    \label{fig:subfind}
\end{figure*}

First, we perform this test for the 1,000 halo catalogues from the $N$-body Gadget simulations. As it can be seen in the top plots of Fig.~\ref{fig:subfind}, both the GNN and the analytic expressions provide accurate predictions of $\Omega_{\rm m}$ overall, with mean relative errors of $\sim 8.8 \%$ and $\sim 9.2 \%$, respectively, across the different halo catalogues. However, there are some interesting features to emphasize. First, while the GNN predictions exhibit an offset and a significant lower-tail bias for the halo catalogue generated with a minimum particle threshold of 100, this is a boundary case considering the interval of the minimum particle thresholds used to construct the catalogues. Moreover, the identification of lower mass halos can vary across different halo finders and this can influence the predictions more strongly than the presence of more massive halos which are more likely to be commonly identified in both halo finders. Second, it can be seen that the analytic approximation demonstrate higher accuracy for this boundary case which is another indication of the better extrapolation capabilities of analytic equations over neural networks. 

Likewise, we performed the same test with halo catalogues from the IllustrisTNG hydrodynamic simulations. The results for this are shown in the bottom plots of Fig.~\ref{fig:subfind}. As it can be seen, the mean relative errors are similar between the GNN and the analytic expressions, averaging to be $\sim 9.5 \%$ across the different halo catalogues. While these metrics indicate slightly decreased precision of the predictions, this can be attributed to additional baryonic effects present. Nevertheless, the overall accuracy further demonstrates the generalization ability of the trained network as it is able to extrapolate to both additional hydrodynamic simulations and varying halo definitions. This agrees with the results discussed in Sections \ref{subsubsec:galaxies_eqns} and \ref{sec:gnn_galaxies_appendix}, where it was also found that both the network and the symbolic approximations are able to obtain robust predictions for catalogues generated from the SWIFT-EAGLE simulations which employ a different halo/subhalo finder (\textsc{VELOCIraptor}).

\section{Additional Plots: Testing GNN on Galaxies}\label{sec:gnn_galaxies_appendix}

\begin{figure*}
    \centering
    \includegraphics[width=0.9\textwidth]{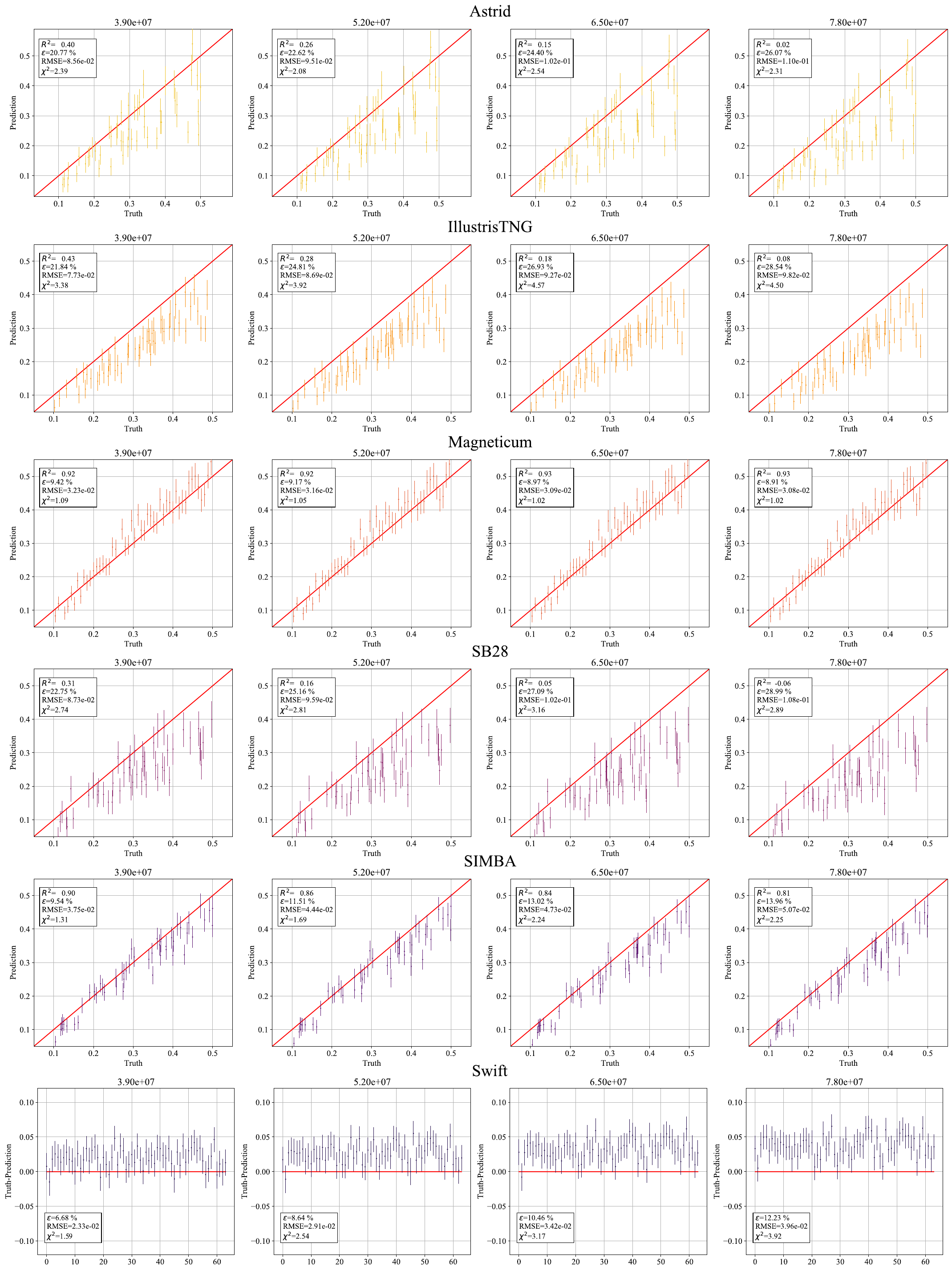}
    \caption{This plot shows the predictions of the GNN trained on halo catalogues from Gadget $N$-body simulations being tested on galaxies from six different hydrodynamic simulations as listed for each row. To construct the galaxy catalogues, we follow the procedure discussed in Section \ref{subsec:catalogues} and use four different stellar mass thresholds which are labeled for each column. For clarity, we plot the predictions for 50 randomly selected catalogues in each panel. As can be seen, the GNN is unable to accurately predict the values of $\Omega_{\rm m}$ as all the predictions exhibit a bias deviating from the true values. This is common across all simulations, which is expected given that there is a nontrivial connection between halo and galaxy distributions. However, as explained in Section \ref{subsubsec:galaxies_eqns}, the analytic equations that approximate the GNN can be easily tuned to avoid this error.}
    \label{fig:galaxies_gnn}
\end{figure*}

As discussed in Section \ref{subsubsec:galaxies_gnn}, we trained a GNN on halo catalogues and tested the learned network on galaxies from six different hydrodynamic simulation suites: Astrid, IllustrisTNG, Magneticum, SB28, Simba, and SWIFT-EAGLE. Here, we present the results for these predictions. As it can be seen in Fig.~\ref{fig:galaxies_gnn}, the GNN is unable to accurately predict the values of $\Omega_{\rm m}$ as all the predictions exhibit a bias deviating from the true values. This is common across all simulations, which is expected given that there is a nontrivial connection between halo and galaxy distributions. On the other hand, as explained in Section \ref{subsubsec:galaxies_eqns}, the analytic equations that approximate the GNN can be tuned to avoid this error.

We believe that these biases are due to the effects of the halo-galaxy connection in addition to the differences in the abundance of galaxies found in the catalogues used for testing and that of halos found in the training dataset. As discussed in Section \ref{subsubsec:galaxies_eqns}, the network is unable to extrapolate to number densities outside of the training range. In the case of galaxy catalogues, as shown in Fig.~\ref{fig:eqns_galaxies_color}, there are many catalogues with galaxy number densities that fall below the range of the halo number densities seen during training, $(1000, 6000)$. However, the under-predicted values of $\Omega_{\rm m}$ cannot be solely attributed to the abundance of galaxies. As discussed in our companion paper, \citet{deSanti_2023}, the full range of galaxy number densities is exhibited for all values of $\Omega_{\rm m}$. Hence, there is no strong correlation between $\Omega_{\rm m}$ and the number of galaxies in each catalogue. This agrees further demonstrates that the biases present in the network predictions are attributed to the intrinsic characteristics of the galaxy population.

% This agrees with the conclusion reached in Section \ref{sec:fixed_ndensity} that the biases present in the network predictions are attributed to the intrinsic characteristics of the galaxy population.

\section{Original Symbolic Regression Equations}\label{sec:sr_eqns}

\begin{figure*}
    \centering
    \includegraphics[width=0.45\textwidth]{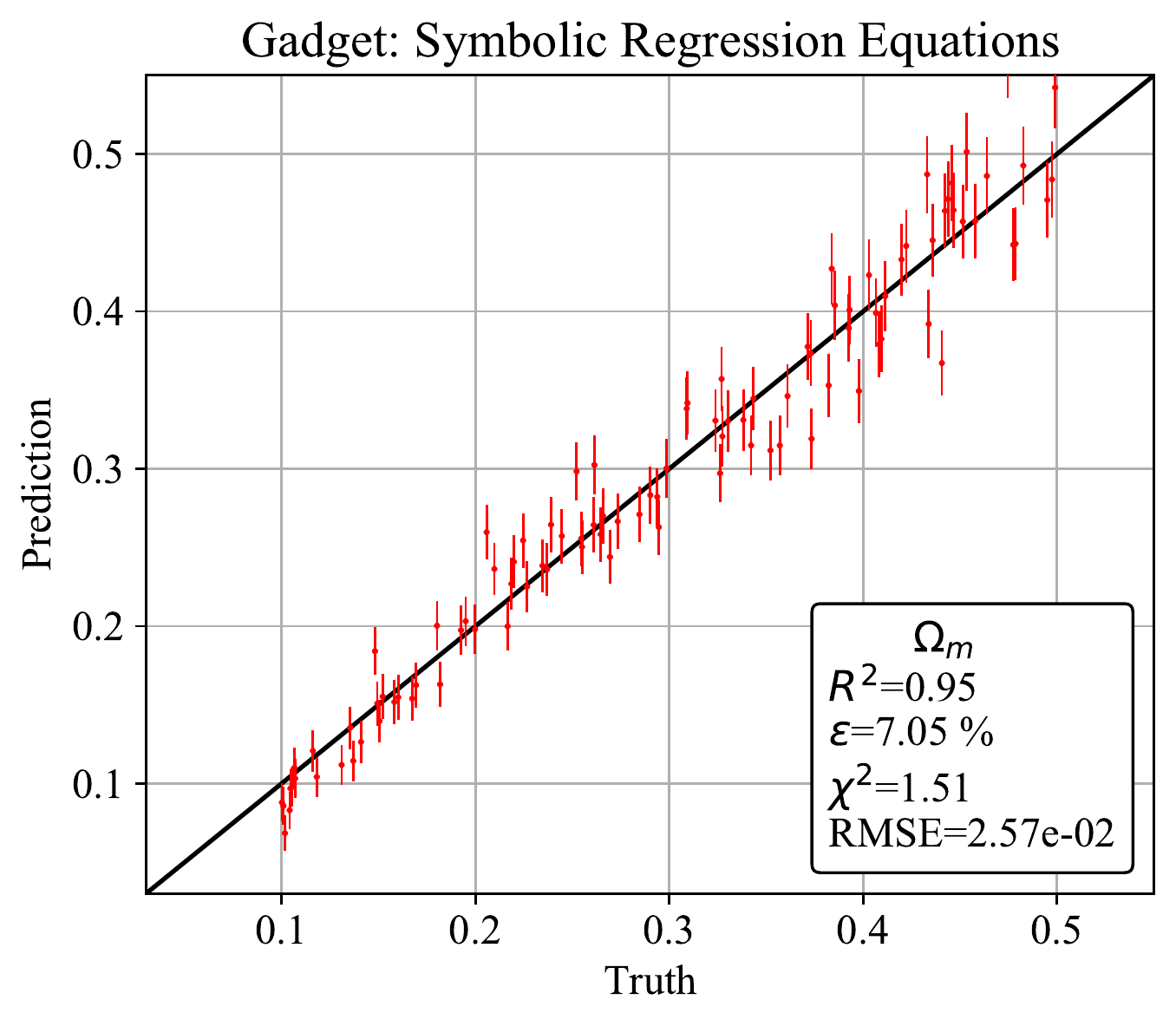}
    \caption{This plot shows the predictions of the original equations found by the symbolic regression algorithm evaluated on the halo catalogues of the Gadget test set. As can be seen, while it achieves similar accuracy to the GNN model, with a mean relative error of ~$7.1 \%$, it is not as accurate as the modified expressions which had an error of ~$6.6 \%$. See Fig. \ref{fig:gadget_individual}.}
    \label{fig:sr_eqns}
\end{figure*}

In Section \ref{sec:sr_eqns}, we presented the equations obtained by the symbolic regression algorithm that were then modified based on motivations of physical principles. Here, in Table \ref{tab:sr_eqns}, we present the equations originally found by the symbolic regression algorithm that we trained following the procedure described in Section \ref{subsec:SR_training}. Note that the edge equations found by the algorithm contained dependencies on the individual velocity modulus of halos in the form of linear combinations of $v_i$ and $v_j$. These terms break the parity between a halo and its neighbor. Moreover, we adapted terms that explicitly reflect differences between the velocity moduli due to the known statistics between pairwise velocities and $\Omega_{\rm m}$, 

We also show the accuracy of these equations when evaluated on halo catalogues from the Gadget test set simulations in Fig.~\ref{fig:sr_eqns}. As it can be seen, these formulas are able to achieve a mean relative error of $\sim 7.1 \%$ which is slightly higher than the error of the modified equations, possibly indicating that the imposed symmetries offer an important constraint on the predictions and play a significant role in achieving accurate inferences (see Fig. \ref{fig:gadget_individual}).

\begin{deluxetable}{lccc}\label{tab:sr_eqns}
\raggedleft
\tablewidth{0pt}
\tabletypesize{\footnotesize}
\tablecaption{This table follows the format of Table \ref{table:modified_eqns} and lists the original analytic formulas found by the symbolic regression algorithm. The key difference is that the edge model equations originally obtained by the algorithm depend on terms $v_i$ and $v_j$, which are the individual halo velocity moduli. As in the equations discussed in Section \ref{subsec:analytic_eqns}. the velocities used in the equations here have also been normalized to aid the model training and to ensure that they are dimensionless: $v_i = \frac{v_i - \mu}{\delta}$, $v_j = \frac{v_j - \mu}{\delta}$. For testing these equations on halo catalogues, we use the fixed values $\mu = 189$ km $\rm s^{-1}$ and $\delta = 129$ km $\rm s^{-1}$ computed from the mean and standard deviation of the velocity moduli for all halos in the training set. The accuracy of these equations are shown in Fig. \ref{fig:sr_eqns}}
\tablehead{
GNN Component & Formula & RMSE}
\startdata
Edge Model: $e^{(1)}_1$  & $1.32|1.05 v_i - v_j + 0.21| - 0.12v_j - 0.12(\gamma_{ij} + \beta_{ij} - 1.73)$ & 0.028 \\
Edge Model: $e^{(1)}_2$ & $|1.53(v_i - 1.06v_j) + 0.45| + 1.93(v_i - 1.02v_j) + 0.55$ & 0.035 \\
\hline
Node Model: $v^{(1)}_1$  & $1.21^{v_i}(0.77^{3.29 \sum_{j\in\mathcal{N}_j} e^{(1)}_1 + \sum_{j\in\mathcal{N}_j} e^{(1)}_2}) + 0.12$ & 0.02 \\
Node Model: $v^{(1)}_1 + v^{(1)}_2$ & $0.78 - \sqrt{\log(0.16 ^ {\sum_{j\in\mathcal{N}_j} e_2+\sum_{j\in\mathcal{N}_j} e_1 - 0.41v_i - 1.05})} + 1.45$ & 0.03 \\
\hline
Final MLP: $\mu$ & $4 \times 10^{-4} \times (-5.5 \sum_{i\in\mathcal{G}} v^{(1)}_2 + 2.21\sum_{i\in\mathcal{G}} v^{(1)}_1 + |0.96 \sum_{i\in\mathcal{G}} v^{(1)}_2 + 0.82\sum_{i\in\mathcal{G}} v^{(1)}_1|) - 0.103\ $ & 0.03\\
\hline
\enddata
\end{deluxetable}

\section{Varying stellar mass thresholds}\label{sec:varying_cuts}

In this section, we discuss the results for testing the analytic equations discussed in Section \ref{subsec:analytic_eqns} on galaxy catalogues constructed with different minimum stellar mass thresholds: $N \times m_*$ for $N \in \{3,4,5,6\}$ where $m_*$ is a fixed mass for a single stellar particle. As explained in Section \ref{subsec:catalogues}, the use of different mass cuts during training of the model and equations enables the models to marginalize over the halo/galaxy number densities found in each simulation due to different halo/galaxy mass functions. Here, we test whether the equations are robust to this using the simulations from ASTRID and SWIFT-EAGLE. To do this, we use the same $\delta$ values optimized for catalogues of the single mass threshold $4\times m_*$ as discussed in Section~\ref{subsubsec:galaxies_eqns}. 

\begin{figure*}
    \centering
    \includegraphics[width=1.0\textwidth]{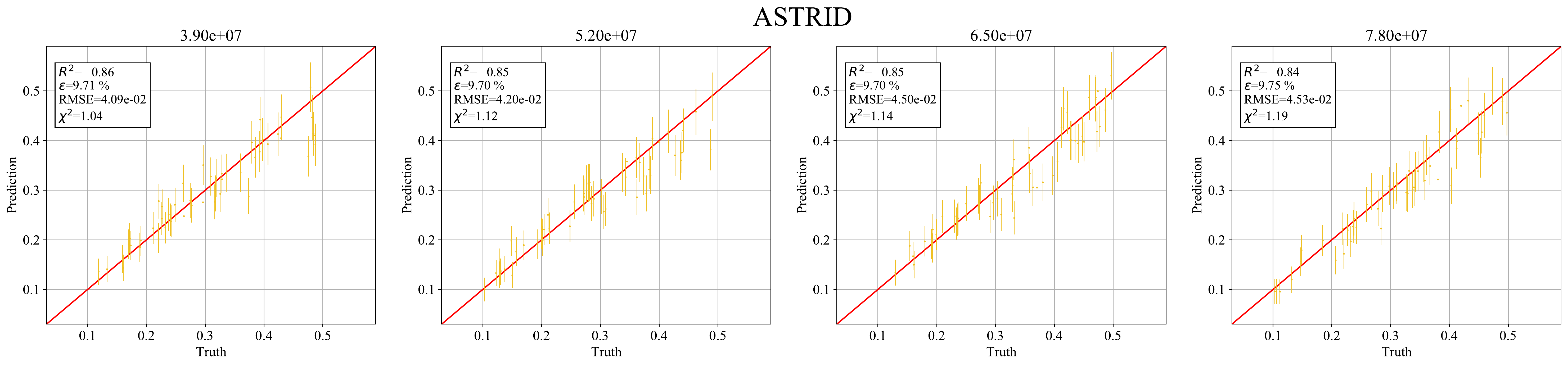}
    \caption{We evaluate the analytic equations discussed in Section \ref{subsec:analytic_eqns} on galaxy catalogues from the Astrid simulation set constructed using four different minimum stellar mass thresholds: $N \times m_*$ for $N \in \{3,4,5,6\}$ where $m_*$ is a fixed mass for a single stellar particle. Each column is labeled with the corresponding mass cut. As it can be seen, the accuracies of the equations are preserved for the different mass thresholds demonstrating that the model has marginalized over the number density of galaxies.}
    \label{fig:astrid_mass_cuts}
\end{figure*}

\begin{figure*}
    \centering
    \includegraphics[width=1.0\textwidth]{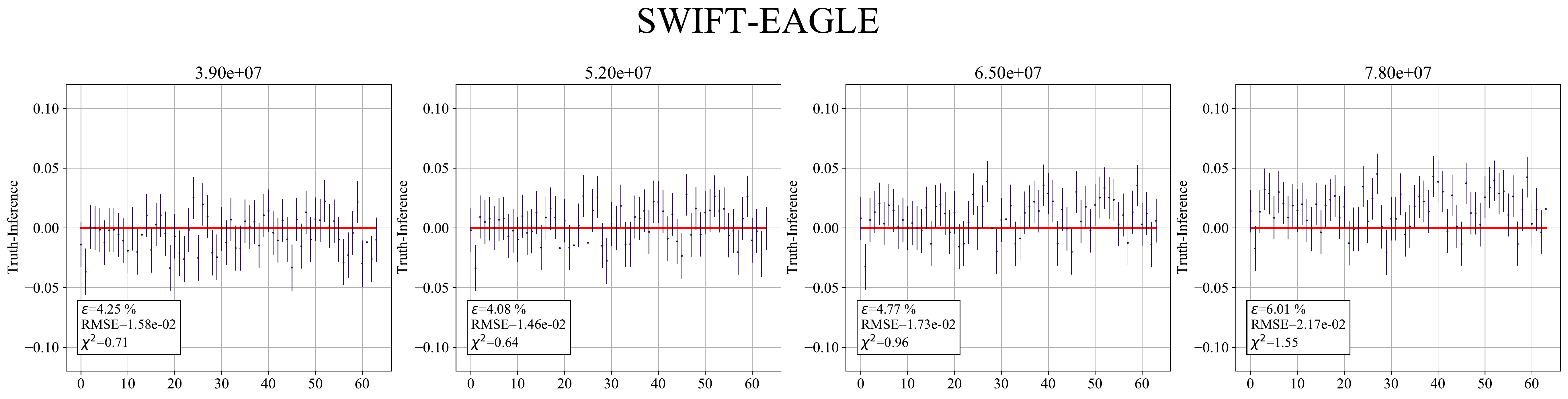}
    \caption{We evaluate the analytic equations discussed in Section \ref{subsec:analytic_eqns} on galaxy catalogues from the SWIFT-EAGLE simulation set constructed using four different minimum stellar mass thresholds: $N \times m_*$ for $N \in \{3,4,5,6\}$ where $m_*$ is a fixed mass for a single stellar particle. Each column is labeled with the corresponding mass cut. We note that since the SWIFT-EAGLE simulations were generated using the same initial random seed, there is a high correlation between the galaxy catalogues of the different stellar mass thresholds for this simulation set that is responsible for the trend of decreasing accuracy as the stellar mass threshold increases.}
    \label{fig:swifteagle_mass_cuts}
\end{figure*}

First, we present the results for ASTIRD, which are shown in Fig.~\ref{fig:astrid_mass_cuts}. As it can be seen, the accuracy of the equations are not largely affected by the different mass thresholds used, as expected. Second, we perform these tests for galaxy catalogues from SWIFT-EAGLE, as shown in Fig.~\ref{fig:swifteagle_mass_cuts}. Here, we explain the apparent trend of increasing scatter in the predicted $\Omega_{\rm m}$ values as the stellar mass threshold increases with the fact that all the SWIFT-EAGLE simulations were run with the same random seed. Hence, the predictions should be considered as highly correlated, which causes the small bias for the catalogue with a larger stellar mass threshold for a fixed $\delta$ value.

\section{Testing with Super-Sample Covariance}\label{sec:supersample_cov}

% {\color{purple} [Natalí: Why do not try to test this effect in another bigger DM halo box? Then your results can be even more power.]}

Here we demonstrate that the analytic equations discussed in Section \ref{subsec:analytic_eqns} are robust to the effects of super-sample covariance. Quantifying how the analytic equations behave in response to super-sample covariance is a critical step towards being able to apply them to observational data from surveys that are sampled with finite volume. This is because in galaxy surveys, the short-wavelength modes that contain information on the non-linear dynamics are coupled to long-wavelength, or super-sample, modes that extend beyond the survey volume \citep{Hamilton_2006, Hu_2003, Takada_2013}. This results in sample variances that dominate the non-linear regime \citep{Sato_2009, Takada_2007, Yu_2011, Putter_2012}. In the analysis that we have performed so far, we have not taken into consideration of this effect because we have used simulations with periodic boundary conditions which are not influenced by background modes that extend outside the simulation box.

To test for these effects, we evaluate the analytic equations on galaxy catalogues constructed from $(25~h^{-1}Mpc)^3$ sub-volumes randomly selected from the IllustrisTNG-300 simulation to match the size of the simulation boxes used for training. As described in Section \ref{subsubsec:hydro_codes}, this simulation has a total volume of $(205~h^{-1}Mpc)^3$ and was run with the cosmology $\Omega_{\rm m} = 0.3089$. It is important to note that unlike the simulations used for training, we do not impose periodic boundary conditions on the sub-volumes used in this test in order to account for the super-sample modes. 

\begin{figure*}
    \centering
    \includegraphics[width=1.0\textwidth]{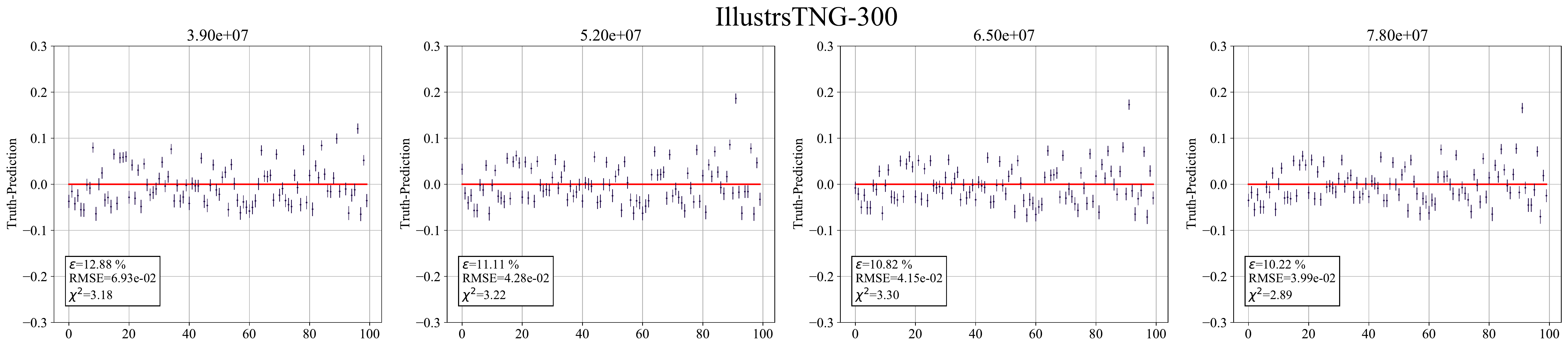}
    \includegraphics[width=1.0\textwidth]{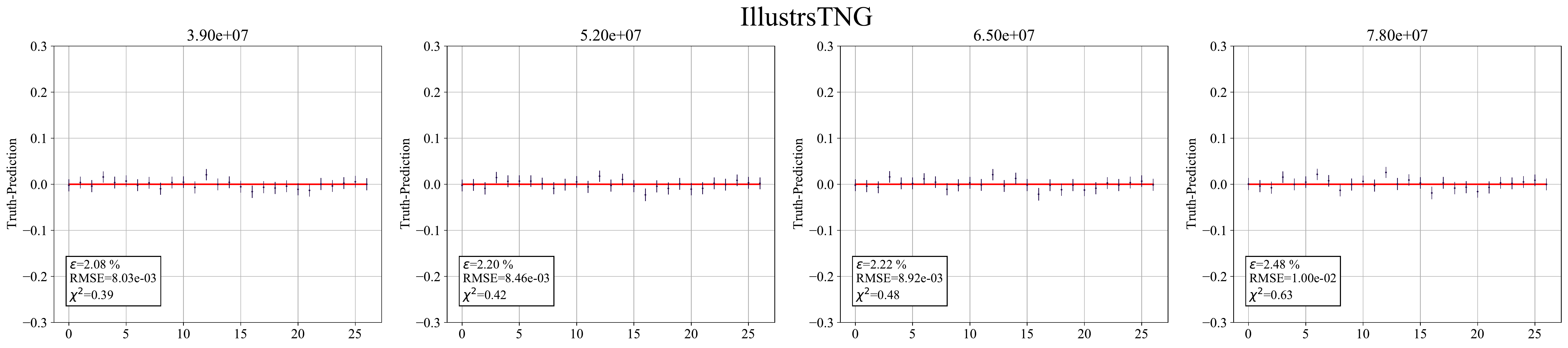}
    \caption{\textbf{Top:} We quantify the behavior of the analytic expressions, discussed in Section \ref{subsec:analytic_eqns}, in the presence of super-sample covariance. We test the analytic expressions on 100 $(25~h^{-1}Mpc)^3$ sub-volumes randomly selected from the IllustrisTNG-300 simulation without imposing periodic boundary conditions. The simulation contains a total volume of $(205~h^{-1}Mpc)^3$ and was run with a cosmology of $\Omega_{\rm m} = 0.308$. The plots depict the difference between the true $\Omega_{\rm}$ value and the predicted for the galaxy catalogues constructed using each of the four stellar mass thresholds as indicated at the top of each column. For all catalogues, the predictions are corrected for their negative offset from the truth by introducing a bias in the analytic expression for the final MLP, $b = -0.19$, which shifts all predictions upwards by a constant. After adjusting for this common offset, the predictions exhibit mean relative errors of $\sim 11.1 \%$, comparable to the predictions for galaxy catalogues from other hydrodynamic simulation codes as discussed in Section \ref{subsubsec:galaxies_eqns}. The common offset can be attributed to the fact that the equations were trained only on simulation volumes of $(25~h^{-1}Mpc)^3$ with periodic boundary conditions and are now being tested on simulations without such conditions. We confirm this reasoning with the results shown in the bottom panel. \textbf{Bottom:} This panel follows the same format as the one above. We show that the analytic equations behave similarly when evaluated on 27 IllustrisTNG simulations from the CV set (see Section \ref{sec:data}) after removing periodic boundary conditions. These simulations were run with the same cosmology $\Omega_{\rm m} = 0.308$. All predictions for galaxy catalogues constructed from these simulations possess a negative offset equal to the one found for the predictions from IllustrisTNG-300 sub-volumes, which was adjusted for by introducing a bias to the final MLP equation: $b = -0.19$. After doing so, the predictions exhibit only a mean relative error of $\sim 2.2\%$. These results indicate that the analytic equations  are able to account for the effects of super-sample covariance if one simply shifts the predictions by a constant bias, $b$, due to the presence of periodic boundary conditions in the training data.}
    \label{fig:ssc}
\end{figure*}

We present the results of this test in the top panel of Fig.~\ref{fig:ssc}. Each plot in the figure depicts the differences between the truth and predicted $\Omega_{\rm m}$ made by the analytic equations for 100 randomly selected sub-volumes. Following the same procedure used to perform the previous tests on galaxies, we construct four catalogues for each sub-volume using the stellar mass thresholds discussed in Section \ref{subsec:catalogues}. Each column is thus labeled with the corresponding stellar mass cut used. We note that the predictions across all catalogues exhibit a common offset from the truth which we account for by introducing to the final MLP equation an additive constant of $b = -0.19$ found using $\chi ^2$ minimization. This common bias is explained by the fact that the equations are being evaluated on sub-volumes that do not contain periodic boundary conditions but were trained only on simulations that contain periodic boundary conditions. After correcting for this, the analytic expressions are able to achieve mean relative errors of $\sim 11.1 \%$.

To confirm that this offset is indeed the consequence of the removal of periodic boundary conditions, we evaluate the analytic expressions on galaxy catalogues constructed from the 27 IllustrisTNG simulations of the CV set as described in Section \ref{sec:data}. These simulations were run with the same cosmology of $\Omega_{\rm m} = 0.3$. We present the results for these simulations in the lower panel of Fig.~\ref{fig:ssc}, which follow the same format the one above. We find that the predictions for these simulations possess the same offset found in the IllustrisTNG-300 sub-volumes. After correcting for this with the bias parameter, $b = -0.19$, in the analytic expressions, we achieve mean relative errors of $\sim 2.2 \%$. This indicates that the offset in the predictions are attributed to the fact that the equations were trained using periodic boundary conditions. This result agrees with the findings of our companion paper, \citet{deSanti_2023}, where a similar offset was found that is common to all predictions made by a GNN model trained on simulations with periodic boundary conditions and tested on simulations without it. Hence, we conclude that the analytic expressions are able to take into account of the effects due to super-sample covariance which is key for applying them to observational data from surveys that contain finite volume. 

\end{document}